\documentclass[12pt]{article}
\usepackage{epsfig,cite}
\usepackage{amssymb,amsmath,amscd}

\catcode`\@=11

\def \be  {\begin{equation}}
\def \ee  {\end{equation}}
\def \ba  {\begin{eqnarray}}
\def \ea  {\end{eqnarray}}
\def \bb  {\begin{thebibliography}}
\def \eb  {\end{thebibliography}}
\def \lab #1 {\label{#1}}
\newcommand\PT{\mathbb{PT}}
\newcommand\cA{\mathcal{A}}
\newcommand\cB{\mathcal{B}}

\newcommand\cH{\mathcal{H}}
\newcommand\cJ{\mathcal{J}}
\newcommand\cL{\mathcal{L}}

\newcommand\cO{\mathcal{O}}

\newcommand\T{\mathbb{T}}
\newcommand\cN{\mathcal{N}}
\newcommand\C {\mathbb{C }}
\newcommand\R {\mathbb{R }}
\newcommand\bP{\mathbb{P}}
\newcommand\CP {\mathbb{CP}}
\newcommand\RP {\mathbb{RP}}

\newcommand\rd{\mathrm{d}}
\newcommand\rD{\mathrm{D}}
\newcommand\e{\mathrm{e}}
\newcommand\im{\mathrm{i}}
\newcommand\la{\langle}
\newcommand\ra{\rangle}
\newcommand\del{\partial}
\newcommand\delbar{\bar{\partial}}
\newcommand\MHVbar{\overline{\mbox{MHV}}}
\newcommand\sgn{\mathrm{sgn}}

\begin{document}


\thispagestyle{empty}

\begin{center}

{\Huge\bf
	 Twistor-Strings, Grassmannians\\ 
\vskip 0.25truecm 
	and Leading Singularities
}\\

\vskip 1truecm

{\bf
	Mathew Bullimore$^1$, Lionel Mason$^2$ and David Skinner$^3$ \\
}

\vskip 0.4truecm

$^1${\it
	Rudolf Peierls Centre for Theoretical Physics,\\
	1 Keble Road, Oxford, OX1 3NP,\\
	United Kingdom}\\
	
\vskip 0.4truecm

$^2${\it 
	Mathematical Institute,\\ 
	24-29 St. Giles', Oxford, OX1 3LB,\\ 
	United Kingdom}\\

\vskip 0.4truecm 

$^3${\it 
	Perimeter Institute for Theoretical Physics,\\ 
	31 Caroline St., Waterloo, ON, N2L 2Y5,\\
	Canada}\\

\end{center}

\vskip 1truecm 

\centerline{\bf Abstract} 

\bigskip 

We derive a systematic procedure for obtaining explicit, $\ell$-loop leading singularities of planar $\cN=4$ super Yang-Mills scattering amplitudes in twistor space directly from their momentum space channel diagram.  The expressions are given as integrals over the moduli of connected, nodal curves in twistor space whose degree and genus matches expectations from twistor-string theory.  We propose that a twistor-string theory for pure $\cN=4$ super Yang-Mills --- if it exists --- is determined by the condition that these leading singularity formul\ae\ arise as residues when an unphysical contour for the path integral is used, by analogy with the momentum space leading singularity conjecture. We go on to show that the genus $g$ twistor-string moduli space for $g$-loop N$^{k-2}$MHV amplitudes may be mapped into the Grassmannian G$(k,n)$. For a leading singularity, the image of this map is a $2(n-2)$-dimensional subcycle of G$(k,n)$ and, when `primitive', it is of exactly the type found from the Grassmannian residue formula of Arkani-Hamed, Cachazo, Cheung \& Kaplan. Based on this correspondence and the Grassmannian conjecture, we deduce restrictions on the
possible leading singularities of multi-loop N$^p$MHV amplitudes. In particular, we argue that no new leading singularities can arise beyond $3p$ loops.


\newpage

\setcounter{page}{1}
\setcounter{footnote}{0}


\section{Introduction}
\label{sec:intro}

We begin with a puzzle. According to twistor-string theory~\cite{Witten:2003nn}, tree level N$^p$MHV amplitudes in $\cN=4$ super Yang-Mills theory are supported on holomorphic curves of degree
\begin{equation}
	d=p+1
\label{twstree}
\end{equation}
and genus zero in dual\footnote{With Penrose conventions for twistor space, MHV amplitudes --- those whose `pure glue' sector involves two \emph{negative} and arbitrarily many \emph{positive} helicity gluons --- are supported on holomorphic lines in \emph{dual} twistor space $\PT^*$. We abuse notation by taking $\PT^*$ to be variously a copy of $\CP^3$, $\CP^{3|4}$ or the neighbourhood of a line in either of these spaces, according to context. We will often describe $\PT^*$ in terms of its homogeneous coordinates $W_\alpha=(\lambda_A,\mu^{A'})$ and $\chi_a$ in the supersymmetric case. This space was called \emph{twistor} space by Witten in~\cite{Witten:2003nn}.} projective twistor space $\PT^*$. By systematically exploiting the recursion relations of Britto, Cachazo, Feng \& Witten~\cite{Britto:2004ap,Britto:2005fq} (organised so as to exhibit dual superconformal invariance~\cite{Drummond:2008vq}), Drummond \& Henn~\cite{Drummond:2008cr} obtained analytic expressions for all such tree amplitudes in momentum space. Working in ultrahyperbolic space-time signature and using Witten's `half Fourier transform'~\cite{Witten:2003nn},  the $\PT^*$ support of this solution was investigated by Korchemsky \& Sokatchev~\cite{Korchemsky:2009jv}, following the earlier work of~\cite{Hodges:2005bf, Hodges:2005aj, Mason:2009sa, ArkaniHamed:2009si} studying BCFW recursion directly in either $\PT^*$ or ambitwistor space.  Korchemsky \& Sokatchev show that each term in the Drummond \& Henn solution for an N$^p$MHV tree amplitude is supported on an arrangement of $2p+1$ intersecting lines in dual twistor space --- a connected (albeit reducible) curve of degree
\begin{equation}
	d=2p+1\ ,
\label{KSdeg}
\end{equation}
in stark contrast to the twistor-string result~\eqref{twstree}.

The puzzle is solved by a more careful examination of the line configurations found in~\cite{Korchemsky:2009jv}; we show that the $2p+1$ intersecting lines in fact form a curve of genus $p$, suggesting that each summand in the solution of~\cite{Drummond:2008cr} is most naturally associated with a $p$-loop amplitude, rather than the tree. Indeed, this interpretation is compatible with the origin~\cite{Britto:2004nc,Britto:2004ap} of the BCF(W) relations from demanding consistency between the infra-red behaviour of planar loop amplitudes and its expansion in terms of a basis of scalar integrals.  In section~\ref{sec:KS} we demonstrate that these individual BCFW contributions to the tree amplitude are really \emph{leading singularities}~\cite{Britto:2004nc,Buchbinder:2005wp,Cachazo:2008dx,Cachazo:2008vp,Cachazo:2008hp} of $p$-loop N$^p$MHV amplitudes, in a momentum space channel that is obtained from knowledge of the $\PT^*$ support. This generalises the well-known fact~\cite{Bern:2004bt, Drummond:2008bq} that the NMHV tree amplitude can be written as a sum of leading singularities of the 1-loop amplitude (where the leading singularities are box coefficients). Returning to the twistor-string, the same calculation that yields~\eqref{twstree} at tree level also states that $\ell$-loop amplitudes are supported on holomorphic curves of degree
\begin{equation}
	d=p+1+\ell\qquad\hbox{and genus}\  \leq \ell
\label{twsloop}
\end{equation}
in $\PT^*$, in complete agreement with the line configurations of~\cite{Korchemsky:2009jv} for $g=p$. Thus, treating each term as a leading singularity resolves the apparent conflict between twistor-strings and the $\PT^*$ support of the Drummond \& Henn expression. 

In section~\ref{sec:twsgenunit} we consider leading singularities in $\PT^*$ more generally. We show that their structure can be understood quite systematically, and obtain a formula that shows how twistor space leading singularities are constructed from gluing together constituent tree subamplitudes, closely reflecting their structure in momentum space. Pleasingly, the gluing formula is simply the standard $\PT^*$ inner product between the two subamplitudes on either side of the cut (see {\it e.g.}~\cite{Penrose:1972ia}). This method yields the twistor support of all the leading singularities we are aware of in the literature, including 1-loop box coefficients~\cite{Britto:2004nc,Bern:2004bt}, leading singularities of multi-loop MHV amplitudes~\cite{Bern:1997nh,Bern:2006vw,Bern:2005iz,Bern:2006ew,Bern:2007ct,Buchbinder:2005wp,Cachazo:2008dx, Cachazo:2008hp}, and the 2-loop NMHV and N$^2$MHV leading singularities uncovered in~\cite{ArkaniHamed:2009dn}. Many of these leading singularities are of a special subclass that we call \emph{primitive}; an $\ell$-loop primitive leading singularity is defined by having precisely $4\ell$ distinguished propagators in the momentum space channel diagram, and tree subamplitudes that are exclusively MHV or 3-point $\MHVbar$. They include all the KS configurations. A leading singularity containing subamplitudes of higher degree may be decomposed into such primitive ones by the use of the Drummond \& Henn expression for N$^p$MHV trees with $p\geq1$.   We show explicitly how to obtain integral formulae for such primitive leading singularities and how they reduce to those given by Korchemsky \& Sokatchev for the KS figures.

General leading singularities may be associated with particular codimension $4g$ boundary components of the moduli space $\overline{M}_{g,n}(\PT^*,d)$ of degree $d$ holomorphic maps from a genus $g$ worldsheet --- the moduli space over which one takes the twistor-string path integral, reviewed in section~\ref{sec:tws}. These boundary components correspond to maps from nodal worldsheets whose images in $\PT^*$ are exactly the nodal curves constructed in section~\ref{sec:twsgenunit} (this also connects with the work of Vergu~\cite{Vergu:2006np}, who showed that multiparticle singularities of tree level amplitudes correspond to nodal twistor-strings at genus zero). It is striking that the $\PT^*$ support of {\it $g$-loop} leading singularities agrees with the twistor-string prediction~\eqref{twsloop}, despite the fact that the twistor-string models of~\cite{Witten:2003nn, Berkovits:2004hg, Mason:2007zv} contain conformal supergravity~\cite{Berkovits:2004jj}.   Furthermore, generalised unitarity in twistor space leads to formulae for leading singularities in terms of integrals of explicit volume forms over the moduli space of such nodal curves.  Such integrals are not clearly defined for twistor-string theory.

A long term aim underlying this work --- as yet unrealized --- is the
construction of a twistor-string theory for pure $\cN=4$ super
Yang-Mills, {\it i.e.}, without contributions from conformal
supergravity (see~\cite{Dolan:2007vv} for a loop calculation in
twistor-string theory that is expected to include conformal
supergravity). A key part of the task of building such a theory is the
construction of an appropriate top holomorphic form on certain moduli
spaces of line bundles over Riemann
surfaces~\cite{Witten:2004cp,Movshev:2006py}. We argue that the
ability to recover the correct leading singularities of $\cN=4$ SYM is
a key criterion for any such choice. Just as one computes leading
singularities in momentum space by choosing a contour that localises
the loop integral on a discrete set of solutions where various
propagators in the loop are forced to be
on-shell~\cite{Cachazo:2008dx}, we expect that one computes $g$-loop
leading singularities in twistor-string theory by choosing a contour
that localises the genus $g$ worldsheet path integral on codimension
$4g$ boundary components of $\overline{M}_{g,n}(\PT^*,d)$. In order
for this localisation to be possible, the integrand of the path
integral must have simple poles on these boundaries. Remarkably,
Gukov, Motl \& Neitzke~\cite{Gukov:2004ei} showed this is true for the
single trace contribution\footnote{In other words, at $g=0$ the
  requirement that one keeps only the single-trace term is equivalent
  to the requirement that one keeps only the contribution to the path
  integral with a simple pole on the boundary of moduli space.} to the
$g=0$ path integral of the original formulations of twistor-strings,
in the context of relating the twistor-string to the MHV
formalism~\cite{Cachazo:2004kj}.  However, at higher genus it is not
clear whether the original twistor-string path integral is
well-defined even by physics standards\footnote{For example, $\cN=4$
  conformal supergravity is known to be anomalous unless coupled to
  $\cN=4$ SYM with a dimension 4 gauge
  group~\cite{Fradkin:1985am}. Berkovits \&
  Witten~\cite{Berkovits:2004jj} suggested that this stringent
  requirement on the gauge group could be made compatible with the
  central charge condition by including a copy of the monster CFT on
  the twistor-string worldsheet, but it is not known whether this
  actually leads to a consistent theory. More generally, it seems
  unlikely that $\cN=4$ conformal supergravity is ever unitary. If
  not, should an equivalent, well-defined string theory exist?}, much
less whether one can decouple conformal supergravity. It is more
practical to attempt to reverse-engineer a twistor-string theory for
pure $\cN=4$ SYM from the known leading singularities.

Following on from recent developments in the twistor description of leading singularities, a key handle on the structure of any twistor-string theory should be a dual description in terms of the Grassmannian G$(k,n)$ of $k$-planes in $\C^n$. 
This is because G$(k,n)$ is the arena for a conjecture of Arkani-Hamed, Cachazo, Cheung \& Kaplan~\cite{ArkaniHamed:2009dn}, that followed earlier work~\cite{Hodges:2005bf, Hodges:2005aj, Mason:2009sa, ArkaniHamed:2009si} on interpreting the BCFW recursion procedure in twistor space. Arkani-Hamed {\it et al.} claim that \emph{all-loop} leading singularities of planar N$^{k-2}$MHV amplitudes in $\cN=4$ SYM may be obtained from the residues of a certain meromorphic form on G$(k,n)$, localised on various $2(n-2)$-dimensional subcycles. We explore this duality in section~\ref{sec:twsgrass}, where we show that the genus $g$, N$^{k-2}$MHV twistor-string moduli space has a natural map onto  G$(k,n)$, generalising the map studied by Spradlin \& Volovich~\cite{Spradlin:2009qr} and Dolan \& Goddard~\cite{Dolan:2009wf} at genus zero. If one restricts the twistor-string to the boundary components of $\overline{M}_{g,n}(\PT^*,d)$ that describe leading singularities, the image of the map to G$(k,n)$ is a $2(n-2)$-dimensional subcycle. Restricting further to \emph{primitive} leading singularities, we find precisely the cycles that arise in the work of~\cite{ArkaniHamed:2009dn}.
This also shows that the Grassmannian residue formula does indeed contain all leading singularities of KS type.

We then return to our starting point by showing how the choice of $2(n-2)$ cycle in G$(k,n)$ determines the support of the leading singularity in $\PT^*$. Combined with the results of the first part of the paper, this gives a simple means to relate leading singularities to specific subcycles of the Grassmannian without performing a detailed residue calculation. As it turns out, the contour choices that yield KS configurations ({\it i.e.} $p$-loop leading singularities appearing in the tree amplitude) are not generic; more general contour choices are shown to correspond to leading singularities of higher-loop amplitudes. This generalises the observations of~\cite{ArkaniHamed:2009dn}, who found that certain residues of their integral formula correspond to leading singularities such as the N$^2$MHV four-mass box coefficient and certain channels in the 2-loop NMHV amplitudes with up to eight external states --- these do not contribute to the tree amplitude, and motivated the conjecture~\cite{ArkaniHamed:2009dn} that the Grassmannian residue formula in fact contains all-loop information. It is straightforward to classify all that can happen at NMHV: assuming the conjecture of~\cite{ArkaniHamed:2009dn}, all leading singularities of arbitrary loop, $n$-particle NMHV amplitudes are determined in terms of their leading singularities at
\begin{itemize}
	\item 1 loop when $n\leq7$,
	\item 2 loops when $7< n<10$ and
	\item 3 loops when $n\geq10$.
\end{itemize}
(The first of these conditions appeared in~\cite{ArkaniHamed:2009dn}.)
More generally, we analyze the structure of generic $2(n-2)$-cycles in the Grassmannian in relation to their twistor support, providing evidence that leading singularities of all-order N$^p$MHV amplitudes are likewise determined in terms of their primitive leading singularities at $3p$ loops and under.  We argue that this restriction arises from the conjecture that, given the MHV and $\MHVbar_3$ tree amplitudes, the twistor support of a primitive leading singularity determines the leading singularity itself.  The conjecture translates into the statement that the geometry of the twistor support cannot be extended by adding further loops beyond $3p$.

Finally, in section~\ref{sec:conclusions} we summarise our work and
discuss some of the many open questions, such as the relationship of
the meromorphic volume form on the Grassmannian to an as-yet-undefined
twistor-string theory for pure $\cN=4$ super Yang-Mills, the role of
infra-red relations, and how they might be understood in this context.


\subsection{On the Choice of Space-Time Signature}
\label{sec:signs}

The Penrose transform~\cite{Penrose:1968me,Eastwood:1981jy} equates solutions of the massless, free field equations for helicity $h$ fields on regions of conformally compactified, complexified space-time with cohomology classes on regions of  twistor space (or dual twistor space). In particular, an on-shell $\cN=4$ Yang-Mills supermultiplet may be represented by a dual twistor field
$$
	a(W,\chi) = g^+(W) + \chi_a \Gamma^a(W) + \frac{1}{2!}\chi_a\chi_b\Phi^{ab}(W) 
	+\frac{1}{3!}\epsilon^{abcd}\chi_a\chi_b\chi_c \tilde\Gamma_d(W) 
	+ \frac{1}{4!}\epsilon^{abcd}\chi_{a}\chi_{b}\chi_{c}\chi_{d}g^-(W)\ ,
$$
where the bosonic components are (Lie algebra valued) elements of  $H^1(\PT^*,\cO(2h-2))$ for $h=1,\frac{1}{2},0,-\frac{1}{2},-1$ respectively, where $\cO(n)$ is the sheaf of holomorphic functions, homogeneous of degree $n$. The portion of $\PT^*$ relevant for $(++--)$ space-time signature is the real slice $\RP^3$, and the Penrose transform becomes~\cite{Atiyah:1979iu} the X-ray transform~\cite{FritzJohn} upon restriction to this real slice. The X-ray transform may be composed~\cite{Mason:2009sa} with the usual Fourier transform from $(++--)$ space-time to momentum space, giving Witten's `half Fourier transform'~\cite{Witten:2003nn} in which $\mu^{A'}$ and $\tilde\lambda_{A'}$ are canonically conjugate. The half Fourier transform directly relates $\PT^*$ fields to fields on the light-cone in momentum space. It is useful because it avoids the redundancy inherent in picking representatives of cohomology classes --- a process that often breaks symmetry. 

By itself, working in an unphysical signature is not too serious a problem at tree level, where momentum space amplitudes are rational functions that extend uniquely over the whole of complexified momentum space. The same applies even at loop level if one merely wishes to translate the \emph{integrand} of a loop expression to a differential form on twistor space. However, if we adhere strictly to the half Fourier transform, the twistor formul\ae\ are cluttered with many conformal symmetry breaking sign functions, both local ({\it e.g.} $\sgn \la \lambda_1\lambda_2\ra$) and non-local (corresponding to sgn$[\tilde \lambda_1\tilde\lambda_2]$ on momentum space).  The non-local signs obscure the $\PT^*$ support of amplitudes or leading singularities (see {\it e.g.}~\cite{Mason:2009sa,Korchemsky:2009jv}).  

Fortunately, it has become increasingly clear that the formul\ae\ obtained by \emph{ignoring} the signs are in fact physically correct when re-interpreted as contour integrals, with $\delta$-functions replaced by Cauchy poles. Firstly, in the original twistor-string calculations~\cite{Roiban:2004vt, Roiban:2004yf}, such a viewpoint was essential in order to obtain the correct tree amplitude: the integral involves contributions from $\delta$-functions of quantities with complex roots, and these roots must be included --- without any sign factors --- if the correct amplitude is to be recovered. Secondly, leading singularities are associated with contour integrals~\cite{Cachazo:2008dx} even if one works purely in momentum space: there are no real Lorentzian or Euclidean solutions to the four cut conditions and although real solutions in $(++--)$ signature exist, one discards a modulus sign that inevitably arises if the propagators are replaced by $\delta$-functions rather than treated as poles in a contour integral. Finally, the Grassmannian residue formula of~\cite{ArkaniHamed:2009dn} obtains leading singularities using a choice of contour that is largely unaffected by whether one chooses to represent external states in momentum space or in twistor space. In these treatments, the sign factors \emph{do} in fact play a role ---  not as part of the integrand, but as data that determines the {\v C}ech cohomology class of this form and thereby helps determine the appropriate contour. This viewpoint is shared by the twistor diagrams of Hodges~\cite{Hodges:2005bf, Hodges:2005aj} where application to Lorentz signature is always to the fore, at the price of dealing with contours that are awkward to specify explicitly.

In this paper we ignore the sign factors, with the understanding that
all our integrals are to be treated as contour integrals.  There
remain a number of questions to be resolved in order to put these
contour integrals on a firmer foundation, both in the characterization
of the contour and the proper cohomological interpretation.  However,
for the leading singularities that are the main focus of this paper,
the computational procedures are clear and these deeper issues are
beyond the scope of the current paper.


\section{Twistor Support of All BCFW Terms}
\label{sec:KS}

We begin by reviewing the results of Korchemsky \&
Sokatchev~\cite{Korchemsky:2009jv} that describe the $\PT^*$ support
of the Drummond \& Henn solution~\cite{Drummond:2008cr} for all tree
amplitudes in $\cN=4$ SYM, obtained from the BCFW recursion
procedure~\cite{Britto:2005fq}. The diagrams
in~\cite{Korchemsky:2009jv} were drawn so as to show simultaneously
both the twistor support of an amplitude, and also the associated
$\alpha$- and $\beta$-planes in space-time. We re-draw their diagrams
in a way that clarifies the $\PT^*$ geometry (although we thereby suppress
the space-time picture). Through so doing, it becomes
clear that each summand in the N$^p$MHV tree amplitude (as
expressed in~\cite{Drummond:2008cr}) is supported on a connected,
nodal curve that not only has degree $2p+1$, but also has genus
$p$. This is natural once one realises that each of these summands has its own identity as a leading singularity of a $p$-loop amplitude, generalising the well-known cases of $p=0,1$ and in line with the 1-loop origin~\cite{Britto:2004nc,Britto:2004ap} of the BCF(W) recursion relations. We give a simple method to identify the relevant 
momentum space channel for the leading singularity directly from the twistor support.

In the following section we consider the construction of leading singularities in $\PT^*$ more
systematically, and will show how all the KS configurations of the
present section --- complete with their explicit integral forms ---
may be straightforwardly recovered.


\subsection{The Three Particle $\overline{\rm MHV}$ Amplitude}
\label{sec:MHVbar3}

In many ways, the most basic amplitude is the three point $\overline{\rm MHV}$. It is the only non-zero amplitude present in the purely self-dual sector of the $\cN=4$ theory and arises~\cite{Boels:2006ir} from the interaction of a holomorphic Chern-Simons theory in $\PT^*$, so is local in $\PT^*$ and has $d=0$. Explicitly, the colour-stripped amplitude for external states localised at fixed points in $\PT^*$ is
\be
\begin{aligned}
	\cA_{\overline{\rm MHV}}(W_1,W_2,W_3)&=\int_{\PT^*}\rD^{3|4}W\ 
	\delta^{3|4}(W;W_1)\,\delta^{3|4}(W;W_2)\,\delta^{3|4}(W;W_3)\\
	&=\  \delta^{3|4}(W_1;W_2)\,\delta^{3|4}(W_1;W_3)\ ,
\end{aligned}
\label{MHVbartws}
\ee
where\footnote{The $\delta$-functions in this and similar formul\ae\
 may be interpreted as Cauchy poles with integration contour an
 appropriate closed curve surrounding the pole, or $\delbar(1/z)$ in a Dolbeault
 framework with the integration being over all of $\C$.} 
\be
	\delta^{3|4}(W;W_i):=\int_{\C} \frac{\rd \xi_i}{\xi_i}\, \delta^{4|4}(W-\xi_iW_i)
\label{elemental}
\ee
is the wavefunction of an elementary state\footnote{A proper cohomological understanding of these wavefunctions is beyond the scope of the current paper.}. Though distributional, this colour-stripped amplitude is manifestly superconformally invariant, and is antisymmetric under the exchange of any two external twistors, balancing the antisymmetry of the colour-factor. 

In space-time, the 3-point $\MHVbar$ amplitude comes from the (supersymmetrization of the) vertex of the Chalmers \& Siegel action~\cite{Chalmers:1996rq} in the self-dual limit, and has the momentum space form
\be
\label{MHVbar3}
	\mathcal{A}_{\overline{\rm MHV}}(\lambda_i,\tilde\lambda_i,\eta_i) 
	=\delta^4\!\left(\sum_{i=1}^{3}\lambda_i\tilde{\lambda}_i \right)
	\frac{\delta^{0|4}(\eta_1[23] + \eta_2[31] + \eta_3[12])}{[12][23][31]} 
\ee 
which vanishes for real momenta in Lorentzian signature (where there are no real self-dual fields). We
will represent the momentum space amplitude by a filled (grey) disc
with precisely three external legs. Note that treating~\eqref{MHVbar3}
as a function of real spinors and Fourier transforming
$\tilde\lambda\to\del/\del\mu$ leads to an expression whose $\PT^*$
support is smeared by a non-local sign operator~\cite{Mason:2009sa},
the transform of $\mathrm{sgn} [2,3]$ from momentum space. However,
following the discussion in section~\ref{sec:signs}, we will take
the correct understanding for amplitudes on complex momentum space
and complex twistor space to be that the support is genuinely point-like
when understood by the appropriate contour integrals or real integrals of the corresponding Dolbeault expressions.


\subsection{MHV Amplitudes}
\label{sec:MHVtree}

\begin{figure}[t]
\begin{center}
	\includegraphics[height=25mm]{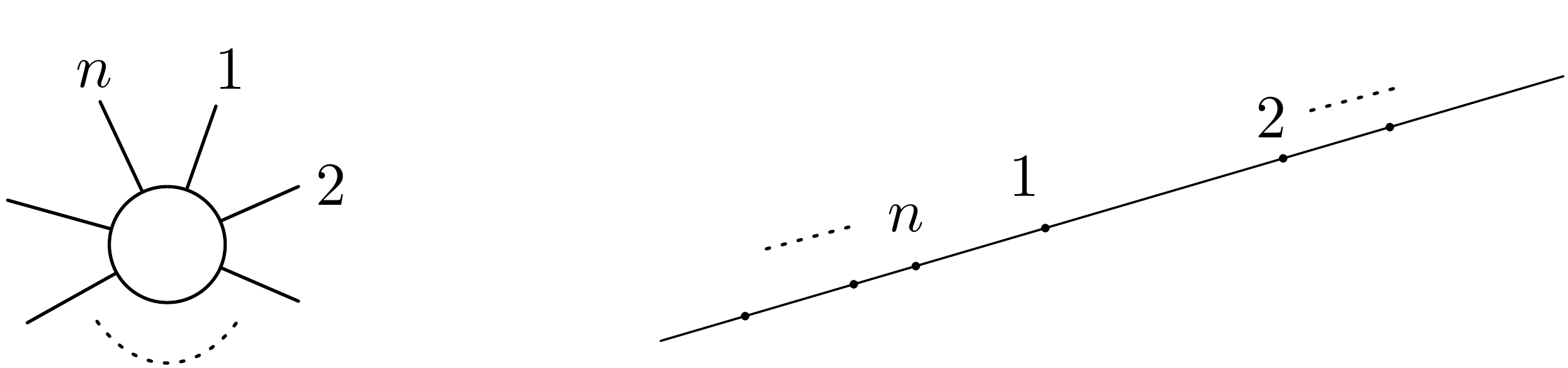}
\end{center}
\caption{{\it The $\rm MHV$ tree amplitude is supported on a line in dual twistor space.}}
\label{fig:MHVtree}
\end{figure}

Nair showed~\cite{Nair:1988bq} that the tree level, $n$-particle MHV amplitude
\be
	\cA_{\rm MHV}^{(0)}(p_1,\ldots,p_n) = \frac{\delta^{4|8}\!
	\left(\sum_{i=1}^n |i\ra[i|\right)}{\la1\,2\ra\la2\,3\ra\cdots\la n\,1\ra}
\ee
is supported on a line --- a degree 1 rational curve --- in dual twistor space. Using homogeneous coordinates $(\lambda_A,\mu^{A'},\chi_a)$ on $\PT^*$, this amplitude may be written as~\cite{Witten:2003nn} 
\be 
	\cA_{\rm MHV}^{(0)}(1,\ldots,n) =\int\frac{\rd^{4|8}x}{\la 1\,2\ra\cdots\la n\,1\ra}\,\prod_{i=1}^n\,
	\delta^{2|4}\!\left(\mu_i-x\lambda_i\right)\ ,
\label{MHVtree}
\ee
where the $\delta$-functions restrict the support to the line $L_{(x,\theta)}\subset\PT^*$ given by
\be\label{line}
\mu^{A'}=x^{AA'}\lambda_A\, , \qquad \chi_a=\theta^A_a\lambda_A\, .
\ee
As in figure~\ref{fig:MHVtree}, we represent this amplitude in momentum space by an empty (white) disc with $n$ external legs.

\subsection{NMHV Amplitudes}
\label{sec:NMHVtree}

The $n$-particle NMHV tree amplitudes may be written as~\cite{Drummond:2008vq}
\be
\cA_{\rm NMHV}^{(0)}(p_1,\ldots,p_n) = \cA_{\rm MHV}^{(0)}\times\sum_{2\leq a,b< n} R_{n;ab}
\label{NMHVtree}
\ee
in on-shell momentum superspace, where
\be
	R_{n;ab} := \frac{\la a\!-\!1\,a\ra\la b\!-\!1\, b\ra \delta^{0|4}\!\left(\Xi_{n;ab}\right)}
	{x_{ab}^2\la n|x_{nb}x_{ba}|a\!-\!1\ra \la n|x_{nb}x_{ba}|a\ra\la n|x_{na}x_{ab}|b\!-\!1\ra\la n|x_{na}x_{ab}|b\ra}\\
\label{Rdef}
\ee
is a dual superconformal invariant, with
\be
	\Xi_{n;ab} := \la n|x_{nb}x_{ba}|\theta_{an}\ra +
	\la n|x_{na}x_{ab}|\theta_{an}\ra\\
\label{Xidef}
\ee
and where $x_{ab} := \sum_{i=a}^{b-1} p_i = \sum_{i=1}^{b-1} |i\ra[i|$ and $\theta_{ab}:=\sum_{i=a}^{b-1} |i\ra\eta_i$. 

\begin{figure}[t]
\begin{center}
	\includegraphics[height=50mm]{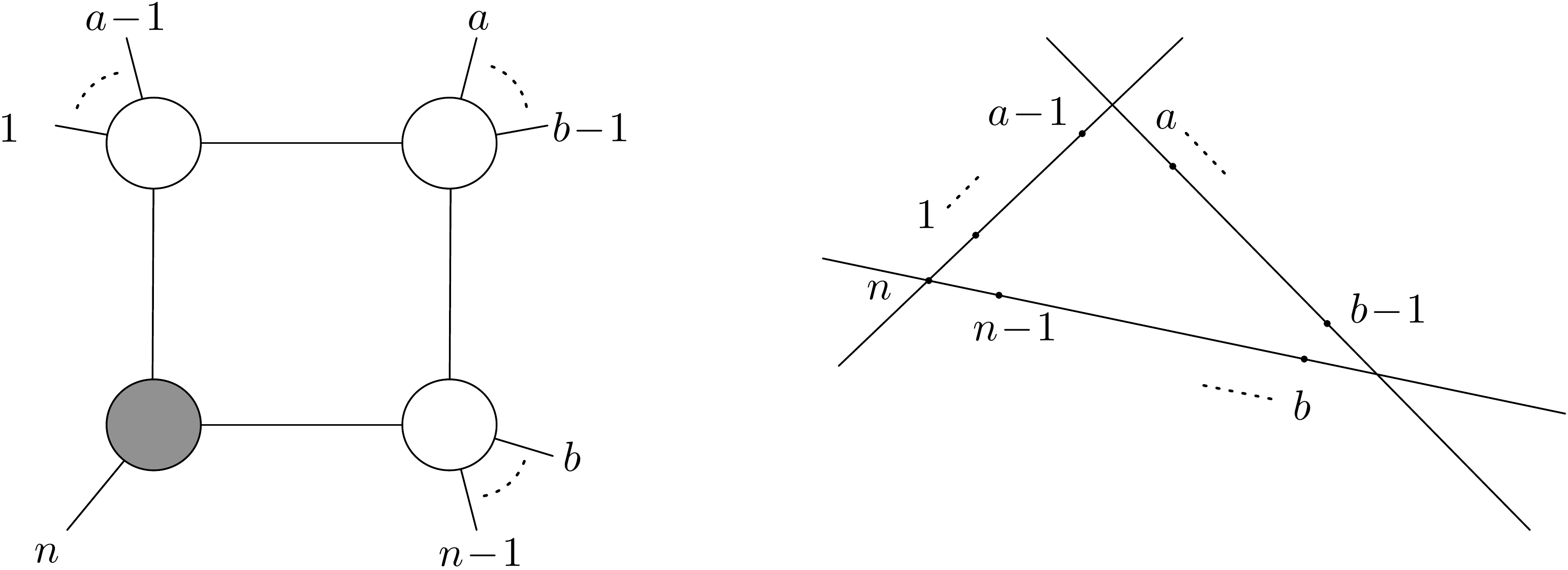}
\end{center}
\caption{{\it  The box coefficient $A^{(0)}_{\rm MHV}R_{n;ab}$ is supported on three, pairwise intersecting lines in $\PT^*$. The curve is connected, but is not irreducible, corresponding to the fact that it gives the twistor support of the leading singularity of a 1-loop amplitude, rather than the loop amplitude itself. The marked points may be located anywhere along the three lines, except that $W_n$ lies at the intersection of two lines, as shown. Note that state $n$ is attached to the 3-particle $\overline{\rm MHV}$ amplitude (denoted by a filled blob in the diagram on the left).}}
\label{fig:3mb}
\end{figure}

The $\PT^*$ support of the each term in~\eqref{NMHVtree} has been computed
many times;  in~\cite{Bern:2004bt,Britto:2004tx} this was done by checking which
combinations of the differential operators introduced by Witten
in~\cite{Witten:2003nn} annihilate the momentum space expression, in~\cite{Mason:2009sa}
it was done by directly solving the BCFW recursion procedure in dual
twistor space, and finally in~\cite{Korchemsky:2009jv} it was achieved by
transforming the momentum space expression to $\PT^*$. The
result is\footnote{Subject to the discussion of section~\ref{sec:signs}.} that each summand in~\eqref{NMHVtree} is supported on three,
pairwise intersecting lines in $\PT^*$, with the external dual
supertwistor $W_n$ located at the intersection point of two of the
lines (see figure~\ref{fig:3mb}). The three lines intersecting lines
form a connected, nodal curve of\footnote{A nodal curve $C$ with $\nu$
irreducible components $C_i$ ($i=1,\ldots,\nu$) each of genus $g_i$, and
$\delta$ nodes has (arithmetic) genus  
$$
	g :=h^1(C,\cO_C)= \left(\sum_{i=1}^\nu g_i\right)+\delta-\nu+1\ .
$$
}
\be
	d=3\qquad\hbox{and}\qquad g=1,
\label{NMHVdegree}
\ee
fitting the general pattern $d=2p+1$ and $g=p$ at N$^p$MHV. Thus, despite the fact that this term arises as a term in the NMHV \emph{tree} amplitude, the twistor space geometry makes it clear that it is far more natural to associate these individual terms with \emph{1-loop} amplitudes. Indeed, it is well-known~\cite{Bern:2004bt,Drummond:2008bq} that each summand in~\eqref{NMHVtree} can also be thought of as a (generically 3-mass box) coefficient in the expansion of a 1-loop NMHV amplitude. The reason the tree amplitude is expressible as a sum of these coefficients can be understood as a consequence of the consistency of the box decomposition with the universal IR behaviour of planar 1-loop amplitudes~\cite{Bern:2004bt,Drummond:2008bq,Britto:2004ap}.

The twistor geometry is closely reflected in the structure of the momentum space cut diagram. Elementary properties of twistor geometry (see {\it e.g.}~\cite{Penrose:1972ia}) show that lines in $\PT^*$ correspond to points in (possibly complex) space-time, and intersection of two twistor lines implies that the two corresponding space-time points are null-separated --- indeed, the space-time conformal structure is determined by the twistor lines in precisely this way. Thus the space-time MHV vertices associated with any pair of lines in figure~\ref{fig:3mb} must be null separated, and likewise the 3-mass box coefficient is extracted by computing the residue of the integrand of the 1-loop NMHV amplitude as the momentum space propagator joining these two MHV subamplitudes goes on-shell.  We investigate the relation between $\PT^*$ support and the momentum space channel in more detail in section~\ref{sec:twsgenunit}, where we will show in detail how to `read off' the momentum channel from the $\PT^*$ support, and {\it vice-versa}. Here, we emphasise that figure does not imply an ordering of the marked points along a given complex line $(\CP^1\subset\PT^*)$, and in particular there is no sense in which these points lie `in between' the intersections of the adjacent lines. However, the configuration of three lines as a whole does know about an ordering, in that which \emph{sets} of points lie on which line is consistent with the colour-ordering of the planar amplitude.


\subsection{N$^2$MHV Amplitudes}
\label{sec:NNMHVtree}

\begin{figure}[t]
\begin{center}
\includegraphics[width=175mm]{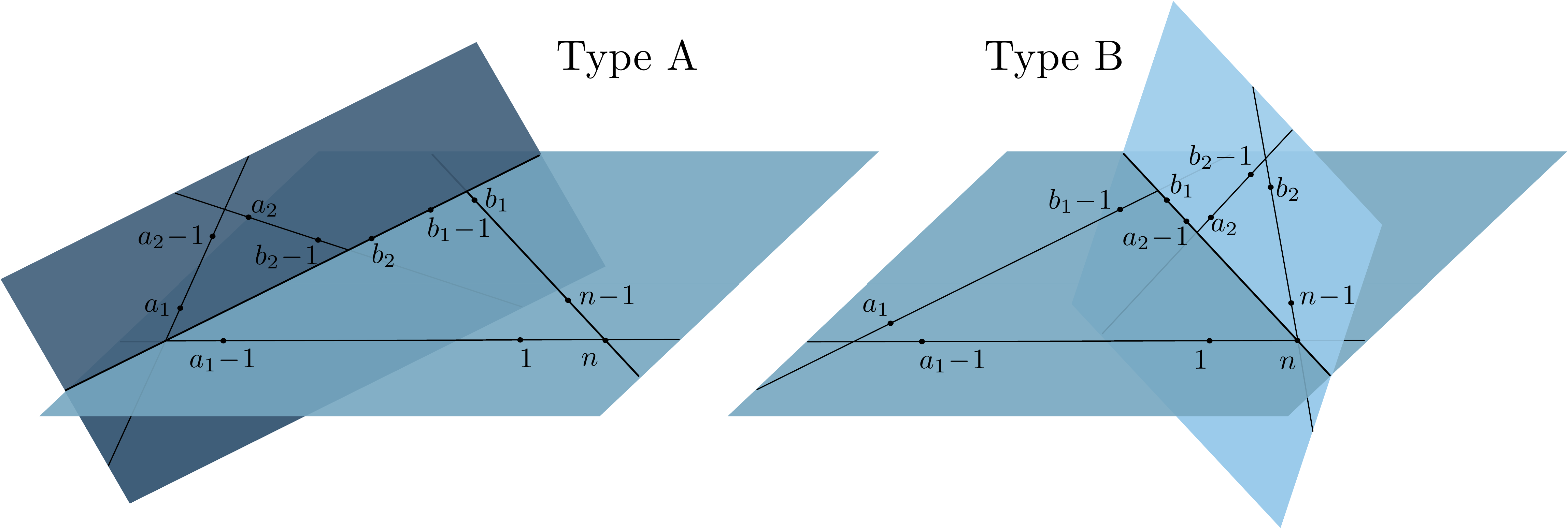}
\end{center}
\caption{{\it The $\PT^*$ support of the two classes of contribution
  to the ${\rm N}^2{\rm MHV}$ tree amplitude. Each term is supported
  on two planes in $\PT^*$, with marked points lying on three
  pairwise intersecting lines in each plane. The intersection of the
  two planes is a common edge of the triangles. We have taken this
  figure from~\cite{Korchemsky:2009jv}, except that we have redrawn it to make the
  $\PT^*$ structure more transparent.}} 
\label{fig:NNMHVAB}
\end{figure}

To obtain the twistor support of a term in the N$^2$MHV tree amplitude~\cite{Drummond:2008cr}
\be
	\cA_{{\rm N}^2{\rm MHV}}^{(0)} = \cA_{\rm MHV}^{(0)} \sum_{2\leq a_1,b_1<n}\!\!\! R_{n;a_1b_1}
	\left[\,\sum_{a_1<a_2,b_2\leq b_1}\!\!\! R^{a_1b_1}_{n;b_1a_1;a_2b_2} 
	+ \sum_{b_1\leq a_2, b_2<n}\!\!\!R^{a_1b_1}_{n;a_2b_2}\right]\ ,
\label{NNMHVtree}
\ee 
two new lines are added to the NMHV configuration of figure~\ref{fig:3mb} --- one through an
existing vertex, the second to make a new triangle with one of the original edges~\cite{Korchemsky:2009jv} (see figure~\ref{fig:NNMHVAB}). There are two ways one is allowed\footnote{See the discussion in section~\ref{sec:NpMHVtree}.} to add lines, corresponding to the two generic (non-boundary) types of term in equation~\eqref{NNMHVtree}. Each of these terms is thus supported on a nodal curve of 
\be 
	d=5\qquad\hbox{and}\qquad g = 2
\label{NNMHVdeg}
\ee
in $\PT^*$, consistent with the formul\ae\ $d=2p+1$, $g=p$ for N$^p$MHV amplitudes.

\begin{figure}[!h]
\begin{center}
\vspace{1cm}
	\includegraphics[width=140mm]{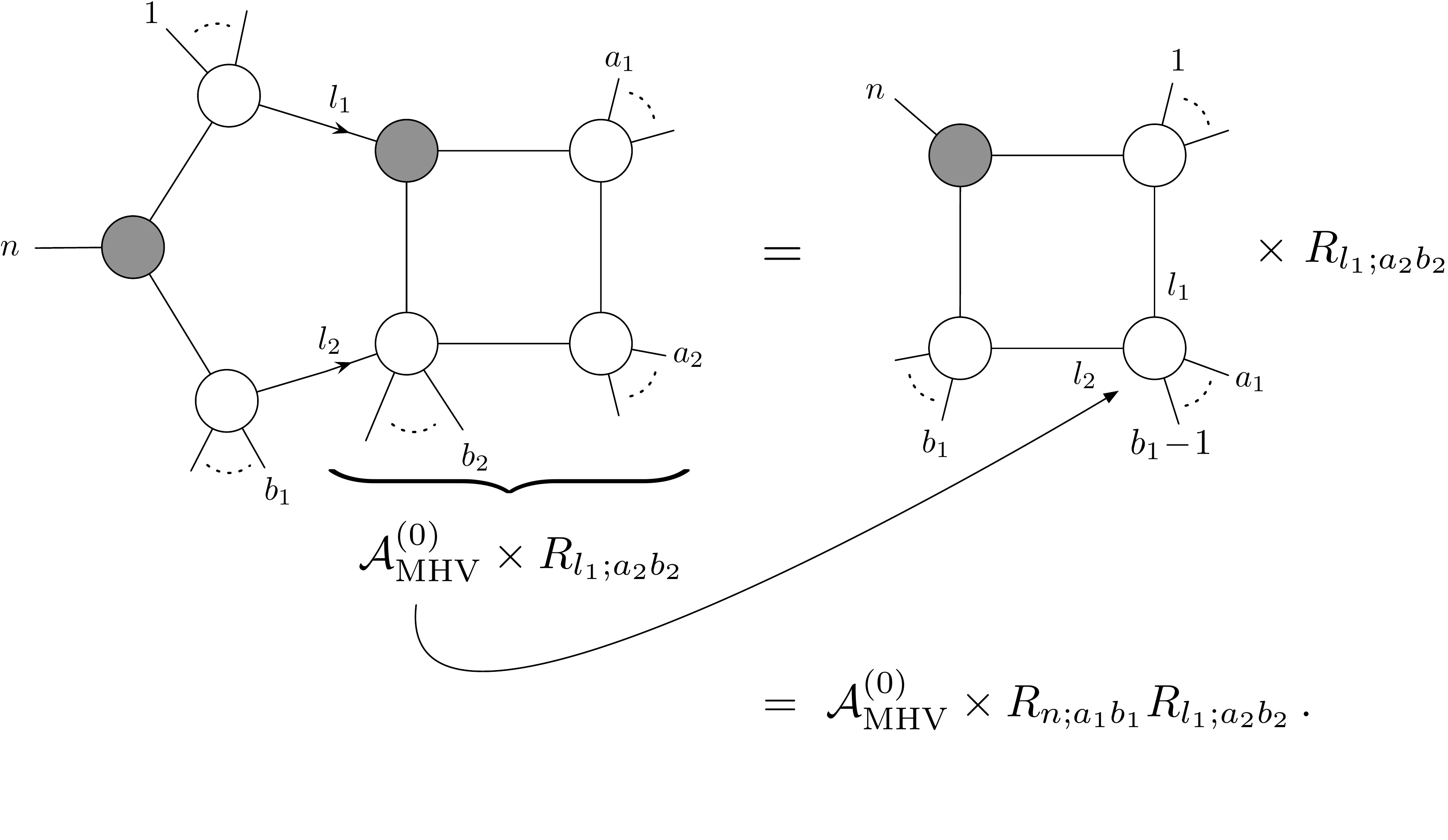}
\end{center}
\vspace{-1.5cm}
\caption{{\it Type A diagrams correspond to a momentum space leading singularity in the pentabox channel shown on the left of this figure. The rest of the figure illustrates the explicit calculation of the leading singularity in this channel.}}
\label{fig:NNMHVleadsing}
\end{figure}

The fact that the twistor support of each of the terms in~\eqref{NNMHVtree} is a curve of genus 2 suggests that each summand corresponds to a leading singularity of a \emph{two loop} N$^2$MHV amplitude. To check this, it is not necessary to through all the (currently unknown) scalar integral topologies that contribute to this $n$-particle, 2-loop amplitude --- as at NMHV, the $\PT^*$ support tells us exactly which integral topology to consider. Let's look first at type A.  The five lines correspond to five MHV vertices. Whenever pairs of lines intersect, the corresponding points in space-time must be null-separated. Space-time null separation corresponds to the requirement that, via the leading singularity contour, we are examining a momentum space channel in which the propagator joining the two MHV subamplitudes is forced to be on-shell. Again, the fact that dual twistor state $n$ lies at the intersection of the line containing points $\{1,\ldots,a_1-1\}$ with the line containing $\{b_1,\ldots,n -1\}$ indicates\footnote{See the discussion of section~\ref{sec:twsgenunit} for a proof.} that the MHV subamplitudes corresponding to these lines are attached to a $\MHVbar$ subamplitude associated with the external state $n$. This leads us to consider the momentum space leading singularity in the channel shown on the left of figure~\ref{fig:NNMHVleadsing} (sometimes called a `pentabox').

We can check that our intuition that the $\PT^*$ support determines the integral topology has not led us astray by actually computing the leading singularity in this channel.  This is straightforward once one notices that the right hand side of the pentabox is a NMHV 3-mass box, whose residue at the poles from the four displayed propagators is $\cA^{(0)}_{\rm MHV}(l_1,a_1,\ldots,b_1\!-\!1,l_2) R_{l_1;a_2b_2}$, where $l_{1,2}$ are the (cut) loop momenta flowing in from the pentagon.  The MHV subamplitude in this partial residue may be re-used to make a further 3-mass NMHV box, so that taking the residue when the remaining four propagators become singular gives 
\be 
	\cA_{\rm MHV}^{(0)}(1,\ldots,n) R_{n;a_1b_1}R_{l_1;a_2b_2}\ ,
\label{Aleadsing}
\ee
where $l_1$ is restricted to the support of the leading
singularity.  In this channel, the on-shell momentum $l_1$ is given by~\cite{Berger:2008sj}
\be
	|l_1\ra \propto x_{a_1b_1}x_{b_1n}|n\ra\ ,
\ee
where $x_{ij}:=p_i+\cdots+p_{j-1}$ are the usual `region momenta'. (The other, complex conjugate solution has vanishing residue because of the $\overline{\rm MHV}$ vertex involving state $n$.) Since $R_{l_1;a_2b_2}$ depends on the cut $l_1$ only through its unprimed spinor component (and is invariant under rescalings of this spinor), $R_{l_1;a_2b_2} = R_{n;b_1a_1;a_2b_2}$ and the leading singularity~\eqref{Aleadsing} is indeed just the type A contribution as promised. 

\begin{figure}[h]
\begin{center}
	\includegraphics[height=55mm]{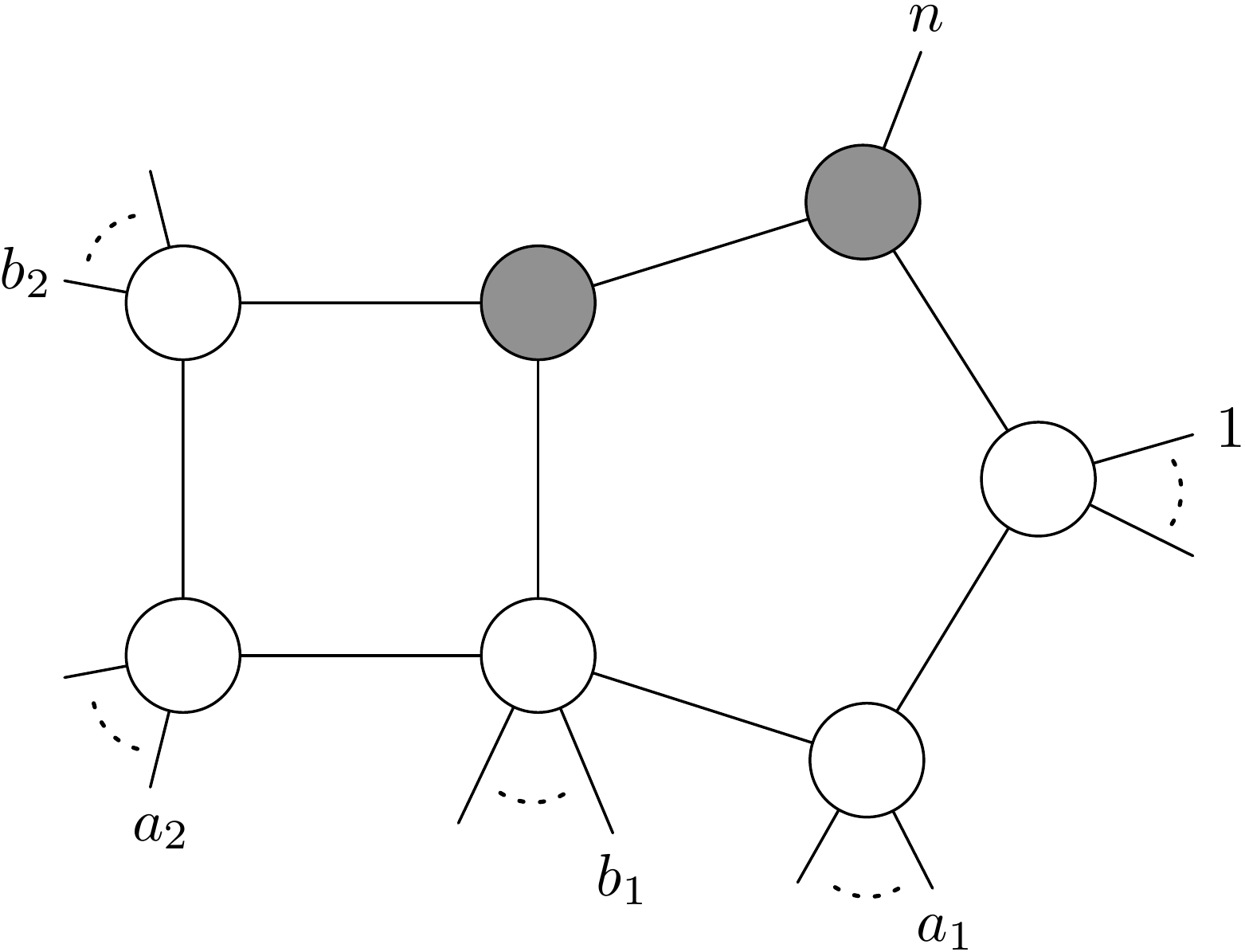}
\end{center}
\vspace{-0.6cm}
\caption{{\it The pentabox corresponding to the type B contributions to the} N$^2$MHV {\it tree amplitude.}}
\label{fig:NNMHVmomB}
\end{figure}

One can similarly show that a type B contribution is the leading singularity in the momentum space channel leads to the pentabox in figure \ref{fig:NNMHVmomB}. Again, we arrived at this momentum space diagram purely by examining the incidence properties of the type B contributions in $\PT^*$ as shown in figure~\ref{fig:NNMHVAB}. Once again, it is readily verified that the leading singularity in the channel shown in figure~\ref{fig:NNMHVmomB} really is $\cA_{\rm
MHV}^{(0)}R_{n;a_1b_1}R_{n;a_2b_2}$.

Finally, note that the boundary terms in~\eqref{NNMHVtree} are really
no different; they just correspond to cases where one of the lines in
figure~\ref{fig:NNMHVAB} has no external states
attached. Specifically, for boundary terms of type A it is the marked
points $\{b_2,\ldots,b_{1}\!-\!1\}$ that are omitted from the diagram,
while for boundary terms of type B, points $\{b_1,\ldots,a_2\!-\!1\}$
should be omitted. In either case, removing these marked points leads
to a line that supports no external states. Consequently, these lines
were omitted in~\cite{Korchemsky:2009jv}, implying that the boundary
terms would be supported on a curve of $d=4$ and $g=0$ in
$\PT^*$. However, the resulting unmarked lines still have three
special points --- their three intersection points with other
lines. These lines correspond to MHV subamplitudes in the momentum
space channel diagrams of
figures~\ref{fig:NNMHVleadsing}-\ref{fig:NNMHVmomB}. There, omitting
the external lines $\{b_2,\ldots b_1\!-\!1\}$ or
$\{b_1,\ldots,a_2\!-\!1\}$ respectively leaves us with a 3-point MHV
subamplitude (with no external legs attached) that cannot
simply be omitted if one wishes to recover the correct leading
singularity contribution. Thus the `unpopulated' lines form an
important part of the picture in dual twistor space, and should be
kept.


\subsection{N$^p$MHV Amplitudes}
\label{sec:NpMHVtree}

\begin{figure}[t]
\begin{center}
\includegraphics[width=175mm]{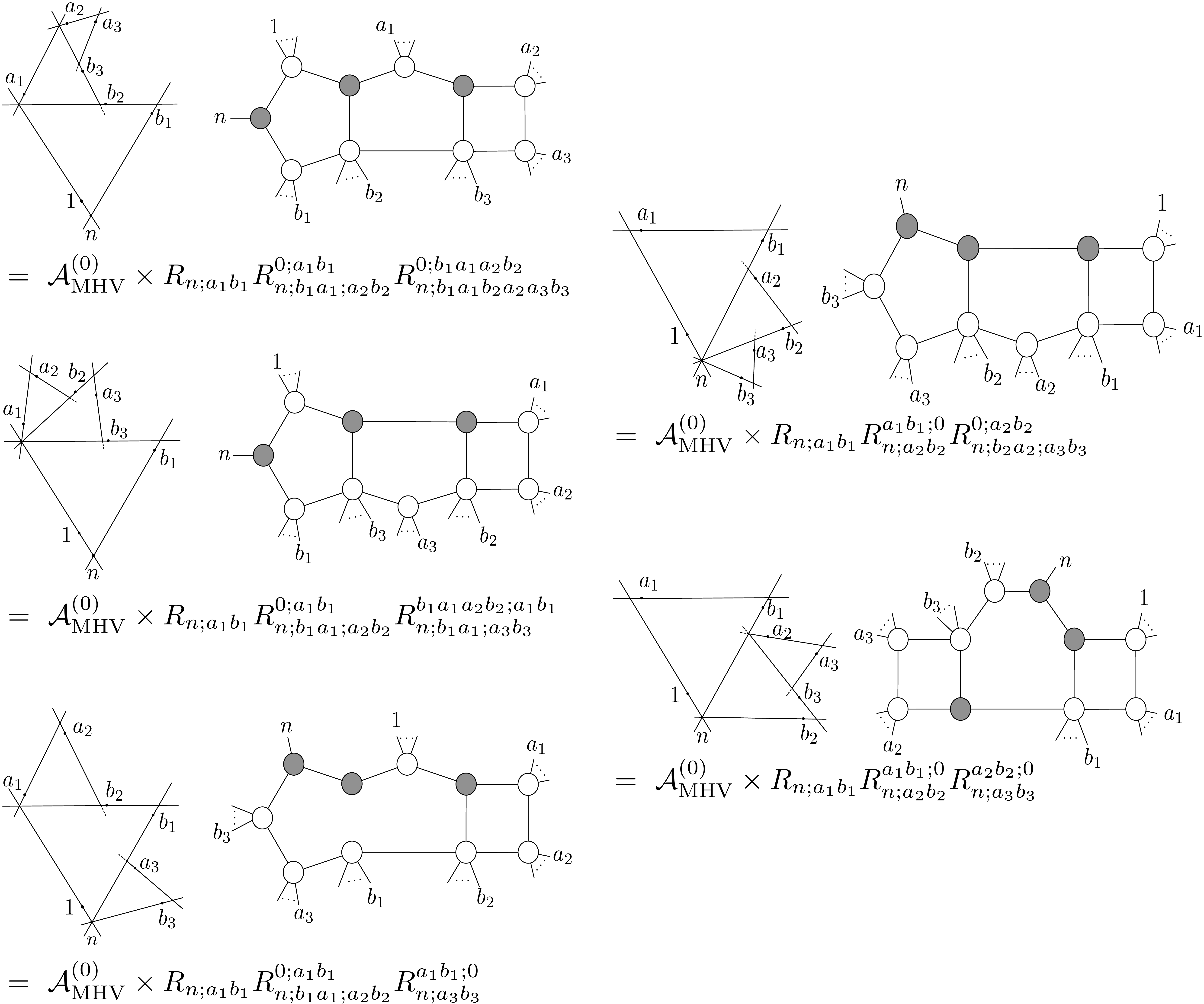}
\caption{{\it Each term in the Drummond \& Henn decomposition of a ${\rm N}^3{\rm MHV}$ tree
    amplitude is a leading singularity of a 3-loop amplitude.}} 
\label{fig:NNNMHV}
\end{center}
\end{figure}

As a final example, the N$^3$MHV tree amplitude may be written as~\cite{Drummond:2008cr}
$$
\begin{aligned}
	\mathcal{A}_{{\rm N}^3{\rm MHV}}^{(0)} &= \mathcal{A}_{\rm MHV}^{(0)}\  \times
	\sum_{2\leq a_1,b_1<n}\hspace{-0.4cm}R_{n;a_1b_1}\\ & 
\times \left\{\sum_{a_1<a_2,b_2\leq b_1}\hspace{-0.4cm}R_{n;b_1a_1;a_2b_2}^{0;a_1b_1}
	\left(\sum_{a_1<a_3,b_3\leq b_2} \hspace{-0.4cm}R_{n;b_1a_1;b_2a_2;a_3b_3}^{0;b_1a_1,a_2b_2}
	\ +\sum_{b_2\leq a_3,b_3\leq b_1}\hspace{-0.4cm}R_{n;b_1a_1;a_3b_3}^{b_1a_1,a_2b_2;a_1b_1} 
	\ +\sum_{b_1\leq a_3,b_3<n}\hspace{-0.4cm}R_{n;a_3b_3}\right)\right.\\
	&\hspace{1cm}\left.+\sum_{b_1\leq a_2,b_2<n}\hspace{-0.4cm}R_{n;a_2b_2}^{a_1b_1;0}
	\left(\sum_{a_2<a_3,b_3\leq b_2}\hspace{-0.4cm}R_{n;b_2a_2;a_3b_3}^{0;a_2b_2}
	\ +\sum_{b_2\leq a_3,b_3<n}\hspace{-0.4cm}R_{n;a_3b_3}^{a_2b_2;0}\right)\right\}\ .
\end{aligned}
$$
The twistor support of each of these terms is shown in figure~\ref{fig:NNNMHV}, as are the momentum space channel diagrams of the corresponding 3-loop N$^3$MHV leading singularities.  The $\PT^*$ support consists of 7 lines arranged as a connected, reducible (nodal) curve of degree 7 and genus 3.  Each
triangle in the figure lies in a different plane in $\PT^*$.  The
marked points are distributed anywhere along the indicated lines,
except those at the intersection of two (or more) lines.  The
correspondence between these terms and the momentum space channel
diagram for the leading singularities is determined as follows (see section~\ref{sec:twsgenunit}). Each
line corresponds to an MHV subamplitude.  When two lines meet at an
unmarked point, there is a (cut) propagator connecting the two corresponding
MHV subamplitudes.  If two lines meet at a marked point, then the two corresponding MHV subamplitudes are joined via an $\MHVbar$ subamplitude with the marked point as an external state.
Additional lines passing through a vertex require an additional
$\MHVbar$ subamplitude (with all three lines internal) to glue them into place.  We have checked explicitly
that the leading singularities in these channels are indeed the
appropriate summand in the tree amplitude.  Once again, the boundary
terms yield figures of exactly the same type, but where certain lines
have no marked points corresponding to external states. In each case,
the resulting line still has three special points --- its intersection
points with three other lines --- and the momentum space channel
diagram requires that this `unpopulated' line be kept.

Korchemsky and Sokatchev give an inductive procedure for building the
N$^{p+1}$MHV figures from those for N$^p$MHV.  This inductive
procedure makes it clear that each summand in the Drummond \& Henn
form of an N$^p$MHV tree amplitude will be supported on a connected
but reducible nodal curve with 
\be
		d=2p+1 \qquad\hbox{and}\qquad g= p\ ,
\label{KSfiguresdeggenus}
\ee 
and whose components are all lines. To see this, first observe that the cyclic ordering induces an
ordering of lines, interleaved by their intersection points (we temporarily ignore all the additional intersections between lines in this ordering). Given an N$^p$MHV term, the inductive procedure starts by
choosing a line in the corresponding $\PT^*$ figure, together with its
intersection point with one of the two lines that are adjacent in the
cyclic ordering.  One then adds an extra pair of intersecting lines,
both of which intersect the chosen line, and one of which intersects
at the pre-existing intersection\footnote{\label{typeC}Korchemsky \& Sokatchev
  impose further conditions on the ordering of the chosen line and
  intersection point so as to ensure compatibility with the ordering
  of the integers $\{a_i,b_i\}$ in the corresponding
  $R$-invariants. This procedure therefore also yields leading
  singularities that do not arise as Drummond \& Henn terms.  This can
  already be seen at N$^2$MHV, where there is no KS figure with two
  lines added to the NMHV triangle both of which intersect one of the
  lines $L_{\{n,1,\ldots,a-1\}}$ or $L_{\{b,\ldots,n\}}$, but neither
  of which pass through the marked point $W_n$. We can call this case
  type C.  These extra figures do
  not alter the conclusion~\eqref{KSfiguresdeggenus}, and in fact
  represent other, perfectly valid leading singularities.}, thus
forming a multiple intersection. Adding these two lines --- not
coplanar with the rest of the diagram --- increases the degree by two
and the genus by one, leading to~\eqref{KSfiguresdeggenus} at
N$^p$MHV.

\begin{figure}[t]
\begin{center}
	\includegraphics[height=85mm]{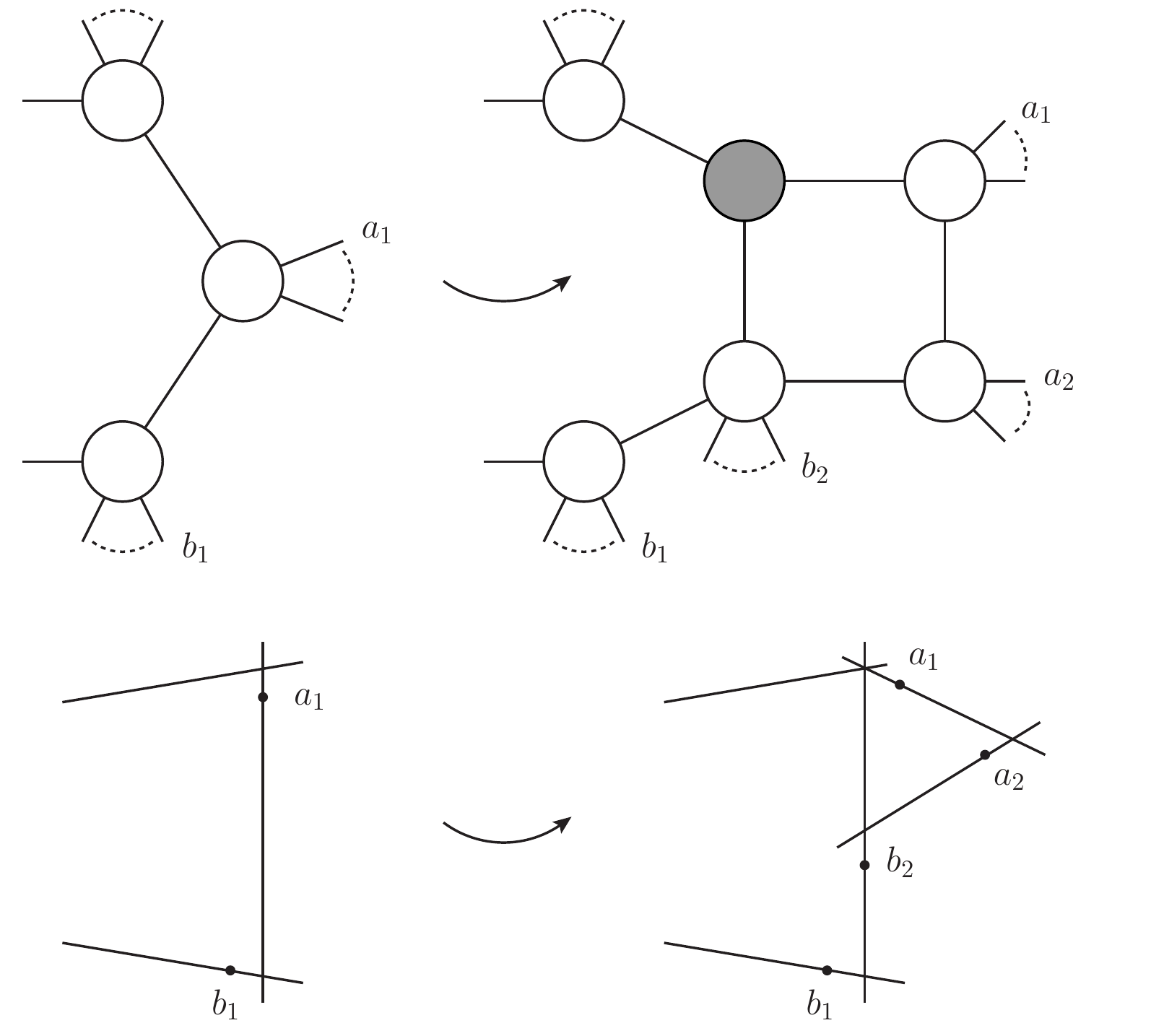}
\caption{{\it Inductive step when new triangle is formed on the unmarked simple vertex between $a_1-1$ and $a_1$.}}
\label{fig:KSinduction1}
\end{center}
\end{figure}

\begin{figure}[!h]
\begin{center}
\includegraphics[height=85mm]{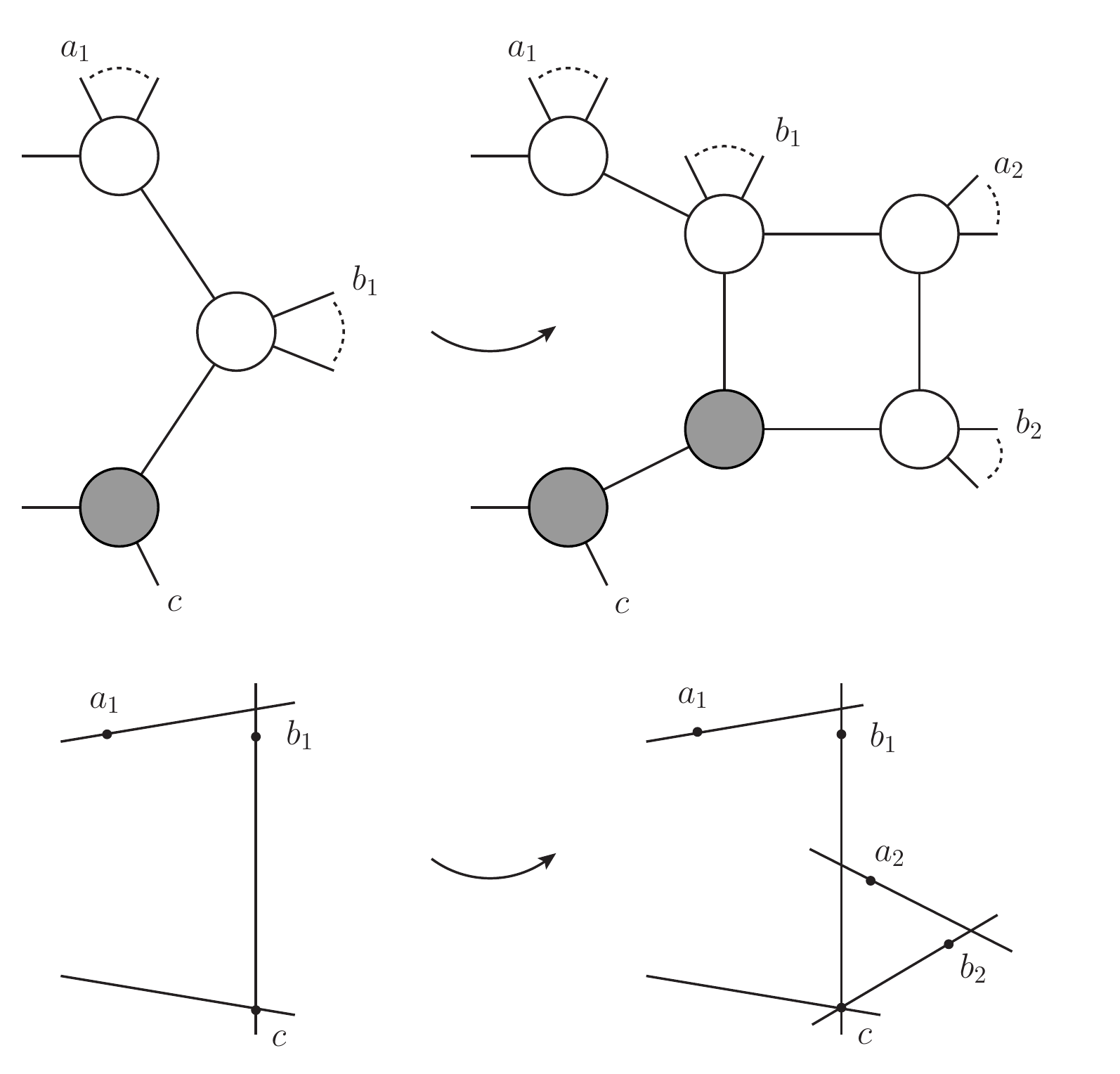}
\caption{{\it Inductive step when new triangle is formed on a marked or multiple vertex at $c$.}}
\label{fig:KSinduction2}
\end{center}
\end{figure}

In the momentum space channel diagram, this process corresponds to inserting two new MHV subamplitudes and
one new $\MHVbar$ subamplitude to make a three mass box including  the
chosen pre-existing MHV vertex as follows. The original choice of
line and intersection point correspond to a choice of MHV subamplitude
together with the propagator\footnote{In the case of building up from
  MHV to NMHV, one simply joins the $\MHVbar$ subamplitude to one of
  the external legs of the original MHV amplitude.} that joins this
MHV subamplitude to the adjacent subamplitude in the cyclic ordering
and that forms part of the boundary of the planar diagram.  
We then insert a new 3-point $\MHVbar$ subamplitude on this propagator, and
build the chosen MHV vertex together with the new $\MHVbar$ vertex
into a three-mass box by adjoining two further MHV vertices.  See
figures \ref{fig:KSinduction1} and \ref{fig:KSinduction2} for two
examples of this. At N$^p$MHV, the momentum space
channel diagrams obtained by this induction have $p$ loops, $p$
$\MHVbar$ subamplitudes and $2p+1$ MHV subamplitudes.  The total
number of propagators is exactly $4p$, and each loop is a polygon with
at least four sides, but just three MHV subamplitudes --- the rest
being $\MHVbar$. Each of these diagrams defines what we will call a
\emph{primitive} leading singularity of a $p$-loop N$^p$MHV
amplitude: one in which all subamplitudes are either MHV
or 3-point $\MHVbar$ and that does not require the use of the `composite'
singularities coming from a Jacobian in the momentum space contour
integral (see {\it e.g.}~\cite{Buchbinder:2005wp, Cachazo:2008dx,
  Cachazo:2008vp}).   

The calculation of the momentum space formula is
also relatively straightforward inductively. The addition of the three
mass box is incorporated by simply multiplying by its associated
R-invariant.  In each of the two figures above, this is simply
$R_{l;a_2b_2}$ where $l$ is the momentum coming into the three mass
box via the $\MHVbar$ vertex. 

We therefore see that each summand in the Drummond \& Henn solution
for N$^p$MHV tree amplitudes is individually a primitive leading
singularity of the $p$-loop N$^p$MHV amplitude. This suggests that an
alternative derivation to the original one of Drummond \& Henn of
their tree amplitude formula should arise from applying the 1-loop IR equation
iteratively on the $p$-loop amplitude; this then provides the rationale for
the appearance of leading singularities in tree amplitudes and
generalises the known case of NMHV~\cite{Bern:2004bt,Drummond:2008bq}.


\section{Generalized Unitarity in Twistor Space}
\label{sec:twsgenunit}

The results of section~\ref{sec:KS} show that the summands in the
Drummond \& Henn solution for tree amplitudes are best understood as
leading singularities of multi-loop amplitudes.  In this section we
show that the twistor support of more general leading singularities
can be understood quite systematically.  Assuming that we have twistor
expressions for the tree amplitudes, we obtain a formula that shows
how leading singularities are created by simply gluing tree
subamplitudes together, much as in momentum space. Remarkably, the
formula is nothing more than a simple inner product between two
twistor wavefunctions.  Thes arguments work for both $\cN=4$
super-Yang-Mills and $\cN=8$ supergravity and the contents of this
section can be applied equally to both.  We will use the notation for
$\cN=4$ super Yang-Mills, but the changes required for gravity are
mostly simply just a matter of replacing the $\cN=4$ by $\cN=8$, and
the SYM tree subamplitudes by their SUGRA counterparts.  Clearly, many
of these ideas also work equally well in a non-supersymmetric theory.


\subsection{Unitarity Cuts in Twistor Space}
\label{sec:twsunitcuts}

\begin{figure}[h]
\centering
\includegraphics[height=25mm]{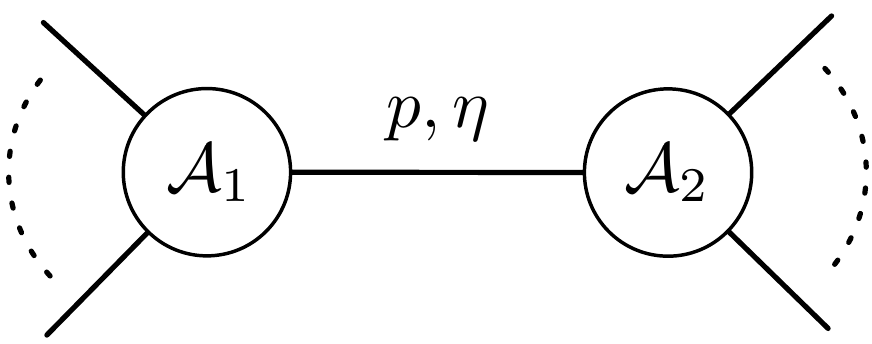}
\caption{{\it A large amplitude factorizes into subamplitudes as an internal propagator goes on-shell. This is the basic building block of leading singularities.}}
\label{fig:InnerProduct}
\end{figure}

Consider computing the residue of a momentum space loop amplitude $\cA$ when one of its internal propagators goes on-shell. Out of all the Feynman diagrams that contribute to $\cA$, only those that contain this propagator contribute to the residue, and standard LSZ arguments ensure that the residue as $p^2\to0$ is simply the product of two subamplitudes on either side of the on-shell propagator, summed over all possible internal states (see figure~\ref{fig:InnerProduct}). This calculation is the cornerstone on which all leading singularity calculations are built.

To rewrite the unitarity cut in terms of amplitudes in twistor space, these individual momentum $\delta$-functions must first be restored, which is easily achieved by distributing the overall momentum $\delta$-function among the subamplitudes, introducing additional momentum integrations to compensate\footnote{See~\cite{Mason:2009sa,ArkaniHamed:2009si} for a similar step in the translation of the BCFW recursion relations.}. Once this has been done, each unitarity cut in the leading singularity takes the form
\be
	\oint \frac{\rd^4p}{p^2}\, \sum\limits_{h} \ \cA_1(\ldots,\{p,h\})\, \cA_2(\{-p,h\},\ldots)\ ,
\ee
where the contour $|p^2|=\varepsilon$ restricts us to the null cone in momentum space, and the sum is over all internal states $h$. In maximally supersymmetric theories, the sum over internal states may be replaced by an integral over the Grassmann coordinate of the on-shell momentum supermultiplet~\cite{Brandhuber:2008pf,Drummond:2008bq,ArkaniHamed:2008gz}. In particular, for $\cN=4$ SYM  we have  
\be
\label{Lightconeintegral1}
\oint \frac{\rd^4p}{p^2}\, \rd^4\eta \ \cA_1(\ldots,\{p,\eta\})\, \cA_2(\{-p,\eta\},\ldots)\ ,
\ee
On the null cone, $p_{AA' }= t \lambda_{A} \tilde{\lambda}_{A'}$,
where in split signature $\lambda$ and $\tilde{\lambda}$ are real and
independent, and $t\in\R^*$ is a scaling
parameter. Thus, dropping an overall factor of
$1/2$,~\eqref{Lightconeintegral1}  reduces 
to~\cite{Cachazo:2004kj} 
\be
\label{Lightconeintegral2}
	\int\limits_{-\infty}^{\infty}  t dt \int \langle \lambda \rd\lambda \rangle [ \tilde{\lambda} \rd\tilde{\lambda} ]\, \rd^4\eta \ 	\cA_1(\ldots,\{t\lambda,\tilde{\lambda},\eta\})\, \cA_2(\{t\lambda,-\tilde{\lambda},\eta\},\ldots)\ ,
\ee
where the $t$ integral will be seen to be convergent. We can replace the momentum space amplitudes by their half Fourier transforms
\be
\begin{aligned}
	\cA_1(\ldots,\{\lambda, -t\tilde{\lambda}, t\eta\}) &=  t^{-2} \int \rd^{2|4}\mu\ 
	\e^{\im t([\mu\tilde{\lambda}]-\chi\cdot \eta)}\,\cA_1(\ldots,\{t\lambda,t\mu,t\chi\})\\
	&= t^{-2} \int \rd^{2|4}\mu\ \e^{\im t([\mu\tilde{\lambda}]-\chi \cdot \eta)}\, \cA_1(\ldots,\{\lambda,\mu,\chi\})
\end{aligned}
\ee
and
\be
	\cA_2(\{\lambda, t\tilde{\lambda}, t\eta\},\ldots) =  t^{-2} \int \rd^{2|4}\mu'\ 
	\e^{-\im t([\mu'\tilde{\lambda}]-\chi' \cdot \eta)}\,\cA_2(\{\lambda,\mu',\chi'\},\ldots)\ ,
\ee
where we have used the fact that the $\cN=4$ superamplitudes have
homogeneity degree zero on $\PT^*$. The integral over $t^{-3}dt$
then combines with $[ \tilde{\lambda} \rd\tilde{\lambda} ]\,\rd^4\eta$
to give $\rd^{2|4}\tilde{\lambda}$ and \eqref{Lightconeintegral2}
becomes 
\be
\begin{aligned}
	& \int \la\lambda\, \rd\lambda \ra\, \rd^{2|4}\tilde{\lambda}\, \rd^{2|4}\mu\, \rd^{2|4}\mu' \ 
	\e^{\im t([\mu-\mu',\tilde{\lambda}] -  (\chi - \chi') \cdot \eta)}\ 
	\cA_1(\ldots,\{\lambda,\mu,\chi\})\,\cA_2(\{\lambda,\mu',\chi'\},\ldots)\\
	=\ & \int \la \lambda\,\rd\lambda \ra\, \rd^{2|4}\mu\, \rd^{2|4}\mu' \  
	\delta^{2|4}(\mu-\mu')\, \cA_1(\ldots,\{\lambda,\mu,\chi\})\,  \cA_2(\{\lambda,\mu',\chi'\},\ldots) \\
	=\ & \int \la \lambda\, \rd\lambda \ra\, \rd^{2|4}\mu\ 
	\cA_1(\ldots,\{\lambda,\mu,\chi\})\, \cA_2(\{\lambda,\mu,\chi\},\ldots)\\
	=\ &\int \rD^{3|4}W\  \cA_1(\ldots,W)\, \cA_2(W,\ldots)\ .
\end{aligned}
\ee
Altogether, we have shown that computing the residue of a momentum space amplitude as an internal propagator goes on-shell amounts to computing the inner product between the corresponding amplitudes in $\PT^*$, {\it i.e.},
\be
	\oint_{|p|^2=\varepsilon} \frac{\rd^4p\,\rd^4\eta}{p^2}\ \cA_1(\ldots,\{p,\eta\})\, \cA_2(\{-p,\eta\},\ldots)
	= \int_{\PT^*}\hspace{-0.2cm}\rD^{3|4}W\, \cA_1(\ldots, W)\, \cA_2(W,\ldots)\ .
\label{Twistorinnerproduct}
\ee
(see {\it e.g.}~\cite{Penrose:1972ia,Mason:2009sa,Eastwood:1981aa} for a discussion of the twistor inner product). The result has a straightforward generalisation beyond the current context of $\cN=4$ SYM; the individual Grassmann components of~\eqref{Twistorinnerproduct} correspond to the exchange of particular helicities in figure~\ref{fig:InnerProduct}, and the different homogeneities of the components of $\cA_1$ and $\cA_2$ balance the integral $\rD^3W$ over the homogeneous coordinates of $\PT^*$. These component terms make sense even in a non-supersymmetric theory.


\subsection{Twistor Space Representation of Primitive Leading Singularities}
\label{sec:twprimleadsing}

In this section we apply equation~\eqref{Twistorinnerproduct} to
derive simple rules for calculating the $\PT^*$ leading singularity
from its channel diagram when the individual tree subamplitudes are
either MHV or the 3-point $\MHVbar$.  We call these \emph{primitive}
leading singularities, and they include all the KS configurations of
section~\ref{sec:KS}.  The primitive leading singularities form a
generating set\footnote{It is not clear whether this will also be the
case for supergravity.}, for all leading singularities for $\cN=4$ super
Yang-Mills, since we may always replace an N$^p$MHV tree subamplitude
by its Drummond \& Henn form --- each term of which is a primitive
leading singularity.

\medskip

\begin{figure}[t]
\centering
\includegraphics[height=14cm]{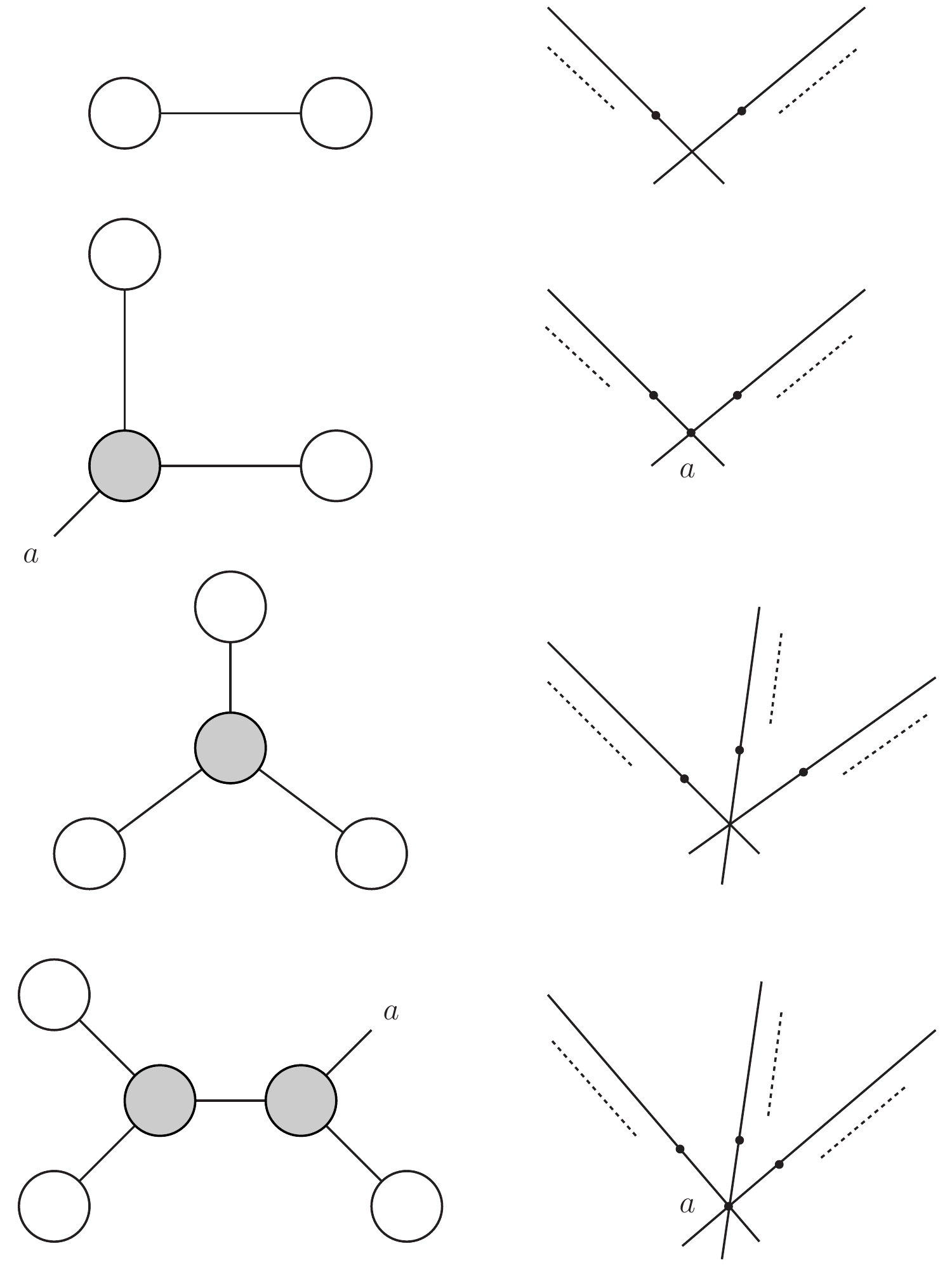}
\caption{{\it The $\PT^*$ support of common components of leading singularities containing} MHV {\it and 3-point} $\MHVbar$ {\it subamplitudes.}}
\label{fig:Twistorsupportrules}
\end{figure}

Given a leading singularity channel in momentum space, its $\PT^*$ support is straightforward to calculate by applying equation~\eqref{Twistorinnerproduct}. For example, consider a single channel connecting two MHV subamplitudes. The cut integration is 
\be
\begin{aligned}
\label{TwoMHVchannel}
	&\oint \frac{\rd^4p\,\rd^4\eta}{p^2}\,\cA_{{\rm MHV}\,1}(\ldots,\{p,\eta\})\,\cA_{{\rm MHV}\,2}(\{-p,\eta\},\ldots)\\
	 &\hspace{3cm}
	 = \int\rD^{3|4}W\, \cA_{{\rm MHV}\,1}(\ldots,W)\,\cA_{{\rm MHV}\,2}(W,\ldots)\ .
\end{aligned}
\ee
We now integrate out $W$ using the $\delta$-functions in the $\PT^*$ form~\eqref{MHVtree} of the MHV tree amplitudes
\be
	\cA_{{\rm MHV}\,i}(W,\ldots) = \int \rd^{4|8}x\, \cB_i(\ldots,\lambda)\,\delta^{2|4}(\mu-x_i\lambda)\ ,
\ee
where the $\cB_i$ depend on $\lambda$ through their (Parke-Taylor) denominators.  Therefore, the right hand side of~\eqref{TwoMHVchannel} becomes
\be
\begin{aligned}
	&\int \rD^{3|4}W\,\rd^{4|8}x_1\,\rd^{4|8}x_2\  
	\delta^{2|4}(\mu-x_1\lambda)\,\delta^{2|4}(\mu-x_2\lambda)\,\cB_1(\ldots,\lambda)\,\cB_2(\lambda,\ldots)\\
	=\ & \int \la \lambda\, \rd \lambda \ra \, \rd^{4|8}x_1\,\rd^{4|8}x_2\ 
	\delta^{2|4}((x_1-x_2)\lambda)\,\cB_1(\ldots,\lambda)\,\cB_2(\lambda,\ldots)\ .
\end{aligned}
\label{2MHVchannel2}
\ee
On the support of $\delta^{2|4}((x_1-x_2)\lambda)$, we can set $(x_1-x_2)^{AA'}=\rho^A\tilde\rho^{A'}$
and $(\theta_1-\theta_2)^{Aa}=\rho^A\eta^a$ where $(\tilde\rho,\eta)$ are arbitrary, but where $\rho\propto\lambda$ so that $\la\rho\,\lambda\ra=0$. The $\delta$-functions $\delta^{2|4}((x_1-x_2)\lambda)$ may then be written as $\delta(\lambda;\rho)\,\delta^{1|4}(x_{12}\lambda)$, where
\be
	\delta(\lambda;\rho):=\int_{\C^*}\frac{\rd\xi}{\xi}\,\delta^2(\lambda-\xi\rho)
\ee
as for the elementary state~\eqref{elemental}, and
\be
	\delta^{1|4}(x_{12}\rho):=\delta(x_{12}^2)\,\delta^{0|4}(\theta_{12}\rho)
\label{delta14def}
\ee
and\footnote{We stress that these are differences of coordinates in \emph{space-time}; they should not be confused with region momenta.} $x_{ij}:=x_i-x_j$ {\it etc}. This $\delta$-function allows us to perform the $\lambda$ integral in~\eqref{2MHVchannel2}. Since the overall expression is homogeneous of degree zero, one simply replaces $\lambda$ by $\rho$, whereupon the right hand side of~\eqref{TwoMHVchannel} reduces to
\be
\label{TwoMHVchannel-fin}
	 \int  \rd^{4|8}x_1\,\rd^{4|8}x_2\ \delta^{1|4}(x_{12}\rho)\,\cB_1(\ldots,\rho)\,\cB_2(\rho,\ldots)\ .
\ee
The $\delta$-function  constraining $(x_1-x_2)^2=0$ ensures that the integrand of equation~\eqref{TwoMHVchannel-fin} vanishes unless the two lines $L_{x_1},L_{x_2}\subset\PT^*$ intersect (see the first diagram in figure~\ref{fig:Twistorsupportrules}).

Now consider the second diagram in figure~\ref{fig:Twistorsupportrules}, consisting of a 3-point $\MHVbar$ subamplitude  connected to two MHV subamplitudes. We expect that the two lines corresponding to the MHV subamplitudes should intersect at the point in $\PT^*$ determined by the $\MHVbar$ subamplitude. To confirm that this is indeed the case,
let us again perform the cut integrations explicitly. Using the form~\eqref{MHVbartws} for $\cA_{\overline{\rm MHV}}(W_1,W_a,W_2)$ we have
\be
\begin{aligned}
	&\int \rD^{3|4}W_1\wedge\rD^{3|4}W_2\ \cA_{\rm MHV}(\ldots,W_1)\,
	\cA_{\overline{\rm MHV}}(W_1,W_a,W_2)\,\cA_{\rm MHV}(W_2,\ldots) \\ 
	=\ &  \int \rD^{3|4}W_1\wedge\rD^{3|4}W_2\ \cA_{\rm MHV}(\ldots,W_1)\,
	\delta^{3|4}(W_a;W_1)\,\delta^{3|4}(W_a;W_2)\,\cA_{\rm MHV}(W_2,\ldots) \\
	=\ &\  \cA_{\rm MHV}(\ldots,W_a)\,\cA_{\rm MHV}(W_a,\ldots)\ .
\end{aligned}
\label{TwoMHV+MHVbarchannel}
\ee
Each of these two MHV subamplitudes is supported only when $W_a$ lies on the respective line, so the product is supported only on pairs of lines that intersect at $W_a$, as expected.

\medskip

The rule for translating between momentum channel diagrams and $\PT^*$ support of leading singularities should now be clear: one simply applies the twistor inner product~\eqref{Twistorinnerproduct} to the known tree level subamplitudes. In fact, this rule apply equally when the subamplitudes have any MHV degree. Figure~\ref{fig:Twistorsupportrules}  illustrates various applications of this rule that frequently occur in primitive leading singularities.

\begin{figure}[t]
\centering
\includegraphics[height=5cm]{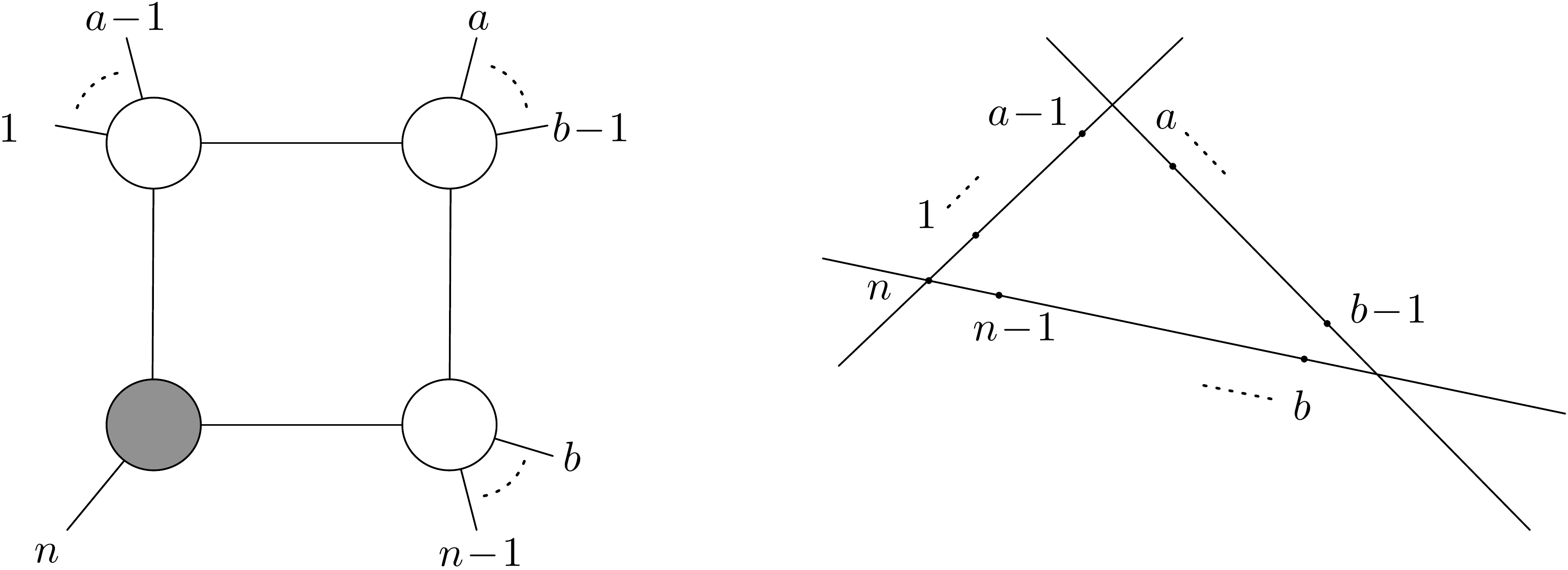}
\caption{{\it The} NMHV {\it 3 mass box coefficient, redrawn from figure~\ref{fig:3mb} for convenience.}} 
\label{fig:NMHVthreemassbox}
\end{figure}

As a non-trivial check on this rule, let us consider the NMHV three\footnote{In the limit that only one external state is attached to the MHV tree subamplitudes, this becomes a 2-mass hard or even 1-mass channel; the $\PT^*$ support is unaffected.} mass box coefficient in more detail than in section~\ref{sec:NMHVtree}. (We have re-shown this in figure~\ref{fig:NMHVthreemassbox} for convenience.)   Each of the three edges $L_e$ ($e=1,2,3$) in the dual twistor picture corresponds to a point $x_e$ in (complexified) space-time, null separated from its neighbour. Because they all intersect, the three lines lie in a common plane ($\CP^2$) in $\PT^*$, so the corresponding points lie in a common $\alpha$-plane\footnote{An $\alpha$-plane is the totally null complex 2-plane given by $x^{AA'}(\kappa) = x_0^{AA'}+\kappa^A\tilde\rho^{A'}$ for some fixed $\tilde\rho$, where $x_0$ is a point on the plane. $\alpha$-planes correspond to points in projective twistor space, or planes in dual projective twistor space. Similarly, $\beta$-planes --- the totally null complex 2-plane $x^{AA'}(\tilde\kappa) = x_0^{AA'}+\rho^A\tilde\kappa^{A'}$ for some fixed $\rho$ --- correspond to points in $\PT^*$ or planes in $\PT$. These space-time planes were indicated by shaded triangles in~\cite{Korchemsky:2009jv}. See {\it e.g.}~\cite{Penrose:1986ca} for an in-depth discussion of twistor geometry.} in space-time. We can thus write
\be
\label{differencesA}
	x_1-x_2=\rho\tilde\rho\, , \quad  x_2-x_3= \sigma\tilde\rho \, , \quad
	x_3-x_1= \lambda_n\tilde\rho
\ee
and a similar set of equations for the Grassmann components, all subject to the identity
\be
	\sigma+\rho+\lambda_n\equiv0\ .
\label{sumtozero}
\ee
Thus, all of the lines $L_e$ are determined in terms of one of them (say $L_1$) and the spinors $\rho$ and $\tilde\rho$. Korchemsky \& Sokatchev~\cite{Korchemsky:2009jv} transformed the standard expression $\cA_{\rm MHV}^{(0)}R_{n;ab}$ given in equations~\eqref{NMHVtree}-\eqref{Xidef} to $\PT^*$, obtaining the $\PT^*$ leading singularity\footnote{Subject to the discussion in section~\ref{sec:signs}.}

\be
	\int \frac {\rd^{4|8}x_1\ \rd^{2|4}\tilde\rho\ \rd^2\rho}
	{\la n\,1\ra\la1\,2\ra\cdots \la a\!-\!1\,\rho\ra\la\rho\,a\ra\cdots\la b\!-\!1\, \sigma\ra 
	\la\sigma \, b\ra\cdots\la n\!-\!1\, n\ra}\ \prod_{i=1}^n\, \delta^{2|4}(\mu_i-  x_{e(i)}\lambda_i )\ ,
\label{KS-3mb-twis}
\ee
where $e(i)=1,2$ or 3, depending on which MHV subamplitude state $i$ lies.

We now check that the above gluing rule does yield this answer. The three MHV vertices are supported on three
lines $L_e\subset\PT^*$, while the $\MHVbar$ subamplitude corresponds to a distinguished point $W_n\in\PT^*$. We must glue these subamplitudes together using~\eqref{TwoMHVchannel-fin} at both $W:=L_1\cap L_2$ and $W':=L_2\cap L_3$, while we should use~\eqref{TwoMHV+MHVbarchannel} to glue $L_2$ to $L_3$ at the marked point $W_n$. Doing so and using the $\delta$-functions in the $\MHVbar$ subamplitude gives
\be
\begin{aligned}
	&\int\rD^{3|4}W\,\rD^{3|4}W'\,\cA_{\rm MHV}(n,1,\ldots,a\!-\!1,W)\,\cA_{\rm MHV}(W,a,\ldots,b\!-\!1,W')\,
	\cA_{\rm MHV}(W',b,\ldots,n\!-\!1,n)\\
	=\ & \int \prod_{e=1}^3\,\rd^{4|8}x_e\,
	\frac{\delta^{1|4}(x_{12}\rho)\,\delta^{1|4}(x_{23}\sigma)\,
	\delta^{2|4}(x_{31}\lambda_n)}{\la n\,\rho\ra \la\rho\,\sigma\ra \la \sigma\,n\ra}\ K \ ,
\end{aligned}
\label{GenUnNMHV}
\ee
where the $\delta^{1|4}$-functions are defined in~\eqref{delta14def}. On the support of these $\delta$-functions, we have 
\be
	\rho\tilde\rho:=x_1-x_2\qquad\hbox{and}\qquad \sigma\tilde\sigma:=x_2-x_3\ ,
\ee
and where $K$ is the Korchemsky-Sokatchev integrand
\be
	\frac{1}{\la n\,1\ra \cdots \la a\!-\!1\,\rho\ra \la\rho\,a\ra\cdots\la b\!-\!1\,\sigma\ra \la\sigma\, b\ra\cdots \la n\!-\!1\, n\ra}
	\ \times\ \prod_{i=1}^n\,\delta^{2|4}(\mu_i- x_{e(i)}\lambda_i )\ .
\ee
The identity $0\equiv\rho+\sigma+\lambda_n$ uniquely fixes the scale
of the spinors $\rho$ and $\sigma$ in terms of that of $\lambda_n$, and the
$\delta$-function $\delta^{2|4}((x_1-x_3)\lambda_n)$ then implies that
$\tilde\sigma =\tilde\rho$, so that all of the conditions
in~\eqref{differencesA} are recovered. A calculation (which we
suppress) shows that the $\delta$-functions reduce the measure 
\be
	\prod_{e=1}^3\,\rd^{4|8}x_e\,\frac{\delta^{1|4}(x_{12}\rho)\,\delta^{1|4}(x_{23}\sigma)\,
	\delta^{2|4}(x_{31}\lambda_n)}{\la n\,\rho\ra \la\rho\,\sigma\ra \la \sigma\,n\ra}
	= \rd^{4|8}x_1\,\rd^{2|4}\tilde\rho\,\rd^2\rho\ ,
\ee
so that~\eqref{GenUnNMHV} reduces to the Korchemsky-Sokatchev expression~\eqref{KS-3mb-twis}, as
required.

It is straightforward to verify that explicit, $\PT^*$ formul\ae\ for
leading singularities associated to KS configurations, or indeed any
primitive leading singularity, may be computed in exactly the same
manner. In section~\ref{sec:nodal} we give an alternative method of
performing these integrals that ties into the twistor-string
representation, preserves explicit superconformal
invariance and will be more suitable for the correspondence with the
Grassmannian later.


\section{Twistor-Strings}
\label{sec:tws}

In section~\ref{sec:KS}, we saw that each summand in the Drummond \&
Henn solution of a tree level N$^p$MHV amplitude can be identified
with a leading singularity of the $n$-particle, $p$-loop N$^p$MHV
amplitude, and is supported on a connected, but nodal curve in dual
twistor space of degree $2p+1$ and genus $p$. Such algebraic curves
are natural from the point of view of twistor-string
theory~\cite{Witten:2003nn}, where $g$-loop N$^p$MHV amplitudes are
associated with holomorphic maps $W:\Sigma\to\PT^*$ of degree  
\be	
	d = p+1+g
\label{twsdg}
\ee from a genus\footnote{Viewing the twistor-string as either a
  gauged $\beta\gamma$-system (following
  Berkovits~\cite{Berkovits:2004hg}) or as a twisted (0,2) sigma
  model~\cite{Mason:2007zv}, the loop order $\ell$ is
  \emph{identified} with the genus of the string worldsheet $\Sigma$;
  $\ell\equiv g$. The genus of the image curve $W(\Sigma)\subset\PT^*$
  can then be $\leq g$ if the map is a covering of its image; this is
  sometimes forced by the genus and degree.} $g$ worldsheet $\Sigma$.
In particular, for $p$-loop N$^p$MHV amplitudes, we should expect to
consider holomorphic maps of degree $2p+1$ from a genus $p$
worldsheet. Thus, despite the fact that twistor-string theory
describes $\cN=4$ SYM coupled to $\cN=4$ conformal
supergravity~\cite{Berkovits:2004jj}, the algebraic curves expected
from \emph{higher genus} twistor-string theory agree precisely with
the KS configurations, coming from a field theory analysis of pure
$\cN=4$ SYM.

In this section we show how to compute general leading singularities
by gluing together N$^p$MHV tree amplitudes obtained from twistor-string theory.  The
generalized unitarity formula \eqref{Twistorinnerproduct} presents the result as an
integral over a space of nodal curves.  We then examine how it might
be possible to obtain leading singularities from a twistor-string theory at higher genus by localising the path integral on such nodal curves. While it is not known whether the worldsheet path integral defined by the original twistor-string theories makes sense when $g>0$ (and even if it does, it describes a theory that includes conformal
supergravity~\cite{Berkovits:2004jj}), the generalized unitarity arguments provide specific integrals over the space of nodal curves that give $\cN=4$ SYM leading singularties by construction.  Just as the leading singularity
conjecture~\cite{Cachazo:2008hp} in momentum space states that the
integrand of an arbitrary multi-loop process is determined by its
leading singularities, we similarly conjecture that the appropriate
volume form for the twistor-string theory path integral (free from
conformal supergravity) can be determined by requiring that it
reproduces leading singularities of $\cN=4$ SYM. We elaborate this conjecture below.


\subsection{The Twistor-String Path Integral}
\label{sec:pathintegral}

We first review the twistor-string path integral. This is unambiguous at tree level, but when $g>1$ the correct form of twistor-string path integral for pure $\cN=4$ SYM is currently unknown and our discussion is at a sufficiently general level  to be insensitive to these ambiguities. Combining the tree level formul\ae\  with the gluing rule of the previous section allows us to calculate leading singularities as integrals over moduli spaces of nodal curves, as we do in the following subsection.

In twistor-string theory, the worldsheet map $W:\Sigma\to\PT^*$ to
dual $\cN=4$ supertwistor space is represented by worldsheet
fields\footnote{We suppress the dual twistor indices in what follows.}
$W(\sigma)$, defined up to an overall scale --- $W(\sigma)$ is the
pullback to $\Sigma$ of the four bosonic and four fermionic
homogeneous coordinates of the target space. This map is constrained
to be holomorphic, either directly in the gauged first-order formulation
of Berkovits~\cite{Berkovits:2004hg}, or via a twisted worldsheet supersymmetry
in~\cite{Witten:2003nn}, so that 
\begin{equation}
	W(\sigma)\in \C^{4|4}\times H^0(\Sigma,\cL)
\label{holmap}
\end{equation}
where $\cL\simeq\cO_\Sigma(d)$ is the pullback by $W$ of the
hyperplane bundle on $\PT^*$.  Now, by Riemann-Roch,
\be
	h^0(\Sigma,\cL)-h^1(\Sigma,\cL) = {\rm deg}(\cL)+1-g\ ,
\label{RiemannRoch}
\ee
and $h^1(\Sigma,\cL)$ vanishes on a dense open set in the moduli space
provided ${\rm deg}(\cL)\geq g-1$, and vanishes everywhere if ${\rm deg}(\cL)\geq 2g-1$.  The CFT path integral over the
$4h^0(\Sigma,\cL)$ fermionic zero-modes vanishes unless it is
saturated by vertex operator insertions of the fermionic components
$\chi(\sigma)$ of $W(\sigma)$. The N$^p$MHV sector is characterized by having homogeneity $4(p+2)$ in the fermionic momenta. So, given that the dual twistor $\cN=4$ SYM multiplet is
$$
	a(W,\chi) = g^+(W)+\chi_a\Gamma^a(W)+\frac{1}{2!}\chi_a\chi_b\Phi^{ab}(W)
	+\frac{1}{3!}\epsilon^{abcd}\chi_a\chi_b\chi_c\,\tilde\Gamma_d(W)
	+\frac{1}{4!}\epsilon^{abcd}\chi_a\chi_b\chi_c\chi_d\,g^-(W)\, ,
$$
the N$^p$MHV sector receives contributions only from worldsheet
instantons for which $h^0(\Sigma,\cL)=p+2$. For later convenience, we
introduce the shorthand $k\equiv h^0(\Sigma,\cL)$.

The holomorphic sections of $\cL$ depend on the complex structures of
both the worldsheet and $\cL$ itself.  At genus zero, there is a
unique holomorphic line bundle $\cL$ for each degree $d$, but for
higher genus curves they form a $g$-dimensional family (actually, an
Abelian variety) known as the Jacobian $\cJ(\Sigma)$. As we move
around in the moduli space $\overline{M}_{g,n}$ of stable\footnote{`Stability' is the requirement that the curve has only a finite number of automorphisms, and amounts to the condition that each genus zero component of the worldsheet has at least 3 special points (either marked points or nodes), and each genus one component has at least one such point. A stable \emph{map} requires these conditions only on components of the worldsheet that are mapped to the target at degree zero. The stability condition is necessary to obtain a Hausdorff moduli space. See {\it e.g.}~\cite{HarrisMorrison,Vakil:2008cu} for introductions to $\overline{M}_{g,n}$ and {\it e.g.}~\cite{Fulton:1997sm} for an introduction to $\overline{M}_{g,n}(\mathbb{P}^r,d)$.} genus $g$, $n$-pointed curves, we obtain a moduli space\footnote{Note that although the Jacobian varities $\cJ(\Sigma)$ on a fixed worldsheet are independent of $d$,
 the moduli spaces ${\rm Jac}^d_{g,n}$ for different $d$ are in
 general not isomorphic (see {\it e.g.}~\cite{HarrisMorrison} for further
 discussion).} ${\rm Jac}^d_{g,n}$ of stable $n$-pointed curves equipped with a degree $d$ rank one sheaf $\cL$ (that may roughly be treated as a line bundle). A dense open set of Jac$_{g,n}^d$ fibres over a dense open set of $\overline{M}_{g,n}$: 
\begin{equation}
\minCDarrowwidth25pt
	\begin{CD}
	\cJ(\Sigma) @> >>{\rm Jac}_{g,n}^d\\
	@. @V VV\\ 
	@. M_{g,n}
	\end{CD}\quad ,
\end{equation}
the dense open set being where the vertex operators do not collide. In physical terms, $\cJ(\Sigma)$ may be thought of as the moduli space of the worldsheet Abelian gauge field under which the $W(\sigma)$ are
charged, while Jac$_{g,n}^d$ parametrizes in addition the worldsheet
complex structure and the location of the $n$ vertex operators.

The path integral involves an integral over the space $\overline{M}_{g,n}(\mathbb{P}^{3|4},d)$.  A dense open set\footnote{The dense open set is where the curve~\eqref{fibre} does not degenerate, and where $h^1(\Sigma,\cL)=0$.}
in this space can be identified with a dense open set in the total space of the fibration
\be
\minCDarrowwidth25pt
	\begin{CD}
	\mathbb{CP}^{4k-1|4k} @> >> M_{g,n}(\mathbb{P}^{3|4},d)\\
	@. @V VV\\
	@. {\rm Jac}_{g,n}^d
	\end{CD}
\label{CYfibration}
\ee
where (a dense open set of) $\mathbb{CP}^{4k-1|4k}=\mathbb{P}\left(\C^{4|4}\otimes H^0(\Sigma,\cL)\right)$ is the space of worldsheet instantons $W(\sigma)$ up to overall scaling, on a fixed worldsheet with fixed choice of $\cL$. As at genus zero, these fibres are Calabi-Yau supermanifolds~\cite{Movshev:2006py, Witten:2004cp}
and so possess a canonical top holomorphic form $\Omega$ that may be
constructed explicitly as follows. At each point
$(\Sigma_g,\sigma_1,\ldots,\sigma_n;\cL)\in{\rm Jac}_{g,n}^d$, let
$\{s_r(\sigma)\}$ be a basis of $H^0(\Sigma,\cL)$. Then we may expand
\begin{equation}
	W(\sigma)=\sum_{r=1}^{k} Y_r s_r(\sigma)
\label{fibre}
\end{equation}
in terms of $k$ twistors $Y_r$, defined up to an overall
scaling. Thus  
\be
	\Omega=\frac{1}{\rm Vol(GL(1))}\prod_{r=1}^k\,\rd^{4|4}Y_r
\label{CYmeasure}
\ee
and is independent of the choice of basis. We will denote the top holomorphic form on the $(4g-3+n)$-dimensional space Jac$_{g,n}^d$ by $\rd\mu$, and incorporate into it the sections of various determinant bundles that come from the path integral over non-zero modes of the worldsheet fields (these vary holomorphically over Jac$_{g,n}^d$). In principle, this measure is determined by the worldsheet CFT and its coupling to worldsheet gravity; for example, see~\cite{Dolan:2007vv} for a careful treatment at genus $\leq 1$ in the Berkovits approach). However, our eventual hope is to obtain a form of twistor-string theory in which conformal gravity is decoupled, and the correct worldsheet theory for this is currently unknown. Thus, we do not specify $\rd\mu$ here, but hope instead to determine its properties by reverse engineering from the known amplitudes (or their leading singularities) of $\cN=4$ SYM.

As above, an $\cN=4$ Yang-Mills supermultiplet is represented on $\PT^*$ by a field  
$$
	a\in H^1(\PT^*,{\rm End}(E))
$$
describing an infinitesimal perturbation of the complex structure of a holomorphic bundle $E\to\PT^*$. (For applications to perturbation theory, this bundle may be taken to be trivial.) We can pull $n$ such perturbations back to the moduli space $\overline{M}_{g,n}(\mathbb{P}^{3|4},d)$ using the $n$ evaluation maps
\be
\begin{array}{clcl}
	{\rm ev}_i: & \overline{M}_{g,n}(\mathbb{P}^{3|4},d) &\to & \mathbb{P}^{3|4}\\
	&(\Sigma,\sigma_1,\ldots,\sigma_n;W)&\mapsto & W(\sigma_i)\ ,
\end{array}
\ee
and write $a_i(W(\sigma_i))\equiv{\rm ev}_i^*(a_i)$. To compare to the results of the previous sections, as in equation~\eqref{elemental} we use external states\footnote{A dual possibility is to consider the (equally formal) states
$$
	{\rm sgn} (W(\sigma)\cdot Z_i):=\int \frac{\rd\xi_i}{\xi_i}\, \e^{\,{\rm i}\xi_i W(\sigma)\cdot Z_i}
$$
and thus obtain an amplitude depending on fixed points $Z_i$ in \emph{twistor} space (rather than $\PT^*$). It is a non-trivial fact that these formul\ae\ for the amplitudes are simply parity conjugates of one another, proved by
Witten~\cite{Witten:2004cp} even at genus $g$ (at least at the formal level at which we work).}
\be
	a_i(W(\sigma_i)) = \delta^{3|4}(W(\sigma_i);W_i) =\int\frac{\rd\xi_i}{\xi_i}\ \delta^{4|4}(W_i-\xi_iW(\sigma_i))
\label{deltaW}
\ee
that are localised at fixed points $W_i\in\PT^*$ (which we assume are distinct). We thus obtain
the formula
\be	
	\cA(W_1,\ldots,W_n)=\oint_{\overline{M}_{g,n}(\mathbb{P}^{3|4},d)} 
	\hspace{-0.5cm}\rd\mu\wedge\Omega \ \prod_{i=1}^n\, \delta^{3|4}(W(\sigma_i);W_i)\ .
\label{Loop-conn-fin} 
\ee
for the colour-stripped amplitude. 

The Yang-Mills vertex operators~\eqref{deltaW} also take values in the adjoint representation of the gauge group. In the original models, as in the standard heterotic string, the Lie algebra indices are absorbed by coupling to a worldsheet current algebra, whose path integral leads to Green's functions $K(\sigma_i,\sigma_j)$. These Green's functions may be treated as functions on ${\rm Jac}_{g,n}^d$, so if we assume that our twistor-string theory involves such a current algebra, we can write
\be
	\rd\mu = \rd\mu'\, \sum_{\rm perms} {\rm T}\, \prod\, K(\sigma_i,\sigma_j)\ ,
\label{originalmeasure}
\ee
where the sum runs over all ways of performing $n$ contractions in the current correlator, each of which leads to a Green's function $K(\sigma_i,\sigma_j)$ for some $i,j$, together with a colour factor T. In particular, the single trace contribution is given by
\be	
	\left.\cA(W_1,\ldots,W_n)\right|_{\rm single\ trace}=\sum_{\rm non-cyclic} \oint\rd\mu'\wedge\Omega \ 
	\prod_{i=1}^n\, K(\sigma_i,\sigma_{i+1})\,\delta^{3|4}(W(\sigma_i),W_i)
\label{Loop-conn-nearly-fin} 
\ee
(suppressing the overall trace). Both because of the multi-trace terms
at genus zero and the fact that the current algebra contributes a
central charge that depends on the gauge group, it is not clear that
such a current algebra should be part of a twistor-string for pure
$\cN=4$ SYM, and we do not commit ourselves 
to~\eqref{originalmeasure}. At tree level, however,  the leading trace
contribution of the measure~\eqref{originalmeasure} is thought to be
correct. 
In particular, when $g=0$, 
\be
{\mathrm {Jac}}_{0,n}^d\simeq\{\mathrm{point}\}\times\overline{M}_{0,n},
\ee 
a dense open set of which is
$\left(\Sigma^{\times n} -\Delta\right)/{\rm PGL(2)}$, where
$\Sigma\cong\CP^1$ and $\Delta$ are the diagonals describing vertex
operator collisions. There is then no ambiguity in $\rd\mu$. Treating
$\sigma$ as an affine worldsheet coordinate, we have 
\be
	\rd\mu' = \frac{1}{\rm Vol(PGL(2))}\,\prod_{i=1}^n\, \rd\sigma_i
	\qquad\hbox{and}\qquad
	K(\sigma_i,\sigma_{i+1})=\frac{1}{\sigma_i-\sigma_{i+1}}
\ee
so that the single trace contribution of the original twistor-string becomes 
\be\label{tree-twistor-int}
	\left.\cA(W_1,\ldots,W_n)\right|_{\rm single\ trace}=\sum_{\rm non-cyclic}
	\oint\frac{\rd^{4k|4k}Y}{\rm Vol (GL(2))} \wedge\prod_{i=1}^n\, \frac{\rd\sigma_i}{\sigma_i-\sigma_{i+1}}
	\,\delta(W_i\,,W(\sigma_i))\ .
\ee
For evidence that this formula does indeed compute tree level amplitudes in $\cN=4$ SYM, see {\it e.g.}~\cite{Roiban:2004vt, Roiban:2004yf,Spradlin:2009qr,Dolan:2009wf}.

The integrand of~\eqref{Loop-conn-fin} varies holomorphically over $\overline{M}_{g,n}(\mathbb{P}^{3|4},d)$, so (the bosonic part of) the path integral should be treated as a contour integral. In $(++--)$ space-time signature, the contour is determined by introducing real structures $\tau_1:\PT^*\to\PT^*$  and $\tau_2:\Sigma\to\Sigma$  that leave fixed an $\R\mathbb{P}^3$ submanifold of twistor space and (at genus zero) an equatorial $S^1\subset\Sigma$, respectively. The contour is then the locus of real maps (those obeying $\tau_1W=W\tau_2$), as in the Berkovits model~\cite{Berkovits:2004hg}. Contour choices appropriate for other space-time signatures have not yet been constructed.


\subsection{Nodal Curves and Leading Singularities}
\label{sec:nodal}

In section~\ref{sec:twsgenunit} we saw that in $\PT^*$,
leading singularities may be constructed by gluing tree amplitudes
using the inner product~\eqref{Twistorinnerproduct}. If our
tree subamplitudes are built from the twistor-string tree formula
\eqref{tree-twistor-int} above, the we will see in this subsection
that the resulting support is on a connected curve whose degree and
genus are respectively determined by the MHV level and loop order of
the leading singularity. By construction, these curves are not
irreducible --- they have many nodes.  In the next subsection we will
examine how such formul\ae\ should arise as a reduction from a putative
full twistor-string path integral at genus $g$ in which we change the
physical contour to one that picks up residues at poles of the
integrand on the subset of moduli space on which the curves become
nodal.

In more detail, to form a leading singularity, choose a channel diagram with $\nu$ tree subamplitudes and $\delta$ (cut)
propagators joining them together.  Let the tree subamplitudes be represented by \eqref{tree-twistor-int}, an integral over a moduli space of maps of a rational curves $\Sigma_i$, $i=1,\ldots,\nu$ into $\PT^*$.  Using \eqref{Twistorinnerproduct}, the leading singularity is obtained by an integral over the moduli of such all such curves glued together at $\delta$ pairs of marked points leading to a nodal curve $\Sigma$ with one component $\Sigma_i$, $i=1,\ldots,\nu$ for each tree subamplitude and $\delta$ nodes. Clearly, $(\nu,\delta)$ give the number of
subamplitudes and cut propagators respectively in the momentum channel
diagram of the leading singularity. A curve $\Sigma$ with $\delta$ nodes 
and $\nu$ irreducible components $\Sigma_i$, each of
which are rational, has genus 
\be 
	g=\delta-\nu+1\ . 
\ee 
agreeing with the loop order of the leading singularity. Because $\ell$-loop primitive leading singularities involve $4\ell$ 
cut propagators, the corresponding curve has $\delta=4g$ nodes, implying that there are 
\be 
	\nu=3g+1
\label{component bound}
\ee 
irreducible worldsheet components. 

\begin{figure}[t]
\centering
	\includegraphics[height=30mm]{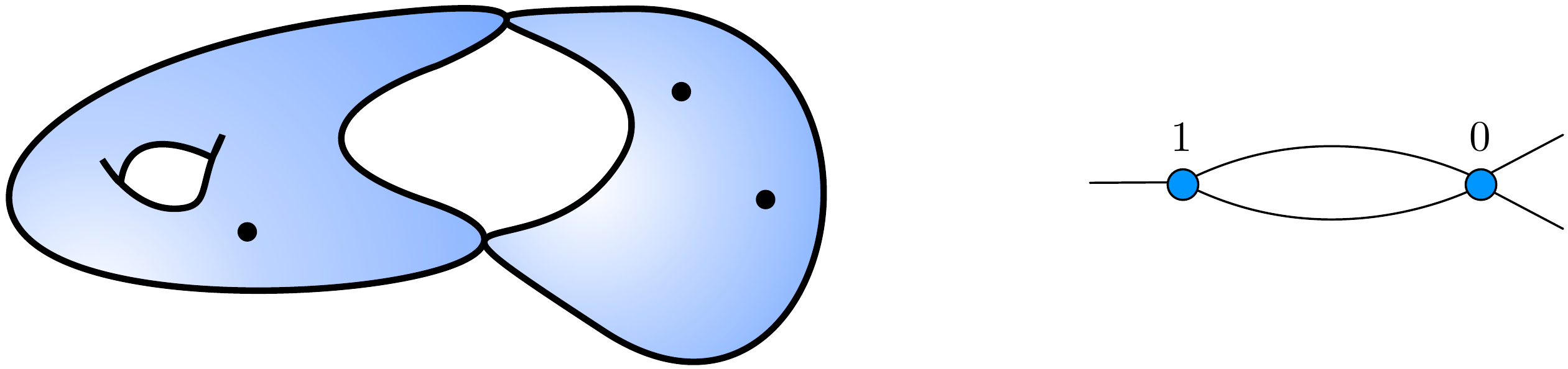}
\caption{{\it An example of a degenerate 3-pointed curve of genus 2, together with its dual graph $\Gamma$. Each vertex of $\Gamma$ represents an irreducible component of the curve, and is labelled by the genus of that component. Such curves live in the boundary stratum $(M_{1,2}\times M_{0,4})/{\rm Sym}\,\Gamma$ of $\overline{M}_{2,3}$, where ${\rm Sym}\,\Gamma$ is the symmetry group of the dual graph (of order 4 in this example). This boundary has codimension two in $\overline{M}_{2,3}$.}}
\label{fig:dualgraph}
\end{figure}

This information may be summarised by the \emph{dual graph} of $\Sigma$. This is defined to be the graph whose vertices correspond to the irreducible components of the curve, and are labelled by the genus of that curve component (although we will typically drop this label as we are mostly interested in the case where all components are rational). Two vertices are connected by an edge if there is a node connecting the two corresponding components.  The marked points --- corresponding to external states --- are represented by external edges joined to the appropriate vertex; see figure~\ref{fig:dualgraph} for an example. 

As in the smooth case, the embedding of such a nodal curve $\Sigma$ into twistor space induces a line bundle $\cL=W^*\cO(1)$ whose degree on each component $\Sigma_i$ determines the MHV degree of that tree subamplitude. The total MHV degree of the leading singularity is still given by~\eqref{twsdg}, as may be seen directly from the gluing argument of section~\ref{sec:twsgenunit} (or the momentum space argument) as follows: For each irreducible, rational component $\Sigma_i$ we have $h^0(\Sigma_i,\cL)=d_i+1$, so that the component represents a tree level N$^{d_i-1}$MHV subamplitude in the momentum channel diagram. Ignoring the gluing at the nodes, $\cL$ has $\sum_{i=1}^\nu d_i+1=d+\nu$ holomorphic sections.  However, consistency at the nodes
gives $\delta$ conditions, just as each propagator in the channel
diagram involves an integral over $\cN=4$ on-shell momentum
superspace, lowering the MHV degree compared to the constituent
subamplitudes. In total, 
\be
	h^0(\Sigma, \cL)=d+\nu-\delta=d+1-g
\label{degree-nodal}
\ee 
(generically) as for the smooth curve.

\begin{figure}[t] 
\centering
	\includegraphics[height=45mm]{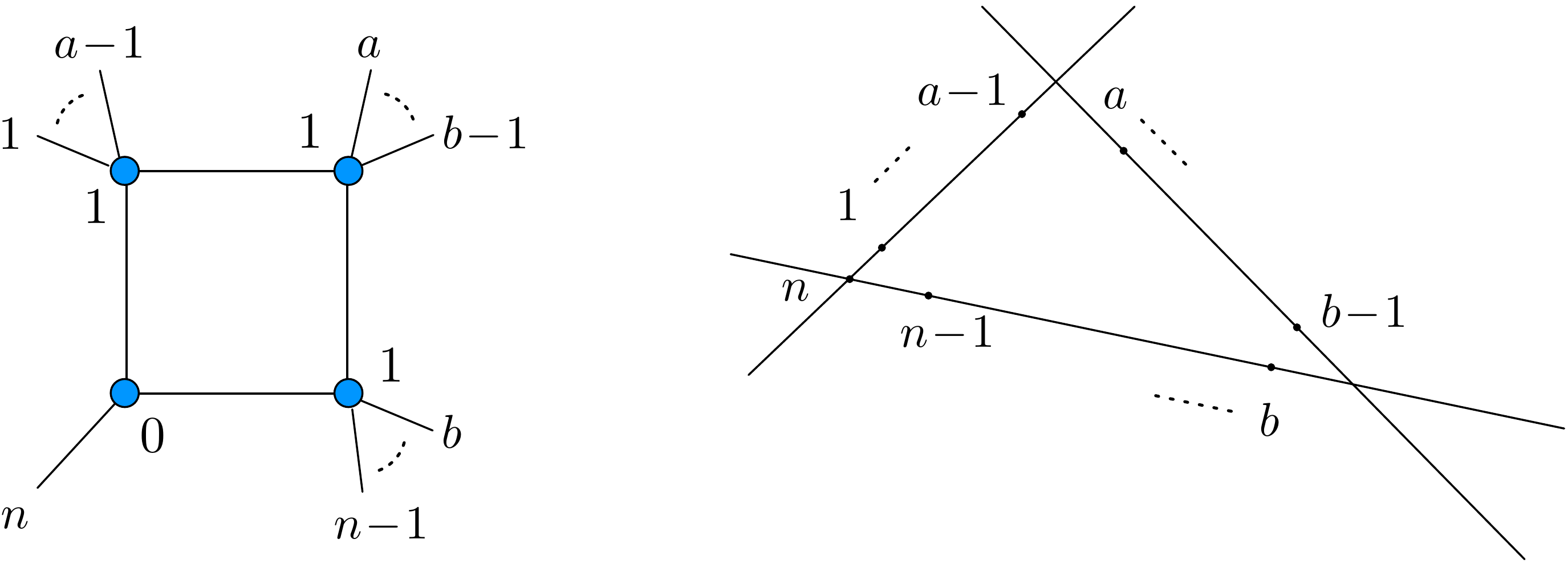}
\caption{{\it Primitive leading singularity channel diagrams in momentum space can also be interpreted as a picture of the dual graph of the twistor-string worldsheet, shown on the left. The} MHV {\it degree of each amplitude is determined by the degree of each component of the map. We illustrate this with the dual graph corresponding to the boundary components of $\overline{M}_{1,n}(\mathbb{P}^{3|4},3)$ (the moduli space for 1-loop} NMHV {\it amplitudes) that are mapped onto the configuration of a 3-mass box coefficient in $\PT^*$. The vertices of the dual graph are labelled by the degree of the map component (each vertex represents a rational curve, so we drop the genus label).}}
\label{fig:dualgraph2}
\end{figure}

If we additionally label the vertices of the dual graph by the degree of the restriction $\cL|_{\Sigma_i}$, then the dual graph becomes precisely the channel diagram; see figure~\ref{fig:dualgraph2} for the example of the three mass box. Thus, the channel diagrams of section~\ref{sec:KS} are all examples of such dual graphs\footnote{In all these pictures, note that the picture of the twistor space support does not properly
represent the original nodal curve, because components of degree
zero map to a point. For example, in the three mass box the original nodal
curve has four components arranged in a square, but the image in
twistor space is just a triangle with the component of degree zero
mapped to the point at the marked point $n$ in figure~\ref{fig:dualgraph2}.}.

\medskip

Thus, although we do not know the correct volume form $\rd\mu$ to use for a pure $\cN=4$ SYM
twistor-string path integral at $g>0$, combining~\eqref{tree-twistor-int} with~\eqref{Twistorinnerproduct}
completely determines the correct formula for leading singularities of arbitrary loop amplitudes.
As an example, we return to the 1-loop NMHV amplitude in the 3 mass box channel.  We give a different parametrization here that brings out the conformal invariance of the operations, and that will be of more use in the discussion of the map to the Grassmannian in section~\ref{sec:twsgrass}. As in~\eqref{GenUnNMHV}, the associated leading singularity is 
\be
\label{3mb-leadsing-twis-gluing}
	\int \rD^{3|4}W'\,\rD^{3|4}W'' \,\cA_{\rm MHV}^{(0)}(n,1,\ldots,W')\,\cA_{\rm MHV}^{(0)}(W',a,\ldots,b\!-\!1,W'')\,
	\cA_{\rm MHV}^{(0)}(W'',b,\ldots,n)\ .
\ee 
Each MHV tree subamplitude may be described in terms of a degree 1 map from a component of the worldsheet, parameterized by\footnote{Strictly, we should also use a different parameter $\sigma$ on each line. We drop this distinction in what follows. In the final formula~\eqref{3mb-leadsing-twis} the $\sigma$s are in any case distinguised by their external particle labels.}
\be
	W(\sigma) =
		\begin{cases}
			 Y_1+\sigma Y_2 \quad&\hbox{for }\ \cA_{\rm MHV}^{(0)}(n,1,\ldots,a\!-\!1,W'),\\
			 U_1+\sigma U_2\quad&\hbox{for }\ \cA_{\rm MHV}^{(0)}(W',a,\ldots,b\!-\!1,W''),\\
			 V_1+\sigma V_2\quad&\hbox{for }\ \cA_{\rm MHV}^{(0)}(W'',b,\ldots,n\!-\!1,n)\ .	 	\end{cases}
\ee 
on each rational component. From~\eqref{tree-twistor-int}, we have
\be
	\cA_{\rm MHV}^{(0)}(n,\ldots,W') = \int \frac{\rd^{4|4}Y_1\,\rd^{4|4}Y_2}{\rm Vol(GL(2))}\wedge 
	\frac{\rd\sigma'\,\delta^{3|4}(W',W(\sigma'))\wedge\prod_{i=n}^{a-1}\rd\sigma_i\,\delta^{3|4}(W_i,W(\sigma_i))}
	{(\sigma'-\sigma_n)(\sigma_n-\sigma_1)\cdots(\sigma_{a-1}-\sigma')}\ ,
\label{MHVtreeconn}
\ee
where $\sigma'$ is the location of the vertex operator corresponding to the auxiliary point $W'$. It is convenient to partially fix the GL(2) redundancy by setting $\sigma_n=0$ and $\sigma'=\infty$ so that~\eqref{MHVtreeconn} becomes
\be
\label{MHV-Y} 
	\cA_{\rm MHV}^{(0)}(n,\ldots,W')=\int \frac{\rd^{4|4}Y_1\,\rd^{4|4}Y_2}{({\rm Vol\, GL(1)})^2}\wedge
	\frac{\rd\xi'}{\xi'}\,\delta^{4|4}(W'-\xi' Y_2)\wedge 
	\frac{\prod_{i=n}^{a-1} \rd\sigma_i\,\delta^{3|4}(W_i\,,W(\sigma_i))}
	{\sigma_1 (\sigma_2-\sigma_1) \ldots (\sigma_{a-1}-\sigma_{a-2})} \ .
\ee 
The remaining symmetry is GL(1)$^2$ --- one copy of GL(1) rescales the $\sigma_i$ (and can be used to fix one of them to unity), while the other rescales the $Y_1,Y_2$. There are similar formul\ae\ for the remaining two MHV tree amplitudes in the product~\eqref{3mb-leadsing-twis-gluing} that use the reference twistors $\{U_1,U_2\}$ and $\{V_1,V_2\}$ in place of $\{Y_1,Y_2\}$. 

In the gluing formula~\eqref{3mb-leadsing-twis-gluing}, each of $W',W''$ and $W_n$ have $\delta$-functions attaching them to two of the subamplitudes. One set of $\delta$-functions in $W'$ and one set in $W''$ may be used to perform the $\rD^{3|4}W'\,\rD^{3|4}W''$ integrations directly. The remaining $\delta$-functions in these variables may then be used to replace $U_1$ by $Y_2$ and $V_1$ by $U_2$. The $\delta$-function $\delta^{4|4}(W_n-\xi_nY_1)$ in~\eqref{MHV-Y} and a similar $\delta$-function in $\cA(W'',b,\ldots,n)$ can then be used to replace $V_2$ by $Y_1$. The remaining auxiliary twistors are then $\{Y_1,Y_2,U_2\}$ which we relabel as $\{Y_1,Y_2,Y_3\}$. Finally, we can use three copies of GL(1) to remove three of the $\xi$ integrals.  After all this, we are left with
\be
\label{3mb-leadsing-twis} 
	\cA^{\rm LS}_{\rm NMHV}(1,\ldots,n)
	=\int \frac{\prod_{r=1}^3\rd^{4|4}Y_r}{{\rm (Vol\,GL(1))}^3}\,\frac{1}{\Delta}\wedge
	\frac{\rd\xi_n}{\xi_n}\,\delta^{4|4}(W_n-\xi_n Y_1)\,\wedge\,
	\prod_{i=1}^{n-1}\rd\sigma_i\wedge\frac{\rd\xi_i}{\xi_i}\,\delta^{4|4}(W_i-\xi_i W(\sigma_i)) 
\ee 
where 
\be
\label{delta-def}
	\Delta:=\sigma_{1}(\sigma_{2}-\sigma_{1}) \ldots(\sigma_{a-1}-\sigma_{a-2}) \sigma_{a}
	(\sigma_{a+1}-\sigma_{a}) \ldots (\sigma_{b-1}-\sigma_{b-2})(\sigma_{b+1} -\sigma_{b}) \ldots (0-\sigma_{n-1})
\ee
is the analogue here of the Parke-Taylor denominator.

\medskip

\begin{figure}[t]
\begin{center}
\includegraphics[height=45mm]{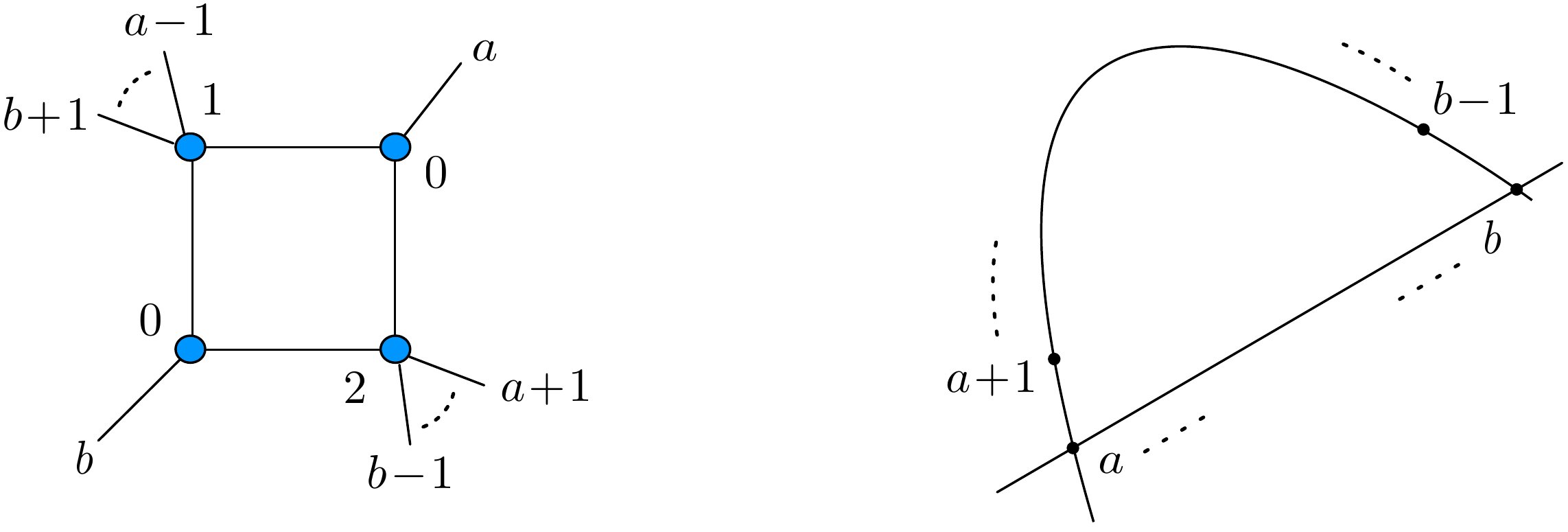}
\end{center}
\caption{{\it A non-primitive 1-loop} NMHV {\it leading singularity, contributing to the two mass easy and one mass box channels. The vertices of the dual graph are rational curves, labelled by the degree of the map component.}}\label{NMHVtwomassboxFig}
\end{figure}

We emphasize that, from the point of view of twistor-strings, the
requirement that the leading singularity contain only tree level MHV
and $\MHVbar_3$ subamplitudes is not at all essential; one can study a
channel whose subamplitudes are arbitrary N$^p$MHV or $\MHVbar_3$ by
using~\eqref{tree-twistor-int} for each of the constituent tree
subamplitudes, appropriately integrated against each other
using~\eqref{Twistorinnerproduct}. For example, consider the 2 mass
easy channel in the 1-loop NMHV amplitude (see
figure~\ref{NMHVtwomassboxFig}; the discussion applies equally to the
`one mass' case in which only one external state is attached to the
MHV subamplitude). In this case, the leading singularity is obtained
by integrating over a space of $g=1$ nodal curves having two
components, one with $d=2$ and the other with $d=1$. This is
analogous to a representation of a tree amplitude that is intermediate
between a smooth degree $d$ curve and $d$ intersecting
lines~\cite{Bena:2004ry}.


\subsection{A Leading Singularity Conjecture for Twistor-Strings}
\label{sec:twsleadsing}

In this subsection we examine how the formul\ae\ for leading
singularities obtained in the previous subsection should arise as a
reduction from a putative full twistor-string path integral at arbitrary genus,
in which the physical contour is replaced by one that picks up residues
at poles of the integrand on the subset of moduli space on which the
curves become nodal.

We first briefly review the compactification of $M_{g,n}(\mathbb{P}^{3|4},d)$ by moduli spaces of maps from nodal
curves.  The moduli space $\overline{M}_{g,n}(\bP^{3|4},d)$ of degree $d$, stable maps from an $n$-pointed, genus $g$ curve $(\Sigma_g;\sigma_1,\ldots,\sigma_n)$ inherits\footnote{At least when $g=0,1$.} its boundaries from the moduli space $\overline{M}_{g,n}$ of stable $n$-pointed curves via the morphism 
\be
	\pi:\overline{M}_{g,n}(\mathbb{P}^{3|4},d)\to\overline{M}_{g,n}\ ,
\label{forgetful}
\ee
obtained by simply forgetting about the map to the target (and contracting any components of the worldsheet that become unstable as a result). The boundary $\overline{M}_{g,n}\backslash M_{g,n}$ describes curves on which a cycle of $\Sigma_g$ has pinched, or where marked points (and hence vertex operators) have collided and bubbled off a new component of $\Sigma$, so that in the limit, the two marked points end up as distinct points on a rational curve that is attached to the rest of the worldsheet at a node. As before, we can conveniently describe the various boundary components of $\overline{M}_{g,n}$ by specifying an associated dual graph (see figure~\ref{fig:dualgraph}), and again, the vertices of the inverse image 
$$
	\pi^{-1}\left(\Gamma\right) \subset \overline{M}_{g,n}(\bP^{3|4},d)\backslash M_{g,n}(\mathbb{P}^{3|4},d)
$$
carry an additional label, denoting the degree $d_i$ of the worldsheet map on each component.  

The configurations derived in sections~\ref{sec:KS}~\&~\ref{sec:twsgenunit} are associated with
codimension $4g$ boundary components of the twistor-string moduli space. The codimension of the space nodal curves with dual graph $\Gamma$ (inside either $\overline{M}_{g,n}$ or $\overline{M}_{g,n}(\bP^{3|4},d)$) is just given by the number of nodes $\delta$ --- the number of propagators in the dual graph. This agrees with the fact that in momentum space, leading singularities freeze each loop momentum at some specific (generically complex) value, each lying on the complexified null cone, so that leading singularities are also codimension $4g$ in the space $\C^{4g}$ of complexified $g$-loop momenta.

More specifically, leading singularities may be computed by analytically extending the $g$-loop momentum space integrand (consisting of field theoretic propagators and vertices  --- the sum of all Feynman diagrams) to a rational function on $\C^{4g}$, and then integrating this over a contour\footnote{The specific contour depends on the channel of leading singularity one wishes to compute.} of topology $T^{4g}\subset\C^{4g}$, rather than the physical contour $(\R^{3,1})^{\times g}\subset\C^{4g}$. Now, we have argued above that any (non self-dual) twistor-string theory must involve a path integral
\be
	\oint \rd\mu\wedge\Omega \ \prod_{i=1}^n\, \delta^{3|4}(W(\sigma_i);W_i)\ .
\label{Loop-contour}
\ee
to be taken over some contour $\Gamma\subset\overline{M}_{g,n}(\bP^{3|4},d)$. In order to extract leading singularities from this path integral, the results above show we must choose a contour that fibres over a $4g$-dimensional torus $T^{4g}\subset{\rm Jac}_{g,n}^{d}$, each factor of which encircles\footnote{One can easily imagine situations in which this prescription is too naive, such as when $4g>4d+n$ or when this contour becomes pinched by the many other singularities of $\overline{M}_{g,n}(\mathbb{P}^3,d)$. Our hope is that it is adequate to obtain at least the primitive leading singularities.} a boundary component of $\overline{M}_{g,n}(\mathbb{P}^{3|4},d)$. So that this contour does indeed localise on the leading singularity configurations, the measure $\rd\mu$ must have a simple pole on each codimension 1 boundary component where the curve develops a node, whose residue is an integral over the space of nodal curves of precisely the type discussed in section~\ref{sec:nodal} above. For example, at one-loop we expect to expand the holomorphic map $W(\sigma)$ in a basis of theta functions (with characteristic determined by the MHV degree). On the leading singularity locus, these theta functions degenerate into polynomials while the residue of the measure $\rd\mu'$ must look like a sequence of Parke-Taylor denominators on each line, as in the explicit construction~\eqref{3mb-leadsing-twis}-\eqref{delta-def} of the 1-loop NMHV leading singularity in the 3 mass box channel.

According to the leading singularity conjecture in momentum
space~\cite{Cachazo:2008dx}, the complete integrand of the full loop
expression is determined by its leading singularities. It is natural
to similarly conjecture that \emph{the correct path integral measure
  --- and thereby the correct worldsheet CFT --- of a twistor-string
  theory for pure $\cN=4$ super Yang-Mills is likewise fixed by
  requiring it reproduces the leading singularities} constructed as in
section~\ref{sec:nodal}\footnote{In particular, because the
  leading singularities of the 1-loop amplitude in $\cN=4$ SYM are all
  in  channels with the topology of a box, the $g=1$ twistor-string
  should be determined by its residue on the sublocus of
  $\overline{M}_{1,n}(\bP^{3|4},d)$ where the worldsheet torus has a
  singularity of Kodaira type $I_4$ (four pairwise intersecting
  rational curves).}.  

Remarkably, at $g=0$, Gukov, Motl \& Neitzke~\cite{Gukov:2004ei} showed that the integrand~\eqref{tree-twistor-int} of the original twistor-string models does indeed have poles on the codimension one components of $\overline{M}_{0,n}(\mathbb{P}^{3|4},d)$ corresponding to nodal curves, {\it i.e.}, where the rational curve breaks into two rational curves meeting at a point. However, while the simple poles do correspond to the pure SYM tree amplitude, there are also higher-order singularities that involve conformal supergravity.  In section~\ref{sec:twsgrass} we study a duality between the twistor-string path integral over $\overline{M}_{g,n}(\mathbb{P}^{3|4},d)$ and the Grassmannian G$(k,n)$. The meromorphic form on G$(k,n)$ introduced by Arkani-Hamed {\it et al.} precisely has simple poles on leading singularities, as required by twistor-string theory.

\begin{figure}[t]
\centering
	\includegraphics[height=80mm]{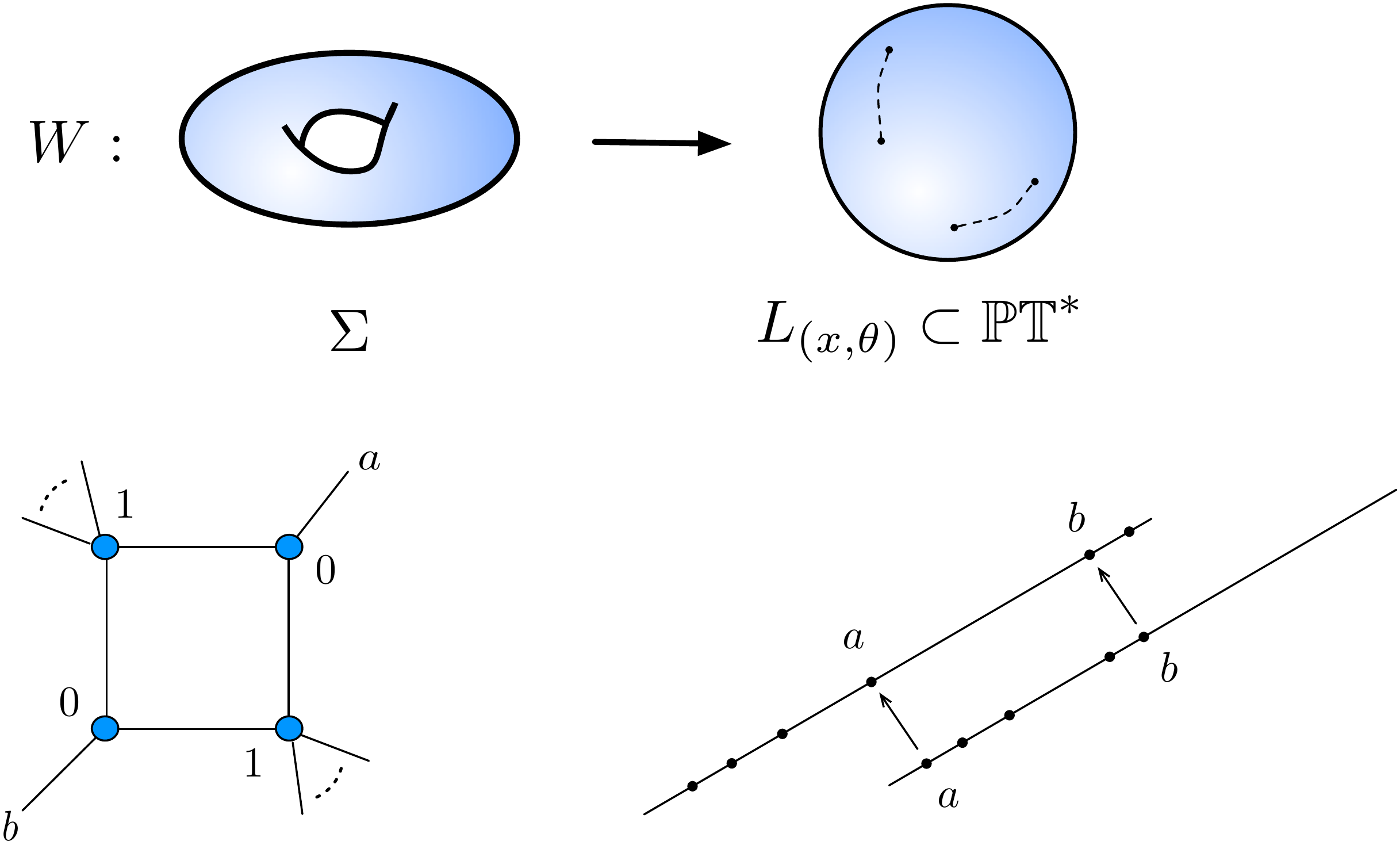}
\caption{{\it In twistor-string theory, 1-loop} MHV {\it amplitudes are associated with a branched cover of a line in $\PT^*$. The boundary component of $\overline{M}_{1,n}(\mathbb{P}^3,2)$ corresponding to the 2-mass easy leading singularity is shown at the bottom left. The support on a double cover of a line can clearly be seen from the dual graph (or equivalently the momentum space channel diagram):  each of the two lines corresponding to the} MHV {\it subamplitudes must support the same two distinct points corresponding to the 3-point $\overline{\rm MHV}$ subamplitudes.}}
\label{fig:branching}
\end{figure}

\medskip

There is one aspect of this conjecture that we can clarify
immediately. It is perfectly possible for a high loop amplitude to
have the same leading singularity as a lower loop amplitude --- one
need look no further than MHV for examples. How is this compatible
with the fact that, for fixed order in N$^p$MHV, the degree $d=p+1+g$
of the map to twistor space depends on the loop level $g$? Let us
consider the MHV case in more detail. Here, twistor-string theory
leads us to expect that the $1$-loop amplitude is obtained from a
degree 2 holomorphic map of a genus 1 worldsheet into
$\PT^*$. However, \emph{all degree 2 curves in $\CP^3$ have genus
  zero}, so the map cannot be an embedding. We should thus consider
degree 2 maps from $\Sigma$ whose image is a double cover of a line
$L_{(x,\theta)}\subset\PT^*$, branched over four points (see
figure~\ref{fig:branching}). Imposing the quadruple cut to extract the
leading singularity presumably then computes the periods of the loop
amplitude as various combinations of external states are taken on
excursions through these branch cuts, so that the leading singularity
itself is a rational function supported on a line in $\PT^*$. In
particular, codimension 4 boundary components of
$\overline{M}_{1,n}(\mathbb{P}^3,2)$ are represented by dual graphs
with four propagators and four vertices. Since the total degree of the
map is 2, at most two of these vertices can be associated with maps
with $d_i>0$ so we obtain the 2-mass easy configuration of
figure~\ref{fig:branching} that indeed corresponds to localisation on
the double cover of a line\footnote{The 2-mass hard configuration is
  ruled out for essentially the same reason as in momentum space: it
  requires that the two twistors at the adjacent massless corners
  \emph{coincide}.}. Similarly, for N$^p$MHV leading singularities, we
can increase the degree (and hence the loop order) whilst keeping the
$\PT^*$ support unchanged by allowing various line components of the
image $W(\Sigma)$ to become multiply covered.


\section{A Twistor-String / Grassmannian Duality}
\label{sec:twsgrass}

According to a conjecture of Arkani-Hamed {\it et al.}~\cite{ArkaniHamed:2009dn},  \emph{all} leading singularities of arbitrary loop, planar, $n$-particle N$^{k-2}$MHV amplitudes in $\cN=4$ SYM may be obtained from the contour integral\footnote{In fact, \cite{ArkaniHamed:2009dn} originally defined $\cL_{k,n}$ in \emph{twistor} space $\PT$ (rather than \emph{dual} twistor space $\PT^*$), writing
$$
	\int\prod_{r=1}^k\,\rd^{4|4}Y_r\,\prod_{i=1}^n\,\delta^{4|4}(W_i-Y_rC_{ri})
	=\int\prod_{i=1}^{n-k}\rd^{4|4} Z_i\ \e^{-\im Z_i\cdot W_i}
	\left\{\prod_{r=1}^k\,\delta^{4|4}\!\left(\sum_{i=1}^nC_{ri}Z_i\right)\right\}\ ,
$$
but the Fourier transformed version~\eqref{G} is more useful for our purposes. Of course, the twistor and dual twistor formulations differ only by parity conjugation, realised here via the duality transformation $$*:{\rm G}(k,\C^n)\to {\rm G}(n-k,{\C^n}^*)$$ that takes a $k$-plane to its orthogonal complement in the dual $\C^n$. In either form, the integrand of $\cL_{k,n}(W_i)$ enjoys manifest superconformal and dihedral invariance, although this may be broken by the choice of contour. Parity invariance is straightforward to demonstrate. It was also shown in~\cite{Mason:2009qx} that a similar integral can also be written in \emph{momentum twistor space}~\cite{Hodges:2009hk}, where dual superconformal invariance~\cite{Drummond:2008vq} is manifest. That the two integrals are equal was proved soon after in~\cite{ArkaniHamed:2009vw}.}
\be
	\cL_{k,n}(W_i) = 
	\oint\frac{{\rm D}^{k(n-k)}C}{(1,2,\ldots, k)\cdots(n,1,\ldots, k\!-\!1)}\,\prod_{r=1}^k\,\rd^{4|4}Y_r \,
	\prod_{i=1}^n\,\delta^{4|4}\!\left(W_i-Y_rC_{ri}\right)\ ,
\label{G}
\ee
where $(i,i+1,\ldots,k+i-1)$ denotes determinant of the $i^{\rm th}$ cyclic minor of the $k\times n$ matrix 
$$
C_{ri}=	\begin{pmatrix}
		C_{11}& C_{12} & &\cdots & &C_{1n}\\
		C_{21}& C_{22} & &\cdots & &C_{2n}\\
		\vdots &\vdots &  & & &\vdots\\
		C_{k1}& C_{k2} & &\cdots &  &C_{kn}
	\end{pmatrix}\ .
$$
This matrix defines\footnote{In order that these $k$ vectors do span a
  $k$-plane, we assume that the $k$ row vectors are linearly
  independent.} a $k$-plane $C\subset\C^n$ through the origin, and the
space of such $k$-planes is the Grassmannian G$(k,n)$. Leading
singularities are associated with the residue form of~\eqref{G} on
$2(n-2)$-dimensional subcycles of G$(k,n)$ on which the denominator
of~\eqref{G} vanishes to order $(k-2)(n-k-2)$.

In this section, we show that the twistor-string moduli space can be mapped into the same Grassmannian G$(k,n)$ as arises in~\eqref{G}. Perhaps surprisingly, this map exists \emph{even at genus $g$}, and even for the moduli space of the complete loop amplitudes. When $g\geq1$, if we restrict attention to the codimension $4g$ boundary components of the moduli space appropriate for a primitive leading singularity, then this map is from a $2(n-2)$-cycle whose image coincides with the cycles defined by poles of~\eqref{G}.


\subsection{The Map to the Grassmannian}
\label{sec:duality}

We begin by sharpening our understanding of the twistor-string vertex operators
\be 
	a_i(W(\sigma_i)) =\delta^{3|4}(W_i\,,W(\sigma_i)) 
	:=\int\frac{\rd\xi_i}{\xi_i}\ \delta^{4|4}(W_i-\xi_iW(\sigma_i))\ .
\label{deltaW2}
\ee 
Each $\xi_i$ must scale so as to compensate the scaling of $W(\sigma_i)$ so that it makes sense to ask  that the product $\xi_iW(\sigma_i)$ equals some fixed point\footnote{Any \emph{common} scaling of $W_i$ and $\xi_i$ drops out of~\eqref{deltaW2} because of cancellation between the bosonic and fermionic $\delta$-functions. The $\xi_i$ must take values in $\left.\cL^{-1}\right|_{\sigma_i}$ because the worldsheet map $W(\sigma)$ and the point $W_i$ do not have to scale in the same way {\it a priori}.} $W_i$ in non-projective dual twistor space.  In other words, the $\xi_i$ are points in the fibres $\cL^{-1}|_{\sigma_i}$ --- the
pullback to $\Sigma$ of the tautological bundle on $\PT^*$, restricted
to the $i^{\rm th}$ marked point. As we move around in
$\overline{M}_{g,n}(\mathbb{P}^{3|4},d)$, the $\cL^{-1}|_{\sigma_i}$
fit together to form the coherent sheaf ${\rm ev}^*_i\cO(-1)$ (that may roughly be treated as a line bundle). Thus, if we include the $\xi_i$ integrals in~\eqref{deltaW2}, the path integral
\be
	\oint\rd\mu\wedge\Omega\wedge\prod_{i=1}^n\, \frac{\rd\xi_i}{\xi_i}\,
	\delta^{4|4}\!\left(W_i - \xi_iW(\sigma_i)\right)\, .  
\label{pathcont}
\ee	
is really taken over (a contour in) the total space of 
\begin{equation}
\minCDarrowwidth25pt
	\begin{CD}
	(\C^*)^n @> >> \bigoplus_{i=1}^n {\rm ev}_i^*\cO(-1) \\
	@. @V VV\\ 
	@.  \overline{M}_{g,n}(\mathbb{P}^{3|4},d)
	\end{CD}
\end{equation}
with holomorphic volume form $\rd\mu\wedge\Omega\wedge\prod_{i=1}^n\rd\xi_i/\xi_i$.

To understand the relation to the Grassmannian, we follow equations~\eqref{fibre}~\&~\eqref{CYmeasure} and write
\be
	W(\sigma)=\sum_{r=1}^kY_rs_r(\sigma)
	\qquad\hbox{and}\qquad
	\Omega=\frac{1}{\rm Vol(GL(1))}\prod_{r=1}^k\,\rd^{4|4}Y_r \ ,
\ee
where $k\equiv h^0(\Sigma,\cL)$ by definition, and contributions to the $g$-loop N$^p$MHV amplitude come from maps with $k=p-2$.  The integrals over the exact same $Y_r$ also appear in the Grassmannian residue formula~\eqref{G}, so we wish to keep these explicit. To do so, we must take the GL(1) to act diagonally on the $\xi_i$s. The remainder of the path integral may then be understood as follows. As discussed in section~\ref{sec:pathintegral}, there is a natural projection 
\be
\begin{array}{lrcl}
	p: &\overline{M}_{g,n}(\bP^{3|4},d)&\to &{\rm Jac}_{g,n}^d\\
	&(\Sigma,\sigma_1,\ldots,\sigma_n;W)&\mapsto &(\Sigma,\sigma_1,\ldots,\sigma_n,\cL)
\end{array}
\ee
that forgets about the $Y_r$s and remembers only the abstract worldsheet, its $n$ markings and the degree $d$ line bundle $\cL$. We can use this projection to push down the ev$_i^*\cO(-1)$ to give rank one sheaves\footnote{Given a map $p:M\to N$ and a sheaf $E$ on $M$,  the direct image sheaf $p_*E$ on $N$ is defined by 
$$
	\left.p_*E\right|_U := H^0(p^{-1}(U),E)
$$
for $U\subset N$ an open set.} $p_*{\rm ev}_i^*\cO(-1)$ on Jac$_{g,n}^d$. Thus, with the $Y_r$ separated out, the remaining path integral variables parametrize the total space of $\bP\left(\bigoplus_{i=1}^np_*{\rm ev}_i^*\cO(-1)\right)$,
where the overall projectivization comes from the overall GL(1) scaling. A dense open set of this space may be thought of as the total space of a fibre bundle with $\CP^{n-1}$ fibres:
\be\label{GC-space}
\minCDarrowwidth25pt
	\begin{CD}
	\CP^{n-1} @> >> \mathbb{P}\left(\bigoplus_{i=1}^n p_*{\rm ev}_i^*\cO(-1)\right)  \\
	@. @V VV\\ 
	@. \mathrm {Jac}^d_{g,n}
	\end{CD}\quad .
\ee 
Since ${\rm dim}\left({\rm Jac}_{g,n}^d\right)=4g-3+n$ and the $\xi_i$ give a further $n$ parameters,  accounting for the overall scaling we have
\be
\begin{aligned}
	{\rm dim}\ \mathbb{P}\left(\bigoplus_{i=1}^n p_*{\rm ev}_i^*\cL^{-1}\right)
	&=(4g-3+n)+n-1\\
	&=2n-4+4g\ ,
\end{aligned}
\label{dimcycle}
\ee
at least generically.

\begin{figure}[t]
\begin{center}
	\includegraphics[height=40mm]{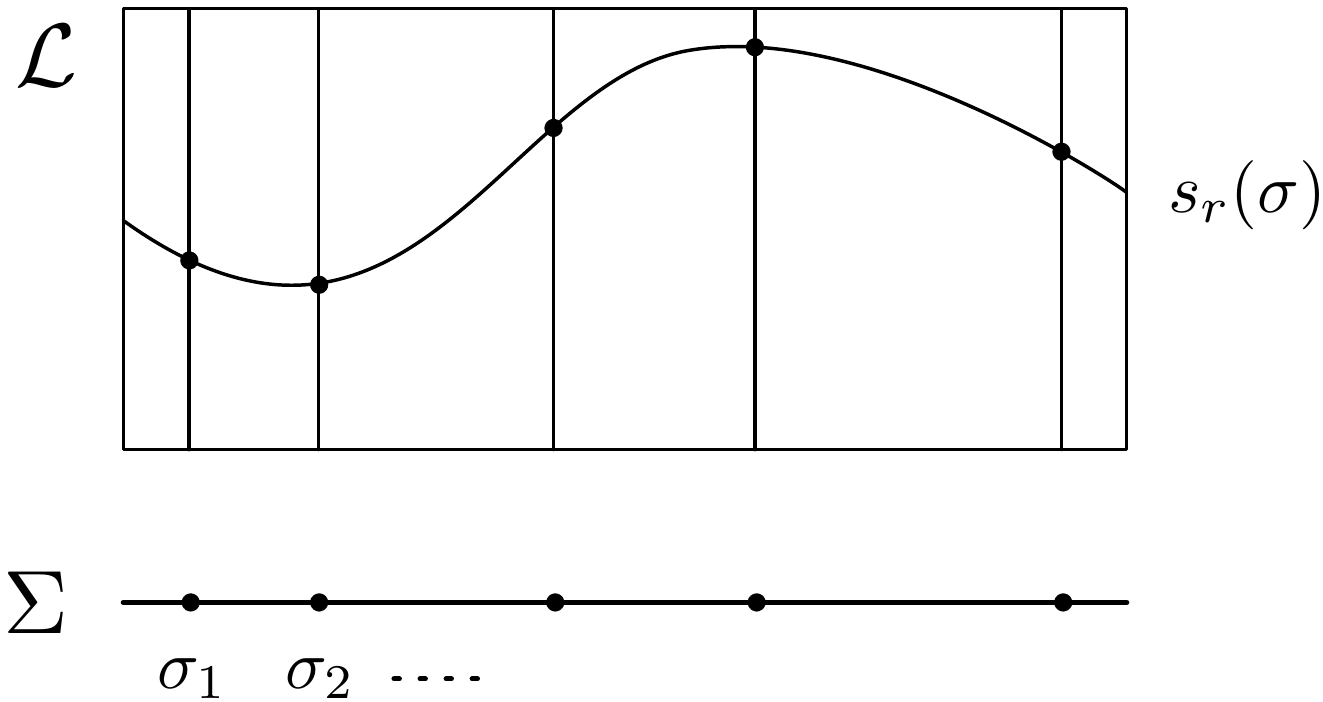}
\end{center}
\caption{{\it The geometry of the map to the Grassmannian. Each of the
    $k$ sections $s_r(\sigma)\in H^0(\Sigma,\cL)$ determines a vector
    in $\C^n$ by restriction to the $n$ marked points and
    multiplication by the scaling parameters $\xi_i$. }} 
\label{fig:section}
\end{figure}

We can now define a map
\be
\begin{array}{lrcl}
	e:&\bP\left(\bigoplus_{i=1}^n p_*{\rm ev}_i^*\cO(-1)\right) &\to &{\rm G}(k,n)\\ \\
	&(\Sigma,\sigma_1,\ldots,\sigma_n,;\cL;\xi_i)&\mapsto &C_{ri}=\xi_is_r(\sigma_i) 
\end{array}
\label{grass-embed}
\ee 
of this moduli space into the Grassmannian, with image $\Gamma$. In fact, the map arises from a standard construction in
algebraic geometry (see {\it e.g.} p. 353 of~\cite{GriffithsHarris}) that may be understood geometrically as follows.  Restricting $s_r(\sigma)\in H^0(\Sigma,\cL)$ to the $n$ marked points $\sigma_i$ gives $n$ complex numbers, or a vector in $\C^n$, so repeating this for $r=1,\ldots,k$ we obtain $k$ such vectors (see figure~\ref{fig:section}). However, by themselves the values of the $s_r(\sigma_i)$ are not meaningful --- they can be changed arbitrarily by bundle automorphisms of $\cL$ (worldsheet gauge transformations). To obtain an invariant result, we multiply each $s_r(\sigma_i)$ by its respective $\xi_i$ to obtain the vector
\be
	\left(\xi_is_r(\sigma_1),\ldots,\xi_ns_r(\sigma_n)\right)\in\bigoplus_{i=1}^n\cO|_{\sigma_i}\cong\C^n
\ee
that is invariantly defined. Thus, on a fixed curve with fixed choice of $\cL$,  there is a map
\be
\begin{array}{lccl}
	\xi:&H^0(\Sigma,\cL)\simeq\C^k &\hookrightarrow &\ \C^n\\
	&s_r(\sigma)\qquad&\mapsto&\xi_is_r(\sigma_i)\ ,
\end{array}
\label{ximap}
\ee
whose image is the $k$-plane $C_{ri}=\xi_i s_r(\sigma_i)$ in $\C^n$ (at least when the parameters are generic).  As the $\xi_i$ and $(\Sigma_g,\sigma_1,\ldots,\sigma_n;\cL)$ both vary, we obtain a family of such $k$-planes parametrized by the base space $\bP\left(\bigoplus_{i=1}^n p_*{\rm ev}_i^*\cO(-1)\right)$.  For future reference, we also set
\be
	\Gamma:=e\left(\bP\bigoplus_ip_*{\rm ev}_i^*\cO(-1)\right)
	\qquad\hbox{and}\qquad
	\rd\mu(\Gamma):=e_* \left(\frac{1}{\rm Vol(GL(1))}\rd\mu\wedge\prod_{i=1}^n\, \frac{\rd\xi_i}{\xi_i}\right)\ ,
\ee
where $e_*$ denotes the pushforward that integrates over the fibres (if any) of the map $e$. Incidentally, note that the $\xi_i$s are not really essential; one could simply have specified a gauge for $s_r(\sigma)$, or else considered a map into the projective Grassmannian of $k-1$ planes in $\bP^{n-1}$ (see~\cite{GriffithsHarris} for further discussion). Thus the map to G$(k,n)$ does not rely on the specific form~\eqref{deltaW2} of the vertex operators, although these vertex operators yield the closest comparison to the residue formula of~\cite{ArkaniHamed:2009dn}.

Since G$(k,n)$ is $k(n-k)$-dimensional, it is clear that for large $g$, the map $e$ cannot be one-to-one; indeed, for the MHV case where $k=2$, it must have fibre dimension at least $4g$. However, both for tree amplitudes and for leading singularities $e$ maps from a $2(n-2)$-dimensional cycle into G$(k,n)$ and we expect the map to be $1:1$ (this is the case in all cases so far computed).  

\medskip

When $g=0$, this map plays a role in~\cite{Spradlin:2009qr,Dolan:2009wf} where it is used to relate the twistor-string tree formula to the Grassmannian formulation of the Drummond \& Henn formula for tree amplitudes. As in equation~\eqref{tree-twistor-int}, the colour-ordered tree amplitude comes from the single trace contribution 
\be
\label{tree-int-grass}
	\int \frac{1}{{\rm Vol(GL(2))}} \prod_{i=1}^n\,\frac{\rd\xi_i}{\xi_i}\frac{\rd\sigma_i}{\sigma_i-\sigma_{i+1}}
	\prod_{r=1}^{k}\,\rd^{4|4}Y_r\prod_{i=1}^n\,\delta^{4|4}\!\left(W_i-\xi_i\sum_{r=1}^{d+1} Y_r\sigma_i^{r-1}\right)
\ee 
of the original twistor-string path integral, where we have combined the GL(1) action on the parameters $\xi_i$ with the PGL(2) automorphism group of the (unmarked) worldsheet. The map to G$(k,n)$ is given by 
\be 
	C_{ri}(\xi_i,\sigma_i)=\xi_i\sigma_i^{r-1}\ . 
\label{Cgenus0}
\ee
This map provides a natural embedding of the GL(2) action on the worldsheet description inside the GL$(k)$ invariance of the Grassmannian, for introducing the coordinates $\sigma_A= \xi^{1/d}(1,\sigma)$, \eqref{Cgenus0} becomes
\be 
	C_{ri}\ \longrightarrow \ C_{(A_1\cdots A_{d});i}=\sigma^{(A_1}_i\ldots\sigma^{A_d)}_i\ , 
\ee 
showing that GL(2) is embedded in GL$(k)$ via the $d^{\rm th}=(k-1)^{\rm st}$ symmetric tensor power. It is easily checked that this map is $1:1$, and hence the image $\Gamma\subset{\rm G}(k,n)$ is a $2(n-2)$-cycle, equipped with the holomorphic form
\be
	\rd\mu(\Gamma)=\frac{1}{{\rm Vol(GL(2))}} 
	\prod_{i=1}^n\,\frac{\rd\xi_i}{\xi_i}\frac{\rd\sigma_i}{\sigma_i-\sigma_{i+1}}\ .
\ee 

For MHV tree amplitudes ($d=1$), this map is simply the standard co-ordinatisation of G$(2,n)$, {\it i.e.}, $\Gamma$ is the fundamental cycle of G$(2,n)$.  At higher degree however, this cycle is \emph{not} found by making contour choices that restrict to a residue in~\eqref{G}. In particular, for generic $\sigma_i$, the cycle implied by~\eqref{Cgenus0} is not a pole of the holomorphic volume form of~\eqref{G} --- with $C_{ri}$ of the form~\eqref{Cgenus0}, the cyclic Pl\"ucker coordinates $(1,\ldots,k)\cdots(n,1,\ldots,k\!-\!1)$ become Vandermonde determinants and do \emph{not} vanish\footnote{Except in the
degenerate limit when the $\sigma_i$ collide.} on the cycle $\Gamma$. By contrast, the cycles defined by vanishing of 
Pl\"ucker coordinates in the denominator of \eqref{G} correspond to the expression of the tree amplitude in Drummond \& Henn form; as explained in section~\ref{sec:KS}, this corresponds to support on degenerate $d=2p+1$, $g=p$ curves in $\PT^*$, rather than the $d=p+1$, $g=0$ curves of the genus zero twistor-string.  The work of~\cite{Spradlin:2009qr,Dolan:2009wf} seeks to find a relationship between the two formulations, not as a direct equivalence of terms, but via a global residue formula.  From the perspective of the present work, the equivalence of N$^p$MHV tree amplitudes in the form~\eqref{tree-int-grass} with certain sums of residues of~\eqref{G} should be viewed as a $p$-loop infra-red equation.

\medskip

$\Gamma$ is also $2(n-2)$-subcycle for $g$-loop leading singularities, associated with integrals over codimension $4g$ boundary components of the moduli space of $n$-pointed, genus $g$ nodal curves $\Sigma$ equipped with a degree $d$ line bundle $\cL$. We will denote such boundary components by (Jac$_{g,n}^{d})_{\rm LS}$; points in (Jac$_{g,n}^d)_{\rm LS}$ represent curves with $\delta =4g$ nodes and $\nu=3g+1$ rational components. The map $e$ into the Grassmannian works exactly as for smooth curves: the $n$ scaling parameters $\xi_i$ define an embedding $\xi:H^0(\Sigma,\cL)\hookrightarrow \C^n$ and so gives a $k$-plane in $\C^n$. Note that the value of $k$ here is the same as for the smooth curve (see equation~\eqref{degree-nodal}).  As we move around in the leading singularity moduli space, we obtain a family of such $k$-planes parametrized by the base 
\be\label{GC-spaceLS}
\minCDarrowwidth25pt
	\begin{CD}
	\CP^{n-1} @> >> \mathbb{P}\left(\bigoplus_{i=1}^n p_*{\rm ev}_i^*\cO(-1)\right)  \\
	@. @V VV\\ 
	@. ({\rm Jac}^d_{g,n})_{\rm LS}
	\end{CD}\quad ,
\ee 
in other words the restriction of the map from the full twistor-string to the leading singularity moduli space. Since this has codimension $4g$, the base of our family is $2(n-2)$-dimensional.

That the cycle has dimension $2(n-2)$ can also be seen explicitly as follows. Recall that the $\PT^*$ leading singularity is defined by combining~\eqref{tree-twistor-int} for the rational components and~\eqref{Twistorinnerproduct} to glue them together at the nodes. Breaking up the curve into its components, each node is a special point on each of the two components that it glues together. To construct $\cL$, we must also give an extra parameter to define how the fibres of $\cL$ on each component are glued together at the nodes. Thus, each node contributes 3 parameters to the count.  Each of the $n$ marked points contributes 2 parameters to the moduli space --- one describing its location on
the rational curve, and another for the $\xi_i$ parameter trivialising $\cL|_{\sigma_i}$.  Finally, each rational component
together with its line bundle has a GL(2) automorphism group, so we
must subtract $-4\nu$ to account for the equivalences. The dimension
of the moduli space is therefore 
\be 
	3\delta +2n-4\nu = 2(n-2)\ , 
\ee
where, as in section~\ref{sec:nodal}, we have used $\delta=4g$ and $\nu=3g+1$ for leading singularities. We expect the map to be 1:1, so the image cycle $\Gamma$ also has dimension $2(n-2)$, this being the case in all examples so far computed. However, as for the higher degree tree amplitude case, we do not expect a direct relationship with the specific residues of~\eqref{G} unless the leading singularity is primitive. The primitive leading singularities include all the Korchemsky-Sokatchev configurations of section~\ref{sec:KS} and, via the Drummond \& Henn expansion of tree
subamplitudes, therefore generate all leading singularities.


\subsection{All-Loop Leading Singularities}
\label{sec:AllLoop}

We now compare the description of leading singularities coming from
the embedding of the twistor-string in G$(k,n)$ with the description
from the residue formula~\eqref{G}. We will do this
explicitly for NMHV and N$^2$MHV leading singularities that
are primitive, {\it i.e.}, each tree subamplitude in the leading
singularity channel is either MHV or $\MHVbar_3$, so that each
component of the worldsheet is mapped with degree either one or
zero. We learn enough from this to obtain a bound on the possible
loop order at which new N$^p$MHV leading singularties arise.  

We first discuss general properties of the residue formula
\be
	\cL_{k,n}(W_i) = 
	\oint\frac{{\rm D}^{k(n-k)}C}{(1,2,\ldots, k)\cdots(n,1,\ldots, k\!-\!1)}\,\prod_{r=1}^k\,\rd^{4|4}Y_r \,
	\prod_{i=1}^n\,\delta^{4|4}\!\left(W_i-Y_rC_{ri}\right)\ .
\label{G2}
\ee
The $\PT^*$ support of a particular residue of~\eqref{G2} may be read off more-or-less directly from the contour choice as follows.  Restoring the dual twistor index, the $\delta$-functions $\delta^{4|4}(W_i-Y_rC_{ri})$ require that
\be
	W_{\alpha i} = \sum_{r=1}^k Y_{\alpha r}C_{ri}
\ee
so that the rank of the matrix $W_{\alpha i}$ --- the dimension of the span of the $n$ points $W_i\in T^*$ --- is bounded by 
\be
	{\rm rk}(W) \leq {\rm min}\left[{\rm rk}(Y),\,{\rm rk}(C)\right] \leq {\rm min}\left[4,\,k\right],
\label{rankfactor}
\ee
or min$\left[3,k-1\right]$ in the projective space.  Thus, for $k=2$ or 3, the $\PT^*$ support is restricted by the G$(k,n)$ formula even for arbitrary $k$-planes $C$ ({\it i.e.}, when $C_{ri}$ is generic). For example, in \emph{any} MHV leading singularity ($k=2$), equation~\eqref{rankfactor} says that the external twistors lie span a subspace of dimension (at most) 2 in $\T^*$, or a line in $\PT^*$. Similarly, the Grassmannian conjecture of~\cite{ArkaniHamed:2009dn} implies  that \emph{any} NMHV leading singularity ($k=3$) is supported on a plane --- the span of $(Y_1,Y_2,Y_3)$ --- in $\PT^*$. However, a generic matrix $C_{ri}$ does not constrain the support of leading singularities at N$^2$MHV level and beyond, nor yields the more refined picture of ({\it e.g.}) NMHV 3-mass box coefficients lying on three, pairwise intersecting lines. To go further, we must examine the effect on the rank of $C$ of a choice of contour that encircles poles in the measure
\be
	\frac{{\rm D}^{k(n-k)}C}{(1,2,\ldots,k)\cdots(n,1,\ldots,k\!-\!1)}\ .
\label{measure}
\ee
We will then  compare the resulting matrices with the specific form of $C_{ri}$ that is found by embedding a twistor-string leading singularity.

\subsubsection{NMHV}

In the NMHV case where $k=3$, suppose we choose a contour that localises on the subvariety (actually, a special Schubert cycle) where the minor $(i\!-\!1,i,i\!+\!1)=0$. This constraint on the 3-plane $C$ leads to a constraint on the $\PT^*$ support of $\cL_{3,n}$ itself, for the matrix
\be
	\left.C\right|_{\{i-1,i,i+1\}}:=
	\begin{pmatrix}
		C_{1\,i-1} & C_{1\,i} & C_{1\,i+1}\\
		C_{2\,i-1} & C_{2\,i} & C_{2\,i+1}\\
		C_{3\,i-1} & C_{3\,i} & C_{3\,i+1}
	\end{pmatrix}
\ee
has rank $\leq3$, so
\be
	{\rm rk}\left(\left.W\right|_{\{i-1,i,i+1\}}\right) 
	\leq {\rm min}\left[{\rm rk}(Y),\, {\rm rk}\left(\left.C\right|_{\{i-1,i,i+1\}}\right)\right] 
	\leq {\rm min}[4,2] = 2\ .
\label{NMHVcollinearcycles}
\ee
Thus, on this cycle in $G(3,n)$, $\cL_{3,n}$ only has support when the points $W_{i-1},W_i$ and $W_{i+1}$ are (at most) collinear in $\PT^*$. In principle, we could consider imposing a further condition on the \emph{same} minor $C|_{\{i-1,i,i+1\}}$, reducing it to rank 1. However, this would force at least two of the points $W_{i-1}$, $W_i$ and $W_{i+1}$ to \emph{coincide} in $\PT^*$, and the amplitude or leading singularity will become singular. We will avoid these singular regions by assuming that no Pl\"ucker coordinate vanishes to higher than $(k-2)^{\rm nd}$ order.

The NMHV Grassmannian has dimension $3(n-3)$, so $n-5$ conditions are required to specify a $2(n-2)$-cycle. Since we do not wish to set any Pl\"ucker coordinate to vanish at second order or above, as in~\cite{ArkaniHamed:2009dn} we must simply impose that $n-5$ of the $n$ denominator factors vanish. We can label our choice by five integers $\{i,j,k,l,m\}$ corresponding to the only five cyclic Pl\"ucker coordinates that remain non-zero\footnote{In our notation, $i$ will correspond to the Pl\"ucker coordinate $(i-1,i,i+1)$.}.  In particular, it was shown in~\cite{ArkaniHamed:2009dn,Mason:2009qx} that the leading singularity in the 3-mass box channel is the residual form of~\eqref{G2} on the cycle $\{n,a-1,a,b-1,b\}$. The geometry of the support in $\PT^*$ is easy to understand from the specification of this cycle: the vanishing of
$$
	(n,1,2),\ (1,2,3), \ \ldots,\ (a-3,a-2,a-1)
$$
implies via~\eqref{NMHVcollinearcycles} that $\{W_n,W_1,\ldots,W_{a-1}\}$ are all collinear. $\{W_a,\ldots,W_{b-1}\}$ and $\{W_b,\ldots,W_n\}$ constrained similarly.  Finally, since we have already seen that any NMHV leading singularity obtained from~\eqref{G2} must necessarily lie in a plane in $\PT^*$, these three lines each intersect, so we have recovered figure~\ref{fig:3mb} from the Grassmannian residue formula.

Going the other way, using the results of sections~\ref{sec:twprimleadsing}~\&~\ref{sec:nodal}, the NMHV 3-mass box leading singularity can be written in twistor-string form as
\be
\label{3mb-leadsing-twis2} 
	\cA^{\rm 3mb}_{\rm NMHV}(1,\ldots,n)
	=\int \frac{\prod_{r=1}^3\rd^{4|4}Y_r}{{\rm (Vol\,GL(1))}^3}\,\frac{1}{\Delta}\wedge
	\frac{\rd\xi_n}{\xi_n}\,\delta^{4|4}(W_n-\xi_n Y_1)\,\wedge\,
	\prod_{i=1}^{n-1}\rd\sigma_i\wedge\frac{\rd\xi_i}{\xi_i}\,\delta^{4|4}(W_i-\xi_i W(\sigma_i)) 
\ee 
where 
\be
	W(\sigma_i) = \begin{cases}
					Y_1+\sigma_i Y_2\ , & i\in\{n,1,\ldots,a-1\}\\
					Y_2+\sigma_i Y_3\ , & i\in\{a,\ldots,b-1\}\\
					Y_3+\sigma_i Y_1\ ,  & i\in\{b,\ldots, n-1,n\}
				\end{cases}
\ee
and $\Delta$ is given in equation~\eqref{delta-def}. This is already in the form required for the embedding into G$(3,n)$, and we  simply read off 
\be
\label{c-triangle}
	C=\begin{pmatrix}
			\xi_1 & \ldots &\xi_{a-1}&0&\ldots&0&\xi_b\sigma_b &\ldots&\xi_{n-1}\sigma_{n-1}&\xi_n\\
			\xi_1\sigma_1 & \ldots &\xi_{a-1}\sigma_{a-1} & \xi_{a}&\ldots&\xi_{b-1} &0&\ldots&0&0\\
			0& \ldots &0 & \xi_{a}\sigma_{a}&\ldots&\xi_{b-1}\sigma_{b-1}&\xi_{b}&\ldots&\xi_{n-1}&0
		\end{pmatrix}\ .
\ee 
One can easily check that all the cyclic minors $(i\!-\!1,i,i\!+\!1)$ of this matrix vanish identically, except for $i\in\{n, a-1, a, b-1, b\}$. The matrix~\eqref{c-triangle} thus provides an explicit parametrization of the cycle $\{n,a-1,a,b-1, a\}$, and the  cycle $\Gamma\subset {\rm G}(3,n)$ defined by the genus 1 nodal twistor-string is \emph{identified} with the cycle used in the Grassmannian residue formula~\eqref{G2}. This is in marked contrast to the twistor-string NMHV tree amplitude, for which the cyclic Pl\"ucker coordinates of the matrix $C_{ri} = \xi_i\sigma_i^{r-1}$ are non-zero everywhere except where vertex operators collide.

We also remark  that the GL(1)$^3$ (that arises in~\eqref{3mb-leadsing-twis2} from the `gauge freedom' of the triangle in $\PT^*$ --- the freedom to rescale the rational parameters on each of the three lines) naturally embeds as the diagonal
subgroup of the GL(3) gauge group of G$(3,n)$. The rest of this GL(3) gauge freedom is fixed by our choice of embedding $\C^2\rightarrow \C^3$ for each $\CP^1$ in the triangle.

\medskip

\begin{figure}[t]
\begin{center}
\includegraphics[height=110mm]{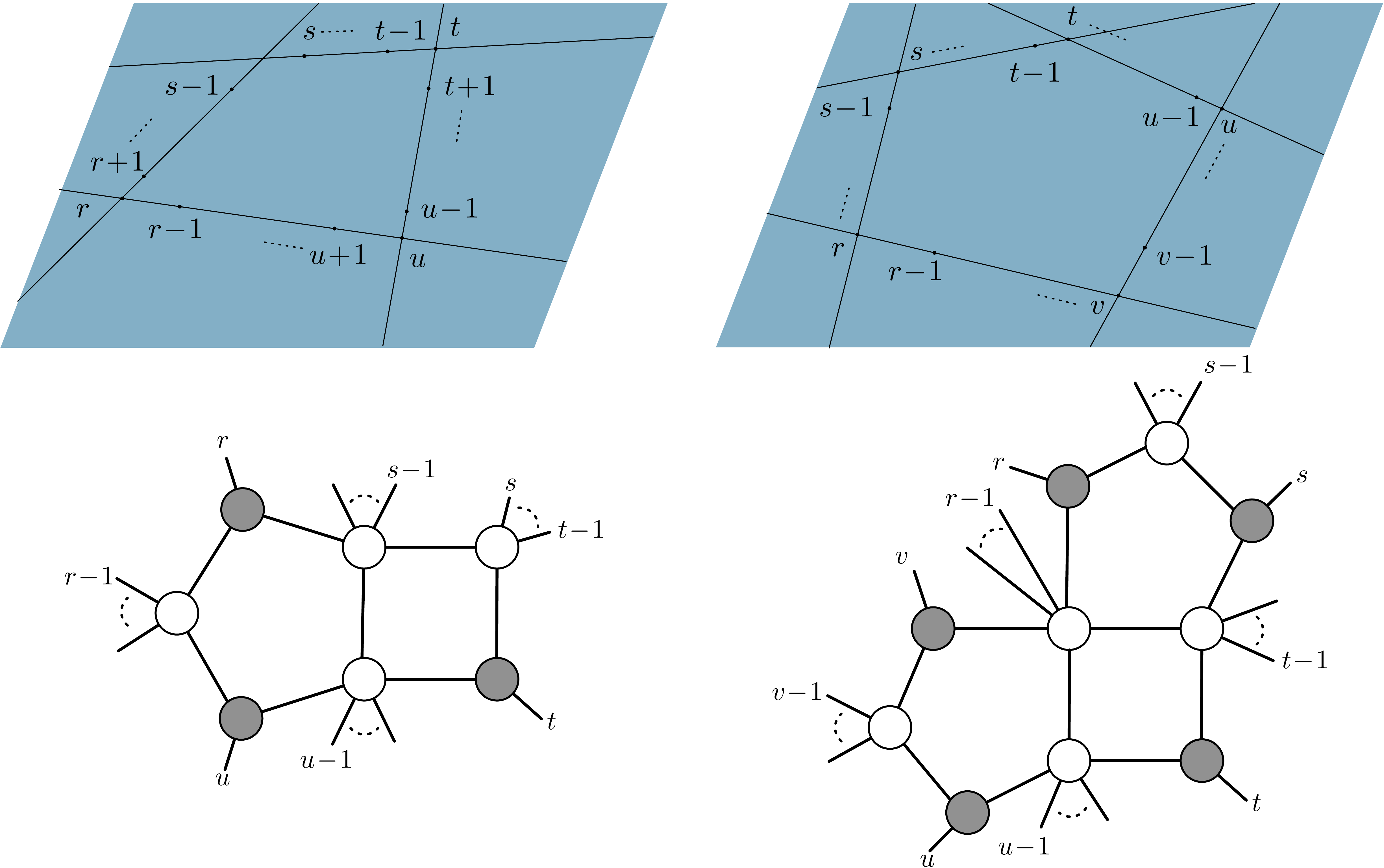}
\end{center}
\vspace{-0.4cm}
\caption{{\it The line configurations in $\PT^*$ of higher-loop} NMHV {\it leading singularities, defined by the cycles $\{r,s-1,s,t,u\}$ and $\{r,s,t,u,v\}$, respectively. In these cases, the momentum channel diagrams are not unique:  the leading singularity is the same in more than one primitive channel at the same loop order. All the possible channel diagrams follow from dihedral transformations on the ones displayed.}}
\label{fig:multiNMHV}
\end{figure}

The choice of cycle $\{n,a-1,a,b-1,b\}$ in which two pairs of non-vanishing Pl\"ucker coordinates are adjacent is clearly not generic --- in general choices the non-vanishing Pl\"ucker coordinates are at generic locations in the cyclic ordering. We can understand the geometry of these more general cycles by considering what happens when, {\it e.g.}, $(i-1,i,i+1)\neq0$, but $(i-2,i-1,i)=(i,i+1,i+2)=0$, so that the non-vanishing minor is isolated. The vanishing conditions imply that $\{W_{i-2},W_{i-1},W_i\}$ and $\{W_i,W_{i+1},W_{i+2}\}$ are each collinear. Since $(i-1,i,i+1)\neq0$, these two lines are not the same, but they intersect at the common marked point $W_i$. Thus, a generic cycle $\{r,s,t,u,v\}$ is supported on a planar pentagon in $\PT^*$, with an external state attached to each vertex. In the intermediate case $\{r,s-1,s,t,u\}$ where only one pair of non-vanishing cyclic minors is adjacent, we obtain a quadrilateral with marked points at all but one of its vertices (see figure~\ref{fig:multiNMHV}).

Superficially, such a pentagon would seem to correspond to a curve of genus 6 --- the standard result for a plane curve of degree 5.  However, many of the intersections in the plane are forced by the other intersections (once a line meets two other lines in the configuration, it is forced to lie in the plane and hence meet all the others).  The loop order of the leading singularity is really determined by the genus of the nodal curve (string worldsheet) \emph{before} it is mapped into $\PT^*$ ---  only the nodes of the abstract worldsheet correspond to factorization channels of the momentum space diagram.  Given a degree $d$ curve in $\PT^*$, the genus of the worldsheet may be determined by the formula
\be
	g=d-p-1
\ee
so the degree 4 curve on the left of figure~\ref{fig:multiNMHV} should be taken to be a two-loop leading singularity, as obtained via `inverse soft limits' in~\cite{ArkaniHamed:2009dn} for the case of eight particles, while the most generic, degree 5 curve corresponds to a three-loop leading singularity. Notice also that on the generic cycle, the nodal worldsheet has 10 components (as may be seen by treating the momentum space channel diagrams as dual graphs). Five of these are mapped to $\PT^*$ with degree 1 and correspond to the five MHV lines, and five are mapped with degree 0, corresponding to the marked points at the vertices of the pentagon. (There are no `internal' degree zero curves joining three lines as there are generically no three-fold intersections.)

Since the most generic $2(n-2)$-cycle in G$(3,n)$ gives a 3-loop leading singularity, we arrive at the striking conclusion that --- assuming that the Grassmannian residue formula generates all leading singularities --- there are no new NMHV leading singularities beyond 3 loops. More specifically, all leading singularities of arbitrary loop, $n$-particle NMHV amplitudes are determined in terms of their leading singularities at
\begin{itemize}
	\item 1 loop when $n\leq7$,
	
	\item 2 loops when $7< n<10$ and
	
	\item 3 loops when $n\geq10$
\end{itemize}
(the first of these conditions appeared in~\cite{ArkaniHamed:2009dn}),
where the $n$-dependence comes from requiring that there are
sufficient cyclic minors for the non-vanishing ones to be
non-adjacent.  

As a final remark, we note that the choice of the five non-vanishing
Plucker coordinates is only a partial classification of these leading
singularities.  At least superficially, there are at least 5 different
subcases with the same specification of non-vanishing pluckers in the
3 loop case as the specific geometry is not cyclically invariant.
This can be seen either directly from the momentum space channel
diagram, or the distribution of the nodes as opposed to coincidental
intersections in the twistor support diagram of figure
\ref{fig:multiNMHV}.

\subsubsection{N$^2$MHV}

In the NMHV case, the location of the reference twistors that describe the KS configuration is clear; they can simply be taken to be the vertices of the triangle. The N$^2$MHV KS configurations are built by gluing a triangle
onto this NMHV triangle, so that the two share an edge and a vertex (see figure~\ref{fig:NNMHVAB}).  Although the resulting figure has 5 vertices, one can characterise it in terms of only four reference twistors, because the common edge supports three vertices, and the location of one of these vertices may be given as a linear combination of the locations of the other two.

There are two types of (non-boundary) contribution to the N$^2$MHV tree amplitude, and correspondingly two KS figures (see section~\ref{sec:NNMHVtree}). The Type A contributions $R_{n;a_1b_1}R_{n;b_1a_1;a_2b_2}$ have $a_1< a_2< b_2\leq b_1$.  The 5 vertices of the KS figure (see figure~\ref{fig:NNMHVAB}) can be parametrized by four dual twistors $Y_1,\ldots Y_4$,  with $Y_1\equiv W_n$, $Y_2$ the intersection point of the lines $L_{\{a_1,\ldots,a_2-1\}}$ and $L_{\{a_2,\ldots,b_2-1\}}$,  $Y_3=L_{\{a_2,\ldots,b_2-1\}}\cap L_{\{b_2,\ldots,b_1-1\}}$ and $Y_4=L_{\{b_2,\ldots,b_1-1\}}\cap L_{\{b_1,\ldots,n-1}\}$. The remaining intersection point $L_{\{n,1,\ldots,a_1-1\}}\cap L_{\{a_1,\ldots,a_2-1\}}$ can be fixed to be $Y_3+Y_4$ by a scaling of these twistors. Overall, the five lines may be described explicitly as
\be
\label{pentagonA}
	W(\sigma_i)=
		 \begin{cases} 
			Y_1+ \sigma_{1i} (Y_3+ Y_4) \ ,	& i\in\{n,1,\ldots ,a_1-1\} \\
 			(Y_3+ Y_4) +\sigma_{2i}Y_2 \ ,		& i\in \{a_1,\ldots, a_2-1\} \\
			Y_2+ \sigma_{3i} Y_3\ ,			& i\in\{a_2,\ldots, b_2-1\}\\
 			Y_3+ \sigma_{4i} Y_4\  ,			& i\in\{b_2,\ldots, b_1-1\}\\
 			Y_4 + \sigma_{5i}Y_1\ , 			& i\in\{b_1,\ldots, n\}
		\end{cases} 
\ee 
where $\sigma_1, \ldots ,\sigma_5$ are parameters on each line, fixed to
be 0 or $\infty$ at the intersection points with lines adjacent in the colour ordering\footnote{In particular, $\sigma_{1n}=0$ while $\sigma_{5n}=\infty$.}.   The parameter $\sigma_4$ is fixed by the requirement that the second vertex is $Y_3+Y_4$, but the others are defined only up to a GL(1) scaling of each line. We often drop the index labelling the different line parameters, since they are in any case determined by the range of marked points on each line.

We can construct the corresponding leading singularity in $\PT^*$ by using the generalized unitarity rules of section~\ref{sec:twsgenunit} to glue the NMHV 3-mass box expression from~\eqref{3mb-leadsing-twis2} to a pair of MHV vertices and an $\MHVbar$ vertex so as to form the N$^2$MHV pentabox configuration shown in figure~\ref{fig:NNMHVleadsing}.   Writing the leading singularities this way again leads to an explicit embedding into the Grassmannian G$(4,n)$, with $C$ given by
\be\label{c-triangle-NNMHVA}
	\left(
		\begin{array}{cccccccccccccccc}
		\xi_1&\ldots &\xi_{a_1-1}&0&\ldots&0&0&\ldots&0&0&\ldots& 0&\sigma_{b_1} &\ldots &\sigma_{n-1}&\xi_n\\
		0&\ldots &0&\sigma_{a_1}&\ldots&\sigma_{a_2-1}&\xi_{a_2}&\ldots&\xi_{b_2-1}&0&\ldots&0&0&\ldots&0&0\\
		\sigma_1& \ldots &\sigma_{a_1-1}&\xi_{a_1}&\ldots &\xi_{a_2-1} &\sigma_{a_2} &\ldots&
		\sigma_{b_2-1} &\xi_{b_2}&\ldots&\xi_{b_1-1}&0&\ldots&0&0\\
		\sigma_1&\ldots &\sigma_{a_1-1}&\xi_{a_1}&\ldots&\xi_{a_2-1}&0&\ldots&0&\sigma_{b_2}&
		\ldots&\sigma_{b_1-1}&\xi_{b_1} & \ldots&\xi_{n-1}&0
		\end{array}
	\right)
\ee 
where we have replaced $\xi_i\sigma_i$ by $\sigma_i$ purely to simplify notation.

The type B contributions $R_{n;a_1b_1}R_{n;a_2b_2}$ have $a_1<b_1<a_2<b_2$.  The corresponding KS configuration (see figure~\ref{fig:NNMHVAB}) can again be parametrized by four reference twistors, where we now choose $Y_1\equiv W_n$, $Y_2=L_{\{n,1,\ldots,a_1-1\}}\cap L_{\{a_1,\ldots b_1-1\}}$, $Y_3=L_{\{b_1,\ldots,a_2-1\}}\cap L_{\{a_2,\ldots,b_2-1\}}$ and finally $Y_4=L_{\{a_2,\ldots,b_2-1\}}\cap L_{\{b_2,\ldots,n\}}$. Again, we can choose the scalings on the line parameters so that the vertex $L_{\{a_1,\ldots,b_1-1\}}\cap L_{\{b_1,\ldots,a_2-1\}}$ is at $Y_1+Y_3$.  The marked points on the pentagon are then 
\be
\label{pentagonB}
	W(\sigma_i)= 
		\begin{cases} 
			Y_1+ \sigma_{1a} Y_2\ , 			& i\in\{n,1,\ldots ,a_1-1\}\\  
			Y_2+\sigma_{2a}(Y_1+ Y_4) \ ,	& i\in \{a_1,\ldots, b_1-1\}\\
			(Y_1+ Y_4) + \sigma_{3a}Y_3\ , 	& i\in\{b_1,\ldots, a_2-1\}\\
			Y_3+ \sigma_{4a} Y_4\ ,			& i\in\{a_2,\ldots, b_2-1\}\\
			Y_4 + \sigma_{5a}Y_1\ ,			& i\in\{b_2,\ldots, n-1\}
		\end{cases} 
\ee 
with similar partial gauge-fixing as in the type A configurations.  The same procedure as above gives the embedding in the Grassmannian by
\be
\label{c-triangle-NNMHVB}
	C=\left(
	\begin{array}{cccccccccccccccc}
		\xi_1&\ldots &\xi_{a_1-1}&\sigma_{a_1}&\ldots&\sigma_{b_1-1}&\xi_{b_1}&\ldots&\xi_{a_2-1}&0&\ldots&
		0&\sigma_{b_2} &\ldots&\sigma_{n-1}&\xi_n\\ 
		\sigma_1&\ldots &\sigma_{a_1-1}&\xi_{a_1}&\ldots&\xi_{b_1-1}&0&\ldots&0&0&\ldots&
		0&0 &\ldots&0&0\\
		0&\ldots&0 &0&\ldots &0 &\sigma_{b_1}&\ldots&\sigma_{a_2-1}&\xi_{a_2}&\ldots&\xi_{b_2-1}
		&0&\ldots&0&0\\
		0&\ldots & 0 &\sigma_{a_1}&\ldots&\sigma_{b_1-1}&\xi_{b_1}&\ldots&\xi_{a_2-1}& \sigma_{a_2}&\ldots&
		\sigma_{b_2-1} &\xi_{b_2} & \ldots &\xi_{n-1}&0 
	\end{array}
	\right)
\ee 
In both types A and B, the $C_{ri}$ depend on the $2n-1$ parameters  $(\xi_1,\ldots ,\xi_n;\sigma_1,\ldots
,\sigma_{n-1})$ explicitly. The remaining gauge freedom in either case is GL(1)$^3$: in the Type A configurations, this is  the subgroup of GL(4) induced by scaling $Y_1, Y_2$ and $Y_3+Y_4$ separately, while in Type B it is the subgroup describing separate scalings of $Y_2, Y_3$ and $Y_1+Y_4$. Accounting for this gauge freedom, the embedding is specified by precisely $2(n-2)$ parameters in each case.

\medskip

Consider the generic case when $n$ is large and the integers $(a_i,b_i,n)$ are each separated by more than 4 --- the more special cases can be understood as degenerations.  It is easily seen from the explicit representations of the $C$ matrices that the $4\times 4$ submatrices built from columns $\{i,i+1,i+2,i+3\}$ have rank two when
\be
	\{i+1,i+2\}\cap 
	\{n, a_1-1,a_1, a_2-1, a_2,b_1-1,b_1,b_2-1,b_2\}
	=\emptyset\ ,
\ee
and otherwise have rank three. Clearly, the rank 2 case corresponds to the four points being collinear, while the rank 3
case corresponds to the points being coplanar in $\PT^*$.  With widely separated $a_i$ and $b_i$, the marked points lie on two lines meeting at an unmarked vertex between either $a_i-1$ and $a_i$ or $b_i-1$ and
$b_i$, or at a marked vertex at $n$ as described in section~\ref{sec:KS} (see
figure~\ref{fig:NNMHVAB}).  It is clear that these degeneracies imply
that the Pl\"ucker coordinates in the denominator of the Grassmannian
integrand~\eqref{G} vanish to various degrees, but it is not so easy
to be as precise as to the degree as in the NMHV case.  The
co-dimension of a cycle of dimension $2(n-2)$ is $2n-12$.  However, to see
such a cycle as arising from a $(2n-12)$-fold self-intersection of the zero set
of the denominator of the volume form in \eqref{G} is hard, as the zero locus of
the determinant of a $4\times 4$ matrix has more strata than the
$3\times 3$ case, and it is not so easy to keep track of their codimension, in part because of the interdependence
arising from Pl\"ucker relations.

\begin{figure}[t]
\centering
\includegraphics[height=90mm]{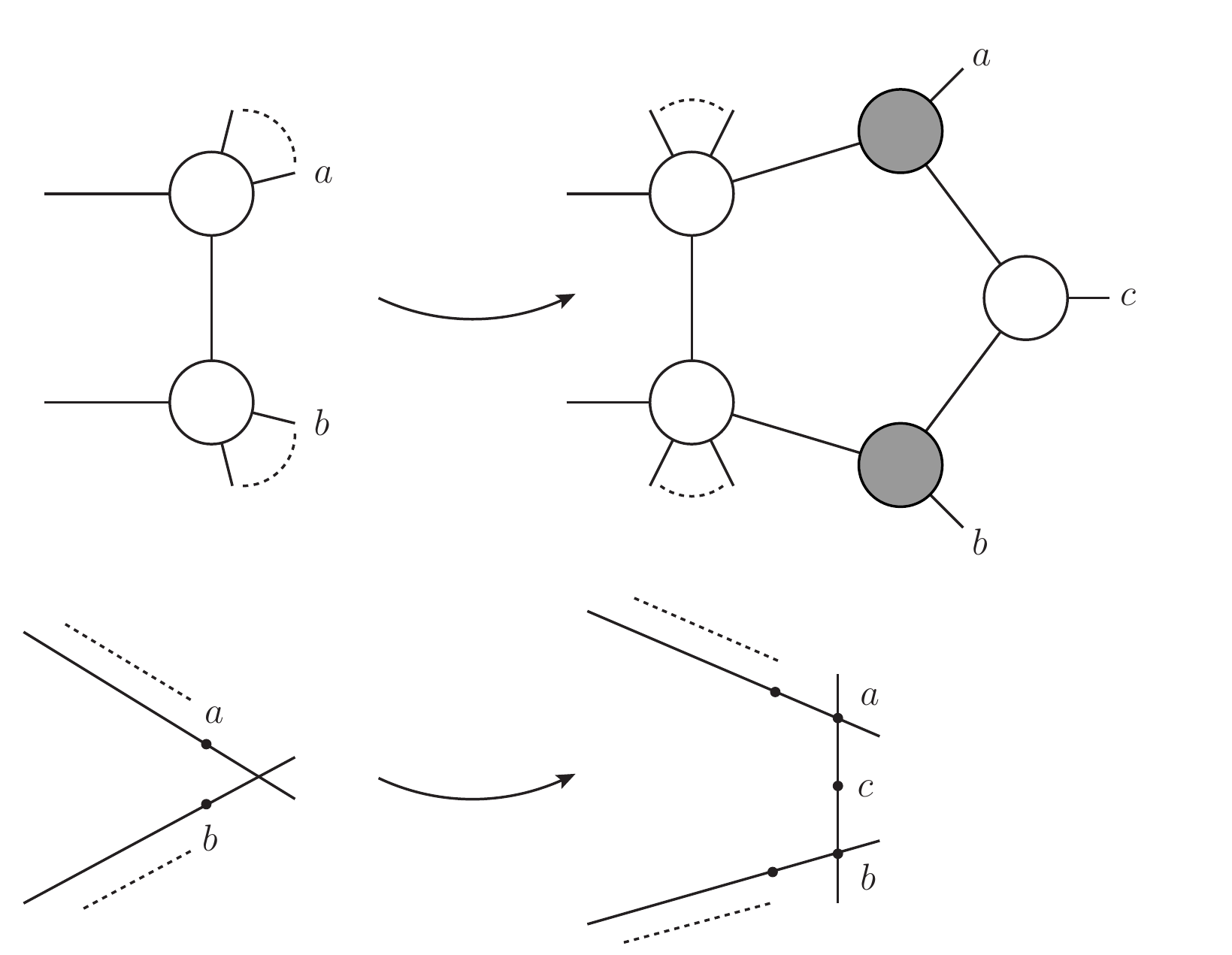}
\vspace{-0.4cm}
\caption{{\it The change in the channel diagram associated to the
    process of cutting across an unmarked vertex by inserting a new
    line with marked points at each end in twistor space.
    } }
\label{fig:ISL}
\end{figure}

We nevertheless argue that, just as at NMHV, the leading
singularities are partially classified by the choice of nine (at N$^2$MHV) 
integers. In the KS configurations these are 
$$
	\{n, a_1\!-\!1,a_1, a_2\!-\!1, a_2,b_1\!-\!1,b_1,b_2\!-\!1,b_2\}\ ,
$$
but more generally we allow all nine integers to be arbitrary --- the adjacencies of the KS choice are again not generic.
Just as in the NMHV case, allowing a pair of integers $(a_i\!-\!1,a_i)$ to become separate corresponds geometrically
to drawing a new line with marked points at each end, bridging across an unmarked vertex.  This is shown in figure \ref{fig:ISL}, where we also show the corresponding change to the channel diagram\footnote{This
corresponds to the action of one of the $\widetilde\cH$-operators of
~\cite{Mason:2009sa}, or one of the `inverse soft limits' of~\cite{ArkaniHamed:2009si,ArkaniHamed:2009dn}. The $\cH$-operator or the other inverse soft limit simply inserts an extra point on one of the lines.}.

\begin{figure}[h]
\centering
\includegraphics[height=55mm]{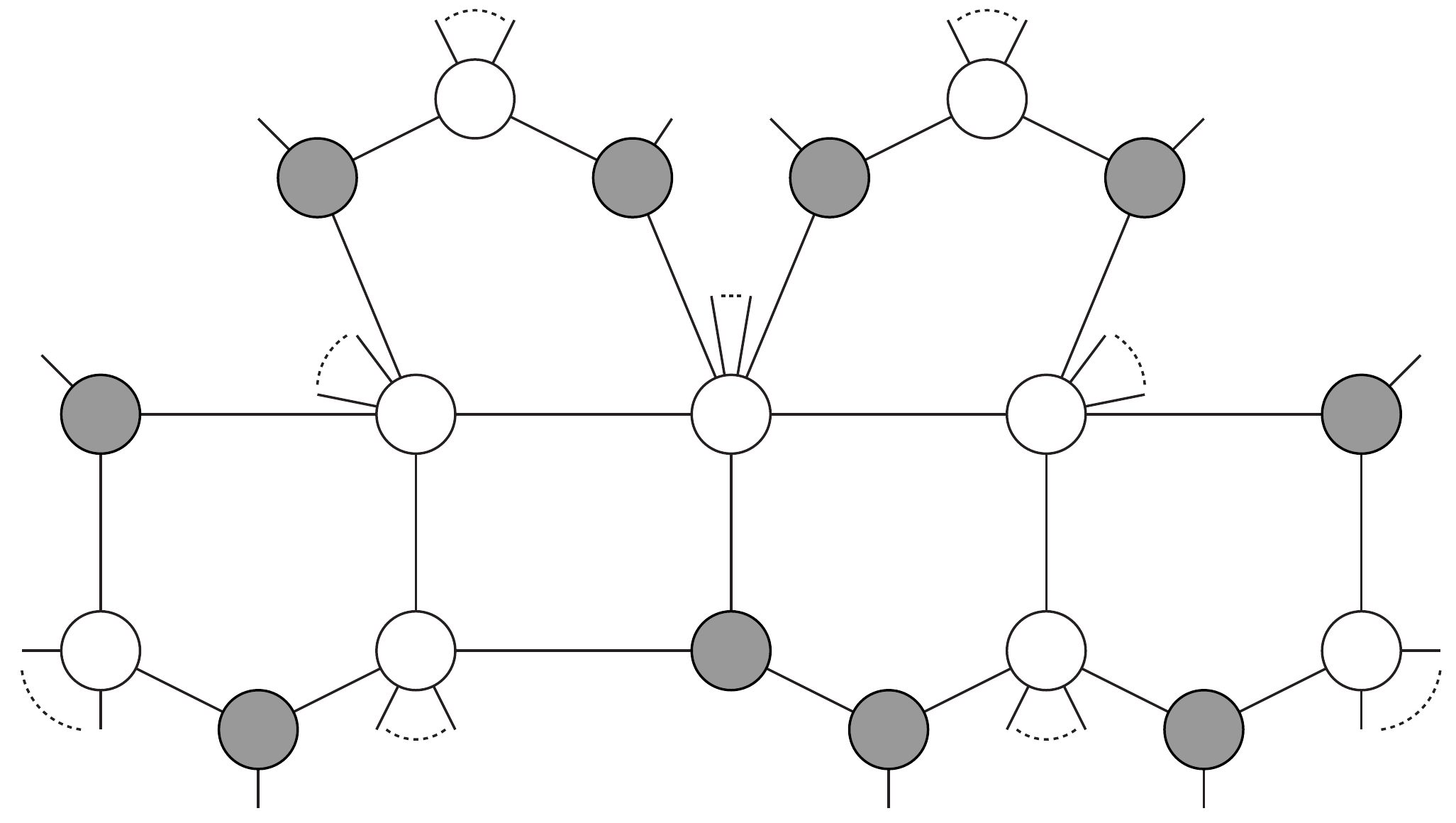}
\vspace{-0.4cm}
\caption{{\it The momentum space channel diagram of a maximal 6-loop}
  N$^2$MHV {\it leading singularity, built from the type B} N$^2$MHV {\it KS
    figure by cutting across each unmarked vertex with a line with
    marked points at each new vertex.} }
\label{fig:multiNNMHV}
\end{figure}

Given an N$^2$MHV KS figure, this procedure may be performed at most four times, yielding a 6-loop leading singularity such as the example shown in figure \ref{fig:multiNNMHV}. Again, it is easily seen that this is a partial classification as we could have started with an N$^2$MHV KS figure of type A or B (or the non KS N$^2$MHV type C described in footnote \ref{typeC}) and performed these four inverse soft limits to obtain different 6-loop leading singularities with the same integers.

In this case our analysis only suggests that one cannot find independent N$^2$MHV leading singularities beyond 6 loops, rather than furnishing a proof as at NMHV (albeit one that depends on the initial Grassmannian conjecture).  However, it is also clear that one can perform no further inverse soft limits that cut across marked vertices to these figures without destroying the planarity property.  Our conjecture then requires that $2(n-2)$-cycles that support a residue of~\eqref{G} with $k=4$ (N$^2$MHV) are partially classified by the choice of nine marked points as described above.  Unmarked vertices in the figure correspond to adjacent pairs of integers, and marked vertices to isolated choices of the integers.  


\subsubsection{N$^p$MHV}

Much of the discussion follows as in the N$^2$MHV case.  At N$^p$MHV,
the KS figures contain $(2p+1)$ lines and vertices, with $2p$ unmarked
vertices and one marked one.  The moduli of the figure are
nevertheless described by just $k=p+2$ reference twistors
$Y_1,\ldots,Y_k$ together with the $2(n-2)$ bosonic parameters which
can be taken to be the $(\sigma_i,\xi_i)$ up to a three-dimensional
gauge freedom.  This is most easily seen from the inductive process
described in section~\ref{sec:NpMHVtree} (or
in~\cite{Korchemsky:2009jv}), which shows that the dual twistors $Y_r$
can be chosen to lie on $k$ of the vertices around the figure by
choosing only the new twistor connecting the two new lines to the rest
of the pre-existing figure at that stage.  (There are many other ways
of choosing $k$ dual twistors to parametrize the figure, but these are
all related by GL$(k)$ transformations.)

We can use such a parametrization to define the embedding into the
Grassmannian (after performing all the generalized unitarity
gluings~\eqref{Twistorinnerproduct} to set up the integral),
represented by a $k\times n$ matrix $C$ as before. Any series of
columns that corresponds to a set of marked points that are all
collinear in the figure will have rank two, while a series of columns
that bridges over a single vertex of the figure will have rank three
(because the corresponding marked points are only coplanar, not
collinear) and a series of columns that bridges over two or more such
vertices will have rank four, and so on. Each KS figure contains only
one vertex that coincides with a marked point --- the remaining $2p$
vertices are unmarked. Unmarked vertices can be recognised from the
Grassmannian description as the case where two triples of columns that
are adjacent in the cyclic ordering ({\it i.e.} columns
$\{i-2,i-1,i\}$ and $\{i-1,i,i+1\}$) each have rank three, while the
marked vertex corresponds to having just a single triple of columns
with rank three (both adjacent ones being of rank two).

Assuming that the parameter count works as for the NMHV case, we
obtain a $2(n-2)$-cycle in G$(p+2,n)$ whenever we have $r$ unmarked
vertices and $4p+1-2r$ marked vertices for N$^p$MHV leading
singularities. Since \be p=d-g-1\ , \ee the maximal loop order for
fixed $p$ comes from curves with the highest degree --- in other
words, the highest number of line components. This happens when all
vertices are marked, so that there are $4p+1$ degree-1 components and
$4p+1$ degree zero components attached to external legs (at each
marked point), and a further $p-1$ internal $\MHVbar$
vertices\footnote{At each inductive step beyond NMHV, there is a new
  internal $\MHVbar$ vertex ({\it i.e.}, one with no external legs
  attached).  This is required to glue the new line to the two or more
  pre-existing lines through the chosen vertex, enforcing concurrency
  of the pre-existing lines with the new one through that vertex in
  the figure.  At N$^p$MHV, there are $p-1$ such multiple
  intersections in total (counted with multiplicity). The marked
  points are distributed on the $2p+1$ lines with an ordering that
  respects the ordering of the lines and vertices.  }.  The total
number of components $\nu$ is $9p+1$, and since $\nu=3g+1$ for
primitive leading singularities, the curve has genus $3p$.

\medskip

We therefore conjecture that the leading singularities of arbitrary
loop N$^p$MHV amplitudes are completely determined in terms of linear
combinations of their leading singularities at $3p$ loops (or below,
for sufficiently few particles).   
As in the N$^2$MHV case, part of our conjecture is that the
$2(n-2)$-cycles in the grassmannian $G(p+2,n)$ that support a residue
of \eqref{G} are partially classified by a choice of $4p+1$ marked
points.  

\medskip

Although there are many of different geometries for both KS
figures~\cite{Korchemsky:2009jv} and their generalizations described
in section~\ref{sec:NpMHVtree} and in this section, these all have a
restricted number of MHV and $\MHVbar$ vertices.  Indeed, the loop
bound suggests that we only need to consider primitive leading
singularities with at most $4p+1$ MHV vertices and $5p$ $\MHVbar$
vertices.  One can however conceive of N$^p$MHV primitive leading
singularity channel diagrams with arbitrary numbers of vertices at
fixed $p$.  Thus, if the Grassmannian conjecture coupled with the
discussion above is true, it seems that all leading singularities can
be all be generated purely by just these generalized KS figures:
{\it i.e.}, those figures obtained from the inductive procedure described in
section~\ref{sec:NpMHVtree} that have been embellished by `inverse soft
limits' that put a line across an unmarked vertex, replacing it with
two marked vertices.


\subsection{Twistor Support and the Loop Bound}

\begin{figure}[t]
\centering
\includegraphics[height=130mm]{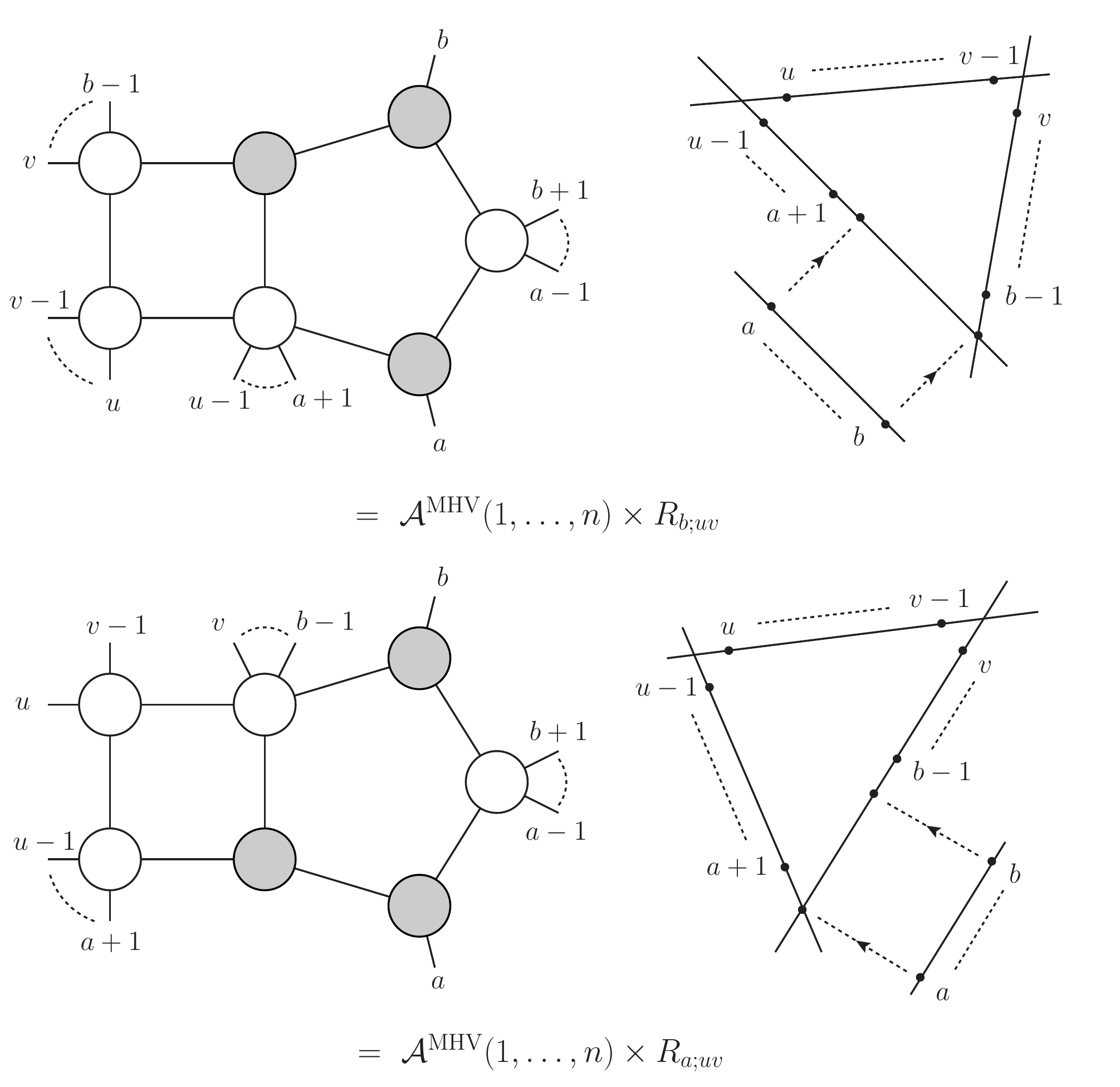}
\vspace{-0.4cm}
\caption{{\it The twistor support of a two-loop} NMHV {\it leading
    singularity collapses down to the one loop three mass box, giving the
    same value for the leading singularity.} }
\label{fig:2loopNMHV-red}
\end{figure}
Given that it is possible to write down perfectly valid channel
diagrams with arbitrary loop order at each fixed N$^p$MHV degree, one
is led to ask what mechanism can be responsible for the fact that no
new leading singularities should be obtained beyond $3p$-loops.  An
explanation comes from the restrictions on the twistor support for
such leading singularities.  This phenomenon is already seen at one
loop for MHV amplitudes where the leading singularities are all two
mass easy boxes.  The twistor support is simply of a pair of lines
glued to each other at two of their marked points (see figure~\ref{fig:branching}).  Two lines glued together at two points must
coincide and in this case, the leading singularity is well known to be
the standard tree level MHV amplitude.   At higher loops, the same twistor
support mechanism will be in play with $\MHVbar$ vertices outnumbering
the MHV vertices: at $\ell$ loops, for a primitive MHV leading
singularity, there will be $3\ell+1$ components to the corresponding
nodal curve, but just $d=\ell+1$ lines and so $2\ell$ $\MHVbar$ vertices.
Each $\MHVbar$ vertex can have at most one external leg, and therefore
the two internal legs will be restricting the support in the rest of
the figure.  Similarly, at NMHV, we can consider the two-loop leading
singularity figure \ref{fig:2loopNMHV-red}.  Here clearly the support
collapses down to that of the 1-loop three mass box as does the value
of the leading singularity.

In general then, the mechanism would seem to be that the value of the
leading singularity is determined by the support of the leading
singularity in twistor space.  For a primitive singularity, the
relation $d=p+1+\ell$ together with the number of subamplitudes being
$3\ell+1$ means that for each additional loop we are adding on one extra
MHV vertex or line, and two extra $\MHVbar$ vertices.  For that line
to be making an extra loop, it needs to be connected to the rest of
the figure in two places and their are severe restrictions as to how
it can do so if the figure is to remain planar.  Thus the validity of
the loop bound for obtaining new leading singularities coming from the
Grassmannian conjecture would seem to follow from a corresponding
conjecture in twistor space that a leading singularity is determined
by its twistor support.


\section{Conclusions and Outlook}
\label{sec:conclusions}

We began this paper by resolving the puzzle of the why the degree of the twistor curves found by Korchemsky \& Sokatchev~\cite{Korchemsky:2009jv} exceeds the twistor-string prediction for the degree of support of tree amplitudes in $\cN=4$ super Yang-Mills.  The resolution was based on the observation that, most fundamentally, the individual terms in the Drummond \& Henn solution for these N$^p$MHV tree amplitudes  should really be thought of as $p$-loop primitive leading singularities, in a channel that can be identified quite systematically from the KS figure.

Spurred on by this, we showed how performing generalized unitarity cuts in twistor space leads to expressions for arbitrary leading singularities in terms of gluing together tree-level amplitudes directly in twistor space. The resulting formula is naturally written as an integral over the moduli of nodal curves in twistor space, each of whose components is rational, with one node for each cut propagator and genus given by the loop order of the leading singularity.  In particular,
for primitive leading singularities the number of nodes is four times the genus.  We have examined how one might obtain such leading singularities by localising the path integral of some form of twistor-string (broadly defined) on such nodal curves.  Part of the data of such a twistor-string is a choice of holomorphic volume form on the moduli space of stable maps.  In parallel with the leading singularity conjecture in momentum space, we conjectured that the correct volume form
for a pure $\cN=4$ super Yang-Mills twistor-string may be determined by the condition that it has the singularities
and residues on the boundary components of the moduli space of stable maps necessary to be consistent with our construction of such integrals for leading singularities.

Given the recent conjecture of Arkani-Hamed {\it et al.}~\cite{ArkaniHamed:2009dn} that all N$^{k-2}$MHV leading singularities may be obtained from residue integrals of a standard meromorphic form on the Grassmann G$(k,n)$, it is natural to ask what relationship this has to the twistor-string inspired description of leading singularities. The answer is that there is a canonical map from the domain of a twistor-string path integral into this Grassmannian. When restricted
to primitive leading singularities, at least in the examples we have studied, the map is 1:1 and the twistor-string integrand identically matches the residue of the meromorphic form on the Grassmannian. This gives weight to the Grassmannian conjecture, in particular for all those primitive leading singularities that arise as
summands in the Drummond \& Henn solution of the BCFW recursion for tree amplitudes.

One can reverse the procedure and determine the nodal curve in twistor
space that arises from a given cycle on the Grassmannian. Combining
this with the translation between twistor support and momentum space
channel diagrams explained in section~\ref{sec:twsgenunit}, one can
determine the leading singularity that is associated to a given cycle,
without the need for a complete residue calculation.  We find that
only primitive leading singularities arise and we have argued that
N$^p$MHV amplitudes involve no new leading singularities beyond $3p$
loops.  Thus there is a considerable duplication amongst all
conceivable leading singularities.  This is consistent with the
experience so far acquired in calculating MHV amplitudes at higher
loops, where all the leading singularities that have been found are
simply the tree amplitude. It suggests that a similar pattern is true
for arbitrary N$^p$MHV amplitudes, but starting at $3p$ loops.
Although there is clearly much work required to complete the evidence
presented in this paper into a proof, the basic mechanisms are clear
and a proof is now in sight.

The close agreement between the twistor-string expectations and the Grassmannian residue formula is suggestive of a key role for this duality with the Grassmannian in the construction of a twistor-string theory for pure $\cN=4$ super Yang-Mills.  The pullback of the specific meromorphic form introduced in~\cite{ArkaniHamed:2009dn} possesses many of the properties one would want for the path integral measure of such a twistor-string, including manifest cyclic symmetry, superconformal invariance and --- importantly --- a simple pole on cycles in G$(k,n)$ that correspond to leading singularities (although it may well not be the unique such form). This cannot be the whole story since, for example, at $g$-loops, the map for MHV amplitudes loses $4g$ dimensions so the pullback will have the wrong degree and be degenerate in these directions.  Nevertheless, it seems likely that the Grassmannian duality should play a crucial role in the construction of a twistor-string theory for $\cN=4$ SYM, not just as a convenient means for recovering the momentum space amplitude, but as an integral part of the definition of the theory.

One of the many interesting aspects of the Grassmannian integral, emphasised by~\cite{ArkaniHamed:2009dn}, is that highly non-trivial identities between different leading singularities ({\it e.g.} those that follow from cyclic symmetry of the tree amplitude, or infra-red consistency conditions on the loop expansion) can be understood via global residue theorems in G$(k,n)$.  Combined with the duality between twistor-strings and G$(k,n)$, these residue theorems resolve a further puzzle of the KS configurations:  from the point of view of twistor geometry, why do such higher genus configurations contribute to the \emph{tree} amplitude?  (Note that the situation is rather different from the equivalence of genus zero twistor-strings and the MHV formalism. There, at N$^p$MHV both the twistor-string moduli space and the moduli space of MHV diagrams contain the moduli space of $p+1$ intersecting lines --- in a configuration with $g=0$ --- as a common boundary~\cite{Gukov:2004ei}. In the present case, while KS configurations lie at the boundary of $\overline{M}_{p,n}(\PT^*,2p+1)$, they are nowhere to be found in the $g=0$ moduli space.)  As mentioned above, in momentum space we expect that multi-loop leading singularities are related to the tree amplitude via IR consistency conditions. The fact that the twistor-string path integrals can all be mapped to the \emph{same} Grassmannian --- depending on the MHV degree but \emph{not} on the genus --- provides a natural way to understand these IR relations in terms of twistor geometry, and the use of the global residue theorems in~\cite{Spradlin:2009qr,Dolan:2009wf} should be seen in this context.


\vspace{1cm}

{\Large\bf\noindent Acknowledgements}

\medskip

\noindent We would like to thank Nima Arkani-Hamed, Freddy Cachazo,
Louise Dolan, Henriette Elvang, Andrew Hodges, Amit Sever, Mark
Spradlin, Cristian Vergu and Anastasia Volovich for useful
discussions. The work MB is supported by an STFC Postgraduate
Studentship. The work of DS is supported by the Perimeter Institute
for Theoretical Physics. Research at the Perimeter Institute is
supported by the Government of Canada through Industry Canada and by
the Province of Ontario through the Ministry of Research $\&$
Innovation. The work of LM and DS was financed in part by EPSRC grant number
EP/F016654, see also {\\ \tt
 http://gow.epsrc.ac.uk/ViewGrant.aspx?GrantRef=EP/F016654/1}. 
\\

\noindent{\it Note added: After an earlier version of this paper was submitted, we discovered that 
Jared Kaplan was also writing a paper with some overlap with
this one~\cite{Kaplan:2009mh}.}

\bibliographystyle{JHEP}
\bibliography{Twistoramplitudesref}

\end{document}